\documentclass[longauth]{aa}  

\usepackage{graphicx}
\usepackage{txfonts}
\usepackage{tabularx}
\usepackage{multirow}
\usepackage{booktabs}
\usepackage{float}
\usepackage{xcolor}
\usepackage{natbib}
\usepackage{comment}
\bibpunct{(}{)}{;}{a}{}{,} 
\usepackage[breaklinks, colorlinks, citecolor=blue, linkcolor=blue]{hyperref}
\makeatletter
\renewcommand*\aa@pageof{, page \thepage{} of \pageref*{LastPage}}
\makeatother

\newlength{\bibitemsep}
\newlength{\bibparskip}
\let\oldthebibliography\thebibliography
\renewcommand\thebibliography[1]{%
  \oldthebibliography{#1}%
  \setlength{\parskip}{\bibitemsep}%
  \setlength{\itemsep}{\bibparskip}%
}

\newcommand{\starname}{HIP~29442}

\begin{document} 

\title{Unveiling the internal structure and formation history of the three planets transiting \starname\ (TOI-469) with CHEOPS\thanks{This article uses data from the CHEOPS Guaranteed Time Observation programme CH\_PR100031.}
}
\titlerunning{\starname}

\author{
    J. A. Egger\inst{1} $^{\href{https://orcid.org/0000-0003-1628-4231}{\includegraphics[scale=0.5]{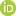}}}$, 
    H. P. Osborn\inst{2,3} $^{\href{https://orcid.org/0000-0002-4047-4724}{\includegraphics[scale=0.5]{figures/orcid.jpg}}}$, 
    D. Kubyshkina\inst{4}, 
    C. Mordasini\inst{1,2}, 
    Y. Alibert\inst{2,1} $^{\href{https://orcid.org/0000-0002-4644-8818}{\includegraphics[scale=0.5]{figures/orcid.jpg}}}$, 
    M. N. Günther\inst{5} $^{\href{https://orcid.org/0000-0002-3164-9086}{\includegraphics[scale=0.5]{figures/orcid.jpg}}}$, 
    M. Lendl\inst{6} $^{\href{https://orcid.org/0000-0001-9699-1459}{\includegraphics[scale=0.5]{figures/orcid.jpg}}}$, 
    A. Brandeker\inst{7} $^{\href{https://orcid.org/0000-0002-7201-7536}{\includegraphics[scale=0.5]{figures/orcid.jpg}}}$, 
    A. Heitzmann\inst{6} $^{\href{https://orcid.org/0000-0002-8091-7526}{\includegraphics[scale=0.5]{figures/orcid.jpg}}}$, 
    A. Leleu\inst{6,1} $^{\href{https://orcid.org/0000-0003-2051-7974}{\includegraphics[scale=0.5]{figures/orcid.jpg}}}$, 
    M. Damasso\inst{8} $^{\href{https://orcid.org/0000-0001-9984-4278}{\includegraphics[scale=0.5]{figures/orcid.jpg}}}$, 
    A. Bonfanti\inst{4} $^{\href{https://orcid.org/0000-0002-1916-5935}{\includegraphics[scale=0.5]{figures/orcid.jpg}}}$, 
    T. G. Wilson\inst{9} $^{\href{https://orcid.org/0000-0001-8749-1962}{\includegraphics[scale=0.5]{figures/orcid.jpg}}}$, 
    S. G. Sousa\inst{10} $^{\href{https://orcid.org/0000-0001-9047-2965}{\includegraphics[scale=0.5]{figures/orcid.jpg}}}$, 
    J. Haldemann\inst{1} $^{\href{https://orcid.org/0000-0003-1231-2389}{\includegraphics[scale=0.5]{figures/orcid.jpg}}}$, 
    L. Delrez\inst{11,12,13} $^{\href{https://orcid.org/0000-0001-6108-4808}{\includegraphics[scale=0.5]{figures/orcid.jpg}}}$, 
    M. J. Hooton\inst{14} $^{\href{https://orcid.org/0000-0003-0030-332X}{\includegraphics[scale=0.5]{figures/orcid.jpg}}}$, 
    T. Zingales\inst{15,16} $^{\href{https://orcid.org/0000-0001-6880-5356}{\includegraphics[scale=0.5]{figures/orcid.jpg}}}$, 
    R. Luque\inst{17}, 
    R. Alonso\inst{18,19} $^{\href{https://orcid.org/0000-0001-8462-8126}{\includegraphics[scale=0.5]{figures/orcid.jpg}}}$, 
    J. Asquier\inst{5}, 
    T. Bárczy\inst{20} $^{\href{https://orcid.org/0000-0002-7822-4413}{\includegraphics[scale=0.5]{figures/orcid.jpg}}}$, 
    D. Barrado Navascues\inst{21} $^{\href{https://orcid.org/0000-0002-5971-9242}{\includegraphics[scale=0.5]{figures/orcid.jpg}}}$, 
    S. C. C. Barros\inst{10,22} $^{\href{https://orcid.org/0000-0003-2434-3625}{\includegraphics[scale=0.5]{figures/orcid.jpg}}}$, 
    W. Baumjohann\inst{4} $^{\href{https://orcid.org/0000-0001-6271-0110}{\includegraphics[scale=0.5]{figures/orcid.jpg}}}$, 
    W. Benz\inst{1,2} $^{\href{https://orcid.org/0000-0001-7896-6479}{\includegraphics[scale=0.5]{figures/orcid.jpg}}}$, 
    N. Billot\inst{6} $^{\href{https://orcid.org/0000-0003-3429-3836}{\includegraphics[scale=0.5]{figures/orcid.jpg}}}$, 
    L. Borsato\inst{16} $^{\href{https://orcid.org/0000-0003-0066-9268}{\includegraphics[scale=0.5]{figures/orcid.jpg}}}$, 
    C. Broeg\inst{1,2} $^{\href{https://orcid.org/0000-0001-5132-2614}{\includegraphics[scale=0.5]{figures/orcid.jpg}}}$, 
    M. Buder\inst{23}, 
    A. Castro-González\inst{21} $^{\href{https://orcid.org/0000-0001-7439-3618}{\includegraphics[scale=0.5]{figures/orcid.jpg}}}$, 
    A. Collier Cameron\inst{24} $^{\href{https://orcid.org/0000-0002-8863-7828}{\includegraphics[scale=0.5]{figures/orcid.jpg}}}$, 
    A. C. M. Correia\inst{25}, 
    D. Cortes\inst{26} $^{\href{https://orcid.org/0009-0003-8464-1306}{\includegraphics[scale=0.5]{figures/orcid.jpg}}}$, 
    Sz. Csizmadia\inst{27} $^{\href{https://orcid.org/0000-0001-6803-9698}{\includegraphics[scale=0.5]{figures/orcid.jpg}}}$, 
    P. E. Cubillos\inst{28,4}, 
    M. B. Davies\inst{29} $^{\href{https://orcid.org/0000-0001-6080-1190}{\includegraphics[scale=0.5]{figures/orcid.jpg}}}$, 
    M. Deleuil\inst{30} $^{\href{https://orcid.org/0000-0001-6036-0225}{\includegraphics[scale=0.5]{figures/orcid.jpg}}}$, 
    A. Deline\inst{6}, 
    O. D. S. Demangeon\inst{10,22} $^{\href{https://orcid.org/0000-0001-7918-0355}{\includegraphics[scale=0.5]{figures/orcid.jpg}}}$, 
    B.-O. Demory\inst{2,1} $^{\href{https://orcid.org/0000-0002-9355-5165}{\includegraphics[scale=0.5]{figures/orcid.jpg}}}$, 
    A. Derekas\inst{31}, 
    B. Edwards\inst{32}, 
    D. Ehrenreich\inst{6,33} $^{\href{https://orcid.org/0000-0001-9704-5405}{\includegraphics[scale=0.5]{figures/orcid.jpg}}}$, 
    A. Erikson\inst{27}, 
    A. Fortier\inst{1,2} $^{\href{https://orcid.org/0000-0001-8450-3374}{\includegraphics[scale=0.5]{figures/orcid.jpg}}}$, 
    L. Fossati\inst{4} $^{\href{https://orcid.org/0000-0003-4426-9530}{\includegraphics[scale=0.5]{figures/orcid.jpg}}}$, 
    M. Fridlund\inst{34,35} $^{\href{https://orcid.org/0000-0002-0855-8426}{\includegraphics[scale=0.5]{figures/orcid.jpg}}}$, 
    D. Gandolfi\inst{36} $^{\href{https://orcid.org/0000-0001-8627-9628}{\includegraphics[scale=0.5]{figures/orcid.jpg}}}$, 
    K. Gazeas\inst{37}, 
    M. Gillon\inst{11} $^{\href{https://orcid.org/0000-0003-1462-7739}{\includegraphics[scale=0.5]{figures/orcid.jpg}}}$, 
    M. Güdel\inst{38}, 
    Ch. Helling\inst{4,39}, 
    K. G. Isaak\inst{5} $^{\href{https://orcid.org/0000-0001-8585-1717}{\includegraphics[scale=0.5]{figures/orcid.jpg}}}$, 
    L. L. Kiss\inst{40,41}, 
    J. Korth\inst{42} $^{\href{https://orcid.org/0000-0002-0076-6239}{\includegraphics[scale=0.5]{figures/orcid.jpg}}}$, 
    K. W. F. Lam\inst{27} $^{\href{https://orcid.org/0000-0002-9910-6088}{\includegraphics[scale=0.5]{figures/orcid.jpg}}}$, 
    J. Laskar\inst{43} $^{\href{https://orcid.org/0000-0003-2634-789X}{\includegraphics[scale=0.5]{figures/orcid.jpg}}}$, 
    B. Lavie\inst{6}, 
    A. Lecavelier des Etangs\inst{44} $^{\href{https://orcid.org/0000-0002-5637-5253}{\includegraphics[scale=0.5]{figures/orcid.jpg}}}$, 
    C. Lovis\inst{6}, 
    A. Luntzer\inst{38}, 
    D. Magrin\inst{16} $^{\href{https://orcid.org/0000-0003-0312-313X}{\includegraphics[scale=0.5]{figures/orcid.jpg}}}$, 
    P. F. L. Maxted\inst{45} $^{\href{https://orcid.org/0000-0003-3794-1317}{\includegraphics[scale=0.5]{figures/orcid.jpg}}}$, 
    B. Merín\inst{46} $^{\href{https://orcid.org/0000-0002-8555-3012}{\includegraphics[scale=0.5]{figures/orcid.jpg}}}$, 
    M. Munari\inst{47} $^{\href{https://orcid.org/0000-0003-0990-050X}{\includegraphics[scale=0.5]{figures/orcid.jpg}}}$, 
    V. Nascimbeni\inst{16} $^{\href{https://orcid.org/0000-0001-9770-1214}{\includegraphics[scale=0.5]{figures/orcid.jpg}}}$, 
    G. Olofsson\inst{7} $^{\href{https://orcid.org/0000-0003-3747-7120}{\includegraphics[scale=0.5]{figures/orcid.jpg}}}$, 
    R. Ottensamer\inst{38}, 
    I. Pagano\inst{47} $^{\href{https://orcid.org/0000-0001-9573-4928}{\includegraphics[scale=0.5]{figures/orcid.jpg}}}$, 
    E. Pallé\inst{18,19} $^{\href{https://orcid.org/0000-0003-0987-1593}{\includegraphics[scale=0.5]{figures/orcid.jpg}}}$, 
    G. Peter\inst{23} $^{\href{https://orcid.org/0000-0001-6101-2513}{\includegraphics[scale=0.5]{figures/orcid.jpg}}}$, 
    D. Piazza\inst{1}, 
    G. Piotto\inst{16,15} $^{\href{https://orcid.org/0000-0002-9937-6387}{\includegraphics[scale=0.5]{figures/orcid.jpg}}}$, 
    D. Pollacco\inst{9}, 
    D. Queloz\inst{48,14} $^{\href{https://orcid.org/0000-0002-3012-0316}{\includegraphics[scale=0.5]{figures/orcid.jpg}}}$, 
    R. Ragazzoni\inst{16,15} $^{\href{https://orcid.org/0000-0002-7697-5555}{\includegraphics[scale=0.5]{figures/orcid.jpg}}}$, 
    N. Rando\inst{5}, 
    H. Rauer\inst{27,49} $^{\href{https://orcid.org/0000-0002-6510-1828}{\includegraphics[scale=0.5]{figures/orcid.jpg}}}$, 
    I. Ribas\inst{50,51} $^{\href{https://orcid.org/0000-0002-6689-0312}{\includegraphics[scale=0.5]{figures/orcid.jpg}}}$, 
    J. Rodrigues\inst{52,53}, 
    N. C. Santos\inst{10,22} $^{\href{https://orcid.org/0000-0003-4422-2919}{\includegraphics[scale=0.5]{figures/orcid.jpg}}}$, 
    G. Scandariato\inst{47} $^{\href{https://orcid.org/0000-0003-2029-0626}{\includegraphics[scale=0.5]{figures/orcid.jpg}}}$, 
    D. Ségransan\inst{6} $^{\href{https://orcid.org/0000-0003-2355-8034}{\includegraphics[scale=0.5]{figures/orcid.jpg}}}$, 
    A. E. Simon\inst{1,2} $^{\href{https://orcid.org/0000-0001-9773-2600}{\includegraphics[scale=0.5]{figures/orcid.jpg}}}$, 
    A. M. S. Smith\inst{27} $^{\href{https://orcid.org/0000-0002-2386-4341}{\includegraphics[scale=0.5]{figures/orcid.jpg}}}$, 
    M. Stalport\inst{12,11}, 
    S. Sulis\inst{30} $^{\href{https://orcid.org/0000-0001-8783-526X}{\includegraphics[scale=0.5]{figures/orcid.jpg}}}$, 
    Gy. M. Szabó\inst{31,54} $^{\href{https://orcid.org/0000-0002-0606-7930}{\includegraphics[scale=0.5]{figures/orcid.jpg}}}$, 
    S. Udry\inst{6} $^{\href{https://orcid.org/0000-0001-7576-6236}{\includegraphics[scale=0.5]{figures/orcid.jpg}}}$, 
    V. Van Grootel\inst{12} $^{\href{https://orcid.org/0000-0003-2144-4316}{\includegraphics[scale=0.5]{figures/orcid.jpg}}}$, 
    J. Venturini\inst{6} $^{\href{https://orcid.org/0000-0001-9527-2903}{\includegraphics[scale=0.5]{figures/orcid.jpg}}}$, 
    E. Villaver\inst{18,19}, 
    N. A. Walton\inst{55} $^{\href{https://orcid.org/0000-0003-3983-8778}{\includegraphics[scale=0.5]{figures/orcid.jpg}}}$
}

\authorrunning{J. A. Egger et al.}

\institute{
    \label{inst:1} Weltraumforschung und Planetologie, Physikalisches Institut, University of Bern, Gesellschaftsstrasse 6, 3012 Bern, Switzerland\\
    \email{jo-ann.egger@unibe.ch} \and
    \label{inst:2} Center for Space and Habitability, University of Bern, Gesellschaftsstrasse 6, 3012 Bern, Switzerland \and
    \label{inst:3} Department of Physics and Kavli Institute for Astrophysics and Space Research, Massachusetts Institute of Technology, Cambridge, MA 02139, USA \and
    \label{inst:4} Space Research Institute, Austrian Academy of Sciences, Schmiedlstrasse 6, A-8042 Graz, Austria \and
    \label{inst:5} European Space Agency (ESA), European Space Research and Technology Centre (ESTEC), Keplerlaan 1, 2201 AZ Noordwijk, The Netherlands \and
    \label{inst:6} Observatoire astronomique de l'Université de Genève, Chemin Pegasi 51, 1290 Versoix, Switzerland \and
    \label{inst:7} Department of Astronomy, Stockholm University, AlbaNova University Center, 10691 Stockholm, Sweden \and
    \label{inst:8} INAF - Osservatorio Astrofisico di Torino, Via Osservatorio 20, I-10025 Pino Torinese, Italy \and
    \label{inst:9} Department of Physics, University of Warwick, Gibbet Hill Road, Coventry CV4 7AL, United Kingdom \and
    \label{inst:10} Instituto de Astrofisica e Ciencias do Espaco, Universidade do Porto, CAUP, Rua das Estrelas, 4150-762 Porto, Portugal \and
    \label{inst:11} Astrobiology Research Unit, Université de Liège, Allée du 6 Août 19C, B-4000 Liège, Belgium \and
    \label{inst:12} Space sciences, Technologies and Astrophysics Research (STAR) Institute, Université de Liège, Allée du 6 Août 19C, 4000 Liège, Belgium \and
    \label{inst:13} Institute of Astronomy, KU Leuven, Celestijnenlaan 200D, 3001 Leuven, Belgium \and
    \label{inst:14} Cavendish Laboratory, JJ Thomson Avenue, Cambridge CB3 0HE, UK \and
    \label{inst:15} Dipartimento di Fisica e Astronomia "Galileo Galilei", Università degli Studi di Padova, Vicolo dell'Osservatorio 3, 35122 Padova, Italy \and
    \label{inst:16} INAF, Osservatorio Astronomico di Padova, Vicolo dell'Osservatorio 5, 35122 Padova, Italy \and
    \label{inst:17} Department of Astronomy \& Astrophysics, University of Chicago, Chicago, IL 60637, USA \and
    \label{inst:18} Instituto de Astrofísica de Canarias, Vía Láctea s/n, 38200 La Laguna, Tenerife, Spain \and
    \label{inst:19} Departamento de Astrofísica, Universidad de La Laguna, Astrofísico Francisco Sanchez s/n, 38206 La Laguna, Tenerife, Spain \and
    \label{inst:20} Admatis, 5. Kandó Kálmán Street, 3534 Miskolc, Hungary \and
    \label{inst:21} Depto. de Astrofísica, Centro de Astrobiología (CSIC-INTA), ESAC campus, 28692 Villanueva de la Cañada (Madrid), Spain \and
    \label{inst:22} Departamento de Fisica e Astronomia, Faculdade de Ciencias, Universidade do Porto, Rua do Campo Alegre, 4169-007 Porto, Portugal \and
    \label{inst:23} Institute of Optical Sensor Systems, German Aerospace Center (DLR), Rutherfordstrasse 2, 12489 Berlin, Germany \and
    \label{inst:24} Centre for Exoplanet Science, SUPA School of Physics and Astronomy, University of St Andrews, North Haugh, St Andrews KY16 9SS, UK \and
    \label{inst:25} CFisUC, Department of Physics, University of Coimbra, 3004-516 Coimbra, Portugal \and
    \label{inst:26} AIRBUS DS Madrid, Paseo John Lennon 2, 28906 Getafe (Madrid), Spain \and
    \label{inst:27} Institute of Planetary Research, German Aerospace Center (DLR), Rutherfordstrasse 2, 12489 Berlin, Germany \and
    \label{inst:28} INAF, Osservatorio Astrofisico di Torino, Via Osservatorio, 20, I-10025 Pino Torinese To, Italy \and
    \label{inst:29} Centre for Mathematical Sciences, Lund University, Box 118, 221 00 Lund, Sweden \and
    \label{inst:30} Aix Marseille Univ, CNRS, CNES, LAM, 38 rue Frédéric Joliot-Curie, 13388 Marseille, France \and
    \label{inst:31} ELTE Gothard Astrophysical Observatory, 9700 Szombathely, Szent Imre h. u. 112, Hungary \and
    \label{inst:32} SRON Netherlands Institute for Space Research, Niels Bohrweg 4, 2333 CA Leiden, Netherlands \newpage \and
    \label{inst:33} Centre Vie dans l’Univers, Faculté des sciences, Université de Genève, Quai Ernest-Ansermet 30, 1211 Genève 4, Switzerland \and
    \label{inst:34} Leiden Observatory, University of Leiden, PO Box 9513, 2300 RA Leiden, The Netherlands \and
    \label{inst:35} Department of Space, Earth and Environment, Chalmers University of Technology, Onsala Space Observatory, 439 92 Onsala, Sweden \and
    \label{inst:36} Dipartimento di Fisica, Università degli Studi di Torino, via Pietro Giuria 1, I-10125, Torino, Italy \and
    \label{inst:37} National and Kapodistrian University of Athens, Department of Physics, University Campus, Zografos GR-157 84, Athens, Greece \and
    \label{inst:38} Department of Astrophysics, University of Vienna, Türkenschanzstrasse 17, 1180 Vienna, Austria \and
    \label{inst:39} Institute for Theoretical Physics and Computational Physics, Graz University of Technology, Petersgasse 16, 8010 Graz, Austria \and
    \label{inst:40} Konkoly Observatory, Research Centre for Astronomy and Earth Sciences, 1121 Budapest, Konkoly Thege Miklós út 15-17, Hungary \and
    \label{inst:41} ELTE E\"otv\"os Lor\'and University, Institute of Physics, P\'azm\'any P\'eter s\'et\'any 1/A, 1117 Budapest, Hungary \and
    \label{inst:42} Lund Observatory, Division of Astrophysics, Department of Physics, Lund University, Box 118, 22100 Lund, Sweden \and
    \label{inst:43} IMCCE, UMR8028 CNRS, Observatoire de Paris, PSL Univ., Sorbonne Univ., 77 av. Denfert-Rochereau, 75014 Paris, France \and
    \label{inst:44} Institut d'astrophysique de Paris, UMR7095 CNRS, Université Pierre \& Marie Curie, 98bis blvd. Arago, 75014 Paris, France \and
    \label{inst:45} Astrophysics Group, Lennard Jones Building, Keele University, Staffordshire, ST5 5BG, United Kingdom \and
    \label{inst:46} European Space Agency, ESA - European Space Astronomy Centre, Camino Bajo del Castillo s/n, 28692 Villanueva de la Cañada, Madrid, Spain \and
    \label{inst:47} INAF, Osservatorio Astrofisico di Catania, Via S. Sofia 78, 95123 Catania, Italy \and
    \label{inst:48} ETH Zurich, Department of Physics, Wolfgang-Pauli-Strasse 2, CH-8093 Zurich, Switzerland \and
    \label{inst:49} Institut fuer Geologische Wissenschaften, Freie Universitaet Berlin, Maltheserstrasse 74-100,12249 Berlin, Germany \and
    \label{inst:50} Institut de Ciencies de l'Espai (ICE, CSIC), Campus UAB, Can Magrans s/n, 08193 Bellaterra, Spain \and
    \label{inst:51} Institut d'Estudis Espacials de Catalunya (IEEC), 08860 Castelldefels (Barcelona), Spain \and
    \label{inst:52} Instituto de Astrofísica e Ciências do Espaço, Universidade do Porto, CAUP, Rua das Estrelas, 4150-762 Porto, Portugal \and
    \label{inst:53} Departamento de Física e Astronomia, Faculdade de Ciências, Universidade do Porto, Rua do Campo Alegre, 4169-007 Porto, Portugal \and
    \label{inst:54} HUN-REN-ELTE Exoplanet Research Group, Szent Imre h. u. 112., Szombathely, H-9700, Hungary \and
    \label{inst:55} Institute of Astronomy, University of Cambridge, Madingley Road, Cambridge, CB3 0HA, United Kingdom
}

\date{Received 22 April 2024; accepted 25 June 2024}

\abstract{
Multiplanetary systems spanning the radius valley are ideal testing grounds for exploring the different proposed explanations for the observed bimodality in the radius distribution of close-in exoplanets. One such system is \starname\ (TOI-469), an evolved K0V star hosting two super-Earths and one sub-Neptune.
We observe \starname \ with CHEOPS for a total of 9.6 days, which we model jointly with 2 sectors of TESS data to derive planetary radii of $3.410\pm0.046$, $1.551\pm0.045$ and $1.538\pm0.049$ R$_\oplus$ for planets~b, c and d, which orbit \starname\ with periods of 13.6, 3.5 and 6.4 days, respectively. For planet~d, this value deviates by more than 3$\sigma$ from the median value reported in the discovery paper, leading us to conclude that caution is required when using TESS photometry to determine the radii of small planets with low per-transit signal-to-noise ratios and large gaps between observations. Given the high precision of these new radii, combining them with published RVs from ESPRESSO and HIRES provides us with ideal conditions to investigate the internal structure and formation pathways of the planets in the system. We introduce the publicly available code \texttt{plaNETic}, a fast and robust neural network-based Bayesian internal structure modelling framework. We then apply hydrodynamic models to explore the upper atmospheric properties of these inferred structures. Finally, we identify planetary system analogues in a synthetic population generated with the Bern model for planet formation and evolution. Based on this analyis, we find that the planets likely formed on opposing sides of the water iceline from a protoplanetary disk with an intermediate solid mass. We finally report that the observed parameters of the \starname \ system are compatible with both a scenario where the second peak in the bimodal radius distribution corresponds to sub-Neptunes with a pure H/He envelope as well as a scenario with water-rich sub-Neptunes.
}


\maketitle

%
\section{Introduction}
\label{sec:introduction}
One of the most notable features in the observed exoplanet population is the so-called radius valley, an underdensity of small exoplanets with radii between $\sim$1.5 and $\sim$2 R$_\oplus$ at orbital periods of less than 100 days.
This bimodality in the radius distribution with two distinct peaks at $\sim$1.3 and $\sim$2.4 R$_\oplus$, with the planets corresponding to the two peaks commonly referred to as super-Earths and sub-Neptunes, was first observationally confirmed by the California-Kepler Survey \citep{Fulton+2017}, after a so-called photoevaporation valley had already been predicted theoretically by several groups \cite[e.g.][]{Owen+Wu2013,Jin+2014,Lopez+Fortney2014}.

The origin of the radius valley is still heavily debated in the literature, and multiple possible explanations have been brought forward. On the one hand, it has been proposed that the observed gap in the radius distribution is caused by planets being stripped of their H/He envelopes through an atmospheric mass loss mechanism and remaining as bare rocky cores, with photoevaporation via high-energy stellar irradiation \citep{Owen+Wu2017,Jin+Mordasini2018} and core-powered mass loss
\citep{Ginzburg+2018,Gupta+Schlichting2019} as the most commonly presented scenarios. On the other hand, planet formation predicts water-rich planets to migrate inwards from beyond the iceline \citep[e.g.][]{Tanaka+2002}, which is inconsistent with a picture where the majority of sub-Neptunes are water-poor and feature pure H/He envelopes. It has therefore been proposed that the second peak of the observed radius distribution is instead populated by planets that are mostly water-rich. 

While this hypothesis was first presented based on mass-radius curves for water-rich planets \citep{Sotin+2007, Valencia+2013, Zeng+2019, Turbet+2020, Mousis+2020}, \cite{Venturini+2020b,Venturini+2020a} then showed for the first time that the observed location of the radius valley can be reproduced using pebble-based combined formation and evolution models when treating water in vapour form uniformly mixed with H/He, with their newer work expanding this for M-dwarf stars \citep{Venturini+2024}. \cite{Izidoro+2022} reach similar conclusions for planets with water in condensed form but also including the gravitational interaction between planets. More recently, \cite{Burn+2024} have been able to naturally reproduce the observed location of the radius valley using a planetesimal-based combined planet formation and evolution model and again assuming water to be in gaseous form mixed with H/He, supporting the scenario of water-rich sub-Neptunes with supercritical steam atmospheres.

Distinguishing between these different scenarios observationally is challenging, as the information we have about these planets is generally very limited. Even if both the mass and radius of a given planet are known in addition to its orbital period, determining the internal structure and composition of the planet is still a highly degenerate problem \citep[e.g.][]{Seager+2007, Rogers+Seager2010}. A wide variety of internal structure modelling approaches can be found in the literature \citep[e.g.][]{Brugger+2017, Dorn+2017, Baumeister+2020, Acuna+2021, Vazan+2022, Unterborn+2023, Haldemann+2023,  Haldemann+2024}, with many of them coupling the internal structure model with Bayesian inference or machine learning approaches in order to determine a probability distribution for the interior composition of an exoplanet. However, these methods only allow us to quantify the degenerate nature of the problem, while the actual degeneracy cannot be broken without additional observational data, such as spectroscopic observations of the planet's atmosphere. Observational evidence for water-rich sub-Neptunes has been presented by \cite{Luque+Palle2022} for planets around M-dwarfs, who show that many planets in their sample are consistent with a 1:1 water-to-rock ratio. However, \cite{Rogers+2023} point out that these observations can also be explained by planets with H/He dominated atmospheres. Furthermore, recent analyses of JWST transmission spectra of sub-Neptunes find high atmospheric metallicities \citep{Benneke+2024, Holmberg+Madhusudhan2024, Madhusudhan+2023}.

When studying the different explanations for the radius valley, multiplanetary systems with planets on either side of the radius valley provide us with an ideal testing ground for the different hypotheses, as they were formed from the same protoplanetary disk and orbit the same star. Currently there are only few such systems with well characterised masses and radii. A search of the NASA Exoplanet Archive\footnote{\url{https://exoplanetarchive.ipac.caltech.edu/}, query from 5~June 2024} revealed 16 radius valley crossing systems with mass errors smaller than 25\% and radius errors smaller than 10\%. Of these, only six systems have mass errors smaller than 15\% and radius errors smaller than 5\% ($\nu^2$ Lupi, \citealp{Delrez+2021, Ehrenreich+2023}; Kepler-10, \citealp{Batalha+2011,Fressin+2011,Bonomo+2023}; Kepler-36, \citealp{Carter+2012, Vissapragada+2020}; LTT 3780, \citealp{Cloutier+2020, Bonfanti+2024}; TOI-1468, \citealp{Chaturvedi+2022}; TOI-561, \citealp{Lacedelli+2022, Patel+2023}). Of particular interest is the TOI-178 system \citep{Leleu+2021, Delrez+2023} with six known transiting planets, two super-Earths and four sub-Neptunes. Multiple studies have investigated the internal structure of the planets in this system \citep{Leleu+2021, Acuna+2022}, and some of its planets have recently been observed with JWST \citep{Hooton+2021}.

Another multiplanetary system with planets on both sides of the radius valley is \starname\ (TOI-469), which consists of three known transiting planets, two super-Earths and an outer sub-Neptune, orbiting an evolved K0 type star on short orbits. The outer sub-Neptune, \starname~b, was first identified as a planetary candidate (TOI-469.01) in data from the Transiting Exoplanet Survey Satellite \citep[TESS; ][]{Ricker+2015} and later validated by \cite{Giacalone+2021}, while the two inner super-Earths, \starname~c and d, were discovered using data from the Echelle SPectrograph for Rocky Exoplanets and Stable Spectroscopic Observations \citep[ESPRESSO; ][]{Pepe+2021} as presented by \cite{Damasso+2023}, who then searched for and found the corresponding transits in the TESS light curves. However, as already pointed out by \cite{Damasso+2023}, these additional transits are extremely shallow, which complicates an accurate determination of the radii for these two planets.

Here we present the results of an extensive photometric follow-up campaign of \starname \ using the CHaracterising ExOPlanet Satellite \citep[CHEOPS; ][]{Benz+2021,Fortier+2024}, driven by the discovery of the additional shallow transits for \starname~c and d in the TESS data. After newly determining values for the host star radius, mass and age in Section~\ref{sec:star}, we present the collected CHEOPS light curves, analyse all available data for this system and summarise our derived planetary parameters in Section~\ref{sec:data+analysis}. In this section, we also compare to the values of \cite{Damasso+2023} and discuss how \starname~b, c and d fit into the larger exoplanet population given our newly determined radius and mass values for all three planets.

In the subsequent sections, we analyse how the \starname \ system fits in with the individual scenarios brought forward to explain the bimodality of the observed radius distribution. In order to do that, we present our improved neural network-based Bayesian internal structure modelling framework \texttt{plaNETic} in Section~\ref{sec:internal_structure} and apply it to the \starname \ planets. In Section~\ref{sec:hydro_modelling}, we employ the 1D hydrodynamic model CHAIN \citep{kubyshkina2024} to study the properties of the upper atmosphere and the atmospheric mass loss for the atmospheric compositions inferred in the previous section. Section~\ref{sec:formation_evolution} then discusses possible formation and evolution pathways of the \starname \ system by identifying planetary system analogues in a variation of the nominal population of the New Generation Planetary Population Synthesis \citep{Emsenhuber+2021a,Emsenhuber+2021b,Emsenhuber+2023,Burn+2024} and analysing the common properties of these synthetic systems. We discuss the outcome and limitations of these different analyses in Section~\ref{sec:discussion}, before finally drawing conclusions and summarising our findings in Section~\ref{sec:conclusion}.

\section{Host star characterisation}
\label{sec:star}

\begin{table}
\renewcommand{\arraystretch}{1.2}
\caption{Stellar properties of \starname.}
\centering
\begin{tabular}{llr}
\hline\hline
\multicolumn{3}{l}{Star names} \\
\hline
\multicolumn{1}{l}{HIP} & \multicolumn{2}{l}{29442} \\
\multicolumn{1}{l}{HD} & \multicolumn{2}{l}{42813} \\
\multicolumn{1}{l}{TOI} & \multicolumn{2}{l}{469} \\
\multicolumn{1}{l}{TIC} & \multicolumn{2}{l}{33692729} \\
\multicolumn{1}{l}{TYC} & \multicolumn{2}{l}{5374-00643-1} \\
\multicolumn{1}{l}{Gaia DR3} & \multicolumn{2}{l}{2993561629444856960} \\
\hline
Parameter & Value & Source \\
\hline
$\alpha$ (J2000) & 06h 12m 13.97s & [1] \\
$\delta$ (J2000) & -14$^\circ$ 39' 00.06'' & [1] \\
G mag & 9.282 $\pm$ 0.003 & [1] \\
T$_\textrm{eff}$ [K] & 5289 $\pm$ 69 & [2] \\
log g [cgs] & 4.39 $\pm$ 0.03 & [2] \\
$[\mathrm{Fe/H}]$ [dex] & 0.24 $\pm$ 0.05 & [2] \\
$[\mathrm{Mg/H}]$ [dex] & 0.26 $\pm$ 0.04 & [2] \\
$[\mathrm{Si/H}]$ [dex] & 0.21 $\pm$ 0.04 & [2] \\
R$_\star$ [R$_\odot$] & 0.980 $\pm$ 0.007 & [3] \\
M$_\star$ [M$_\odot$] & 0.901 $\pm$ 0.043 & [3] \\
t$_\star$ [Gyr] & 11.2 $\pm$ 3.4 & [3] \\
\hline
\end{tabular}
\label{tab:stellar_params}
\tablebib{
[1]~\cite{GaiaCollab2021}; [2] \citet{Damasso+2023}; [3] This work.
}
\end{table}
\renewcommand{\arraystretch}{1}

For the spectroscopic stellar parameters of \starname, we used the values derived by \cite{Damasso+2023}. Their values for the effective temperature, surface gravity and stellar abundances ([Fe/H], [Mg/H] and [Si/H]) are summarised in Table \ref{tab:stellar_params}. They were consistent with our own additional analysis using the spectral analysis package Spectroscopy Made Easy \citep[SME; ][]{Valenti+Piskunov1996, Piskunov+Valenti2017}.

We then utilised a MCMC modified infrared flux method \citep{Blackwell1977,Schanche2020} to determine the stellar radius of HIP\,29442. The stellar spectral parameters and 1$\sigma$ uncertainties from \cite{Damasso+2023} were used as priors in constructing spectral energy distributions (SEDs) using stellar atmospheric models from two catalogues \citep{Kurucz1993,Castelli2003}. Synthetic photometry was computed from these SEDs and compared to the observed broadband photometry in the following bandpasses: \textit{Gaia} $G$, $G_\mathrm{BP}$, and $G_\mathrm{RP}$, 2MASS $J$, $H$, and $K$, and \textit{WISE} $W1$ and $W2$ \citep{Skrutskie2006,Wright2010,GaiaCollaboration2022} to derive the stellar bolometric flux. We used the Stefan-Boltzmann law to convert the bolometric flux to stellar effective temperature and angular diameter. Using the offset-corrected \textit{Gaia} parallax \citep{Lindegren2021}, we then converted the angular diameter into stellar radius. This process was conducted for both atmospheric catalogues with the produced radius posterior distributions combined via a Bayesian modelling averaging in order to correct for atmospheric model uncertainties. Our weighted average radius value $R_{\star}$ is reported in Table~\ref{tab:stellar_params}.

We further determined the isochronal mass~$M_{\star}$ and age~$t_{\star}$ of HIP\,29442 by inputting $T_{\mathrm{eff}}$, [Fe/H], and $R_{\star}$ along with their uncertainties in the isochrone placement routine by \citet{bonfanti2015,bonfanti2016}. In short, the algorithm computes $M_{\star}$ and $t_{\star}$ following interpolation of the input parameters within pre-computed grids of PARSEC\footnote{\textsl{PA}dova and T\textsl{R}ieste \textsl{S}tellar \textsl{E}volutionary \textsl{C}ode: \url{http://stev.oapd.inaf.it/cgi-bin/cmd}} v1.2S \citep{marigo2017} isochrones and evolutionary tracks. As justified in \citet{bonfanti2021}, we inflated the internal uncertainties on mass and age by 4\% and 20\%, respectively, to account for isochrone systematics and we obtained $M_{\star}=0.901\pm0.043~M_{\odot}$ and $t_{\star}=11.2\pm3.4$~Gyr.

\section{Observations and data analysis}
\label{sec:data+analysis}

\subsection{Photometry}

\begin{table*}
    \caption{CHEOPS observation log for \starname.}
    \centering
    \begin{tabular}{ccccccc}
    \toprule
    \toprule
    ID & Start Date [UTC] & Dur [hours] & File Key & Av. eff. [\%] & RMS [ppm] & Planets\\
    \midrule
    1 & 2022-11-13T09:51:58.566 & $ 13.23 $ & PR100031\_TG054401\_V0300 & $ 56 $ & $ 16.6 $ & c\\
    2 & 2022-11-16T15:20:12.607 & $ 22.07 $ & PR100031\_TG054501\_V0300 & $ 56 $ & $ 12.1 $ & c, d\\
    3 & 2022-11-22T01:33:34.389 & $ 18.18 $ & PR100031\_TG054301\_V0300 & $ 57 $ & $ 18.9 $ & b\\
    4 & 2022-11-27T14:08:14.891 & $ 10.13 $ & PR100031\_TG054402\_V0300 & $ 62 $ & $ 22.9 $ & c\\
    5 & 2022-11-29T09:56:19.658 & $ 13.76 $ & PR100031\_TG054502\_V0300 & $ 66 $ & $ 17.7 $ & d\\
    6 & 2022-12-05T15:56:56.577 & $ 18.64 $ & PR100031\_TG054302\_V0300 & $ 72 $ & $ 12.3 $ & b, d\\
    7 & 2022-12-22T06:24:00.725 & $ 11.45 $ & PR100031\_TG054403\_V0300 & $ 62 $ & $ 12.9 $ & c\\
    8 & 2022-12-25T04:12:40.557 & $ 14.84 $ & PR100031\_TG054701\_V0300 & $ 65 $ & $ 23.4 $ & d\\
    9 & 2022-12-25T19:17:41.222 & $ 10.79 $ & PR100031\_TG054601\_V0300 & $ 69 $ & $ 17.5 $ & c\\
    10 & 2022-12-29T08:57:59.537 & $ 10.88 $ & PR100031\_TG054602\_V0300 & $ 70 $ & $ 17.9 $ & c\\
    11 & 2022-12-31T13:33:58.633 & $ 14.24 $ & PR100031\_TG054702\_V0300 & $ 67 $ & $ 11.7 $ & d\\
    12 & 2023-01-02T00:54:58.150 & $ 11.98 $ & PR100031\_TG054603\_V0300 & $ 73 $ & $ 19.1 $ & b, c\\
    13 & 2023-01-06T23:41:52.550 & $ 12.97 $ & PR100031\_TG054703\_V0300 & $ 75 $ & $ 14.5 $ & d\\
    14 & 2023-01-12T11:51:42.819 & $ 10.55 $ & PR100031\_TG054604\_V0300 & $ 73 $ & $ 18.3 $ & c\\
    15 & 2023-01-19T22:52:03.617 & $ 14.95 $ & PR100031\_TG054901\_V0300 & $ 62 $ & $ 26.0 $ & d\\
    16 & 2023-02-06T06:16:57.508 & $ 10.49 $ & PR100031\_TG054801\_V0300 & $ 57 $ & $ 13.6 $ & c\\
    17 & 2023-02-08T04:11:27.804 & $ 12.37 $ & PR100031\_TG054902\_V0300 & $ 56 $ & $ 22.6 $ & d\\
    \bottomrule
    \end{tabular}
    \label{tab:file_keys}
\end{table*}

\begin{figure*}
    \centering
    \includegraphics[width=0.95\textwidth]{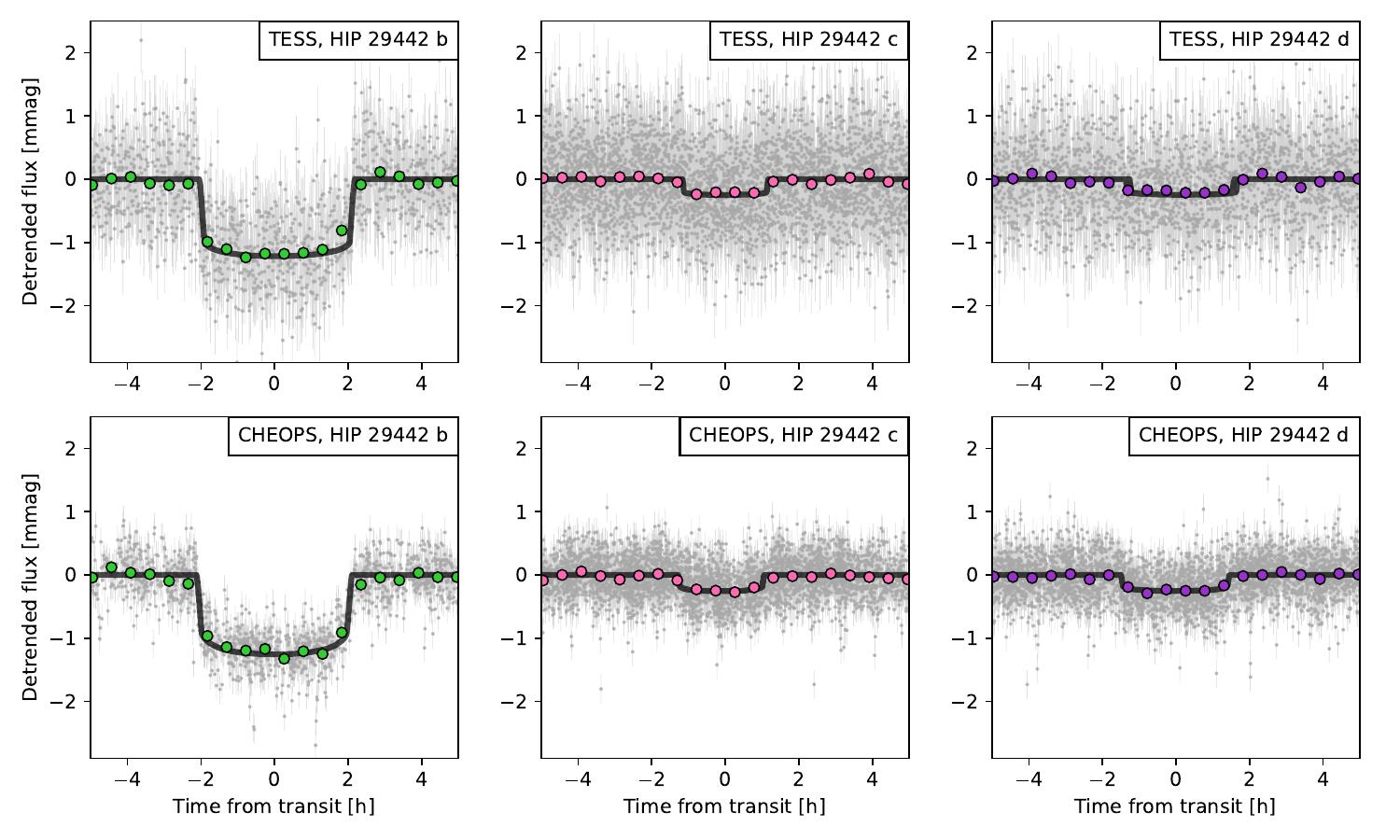}
    \caption{Detrended TESS (top row) and CHEOPS (bottom row) light curves, phase-folded to the orbital periods of \starname~b (left), c (middle) and d (right). The coloured points show the 30-min binned data, the solid black lines show the best-fit transit models.}
    \label{fig:transits}
\end{figure*}

\begin{table*}
\renewcommand{\arraystretch}{1.4}
\caption{Posterior distributions of the planetary parameters for planets b, c and d.}
\centering
\begin{tabular}{lccc}
\hline\hline
Parameter & \starname~b & \starname~c & \starname~d \\
\hline
Epoch, $t_0$ [BJD-2457000] & 2458474.56927 $\pm$ 0.00084 & 2458471.3600 $\pm$ 0.0033 & 2458472.9865 $\pm$ 0.0051 \\
Period, $P$ [d] & 13.6308205 $\pm$ 0.0000090 & 3.5379559 $\pm$ 0.0000082 & 6.429575 $\pm$ 0.000026 \\
Semi-major axis, $a$ [AU] & 0.1110 $\pm$ 0.0020 &  0.04518 $\pm$ 0.00082 & 0.0673 $\pm$ 0.0012 \\
Semi-major axis over stellar radius, $a/R_s$ & 23.88 $\pm$ 0.43 & 9.72 $\pm$ 0.18 & 14.5 $\pm$ 0.26 \\
Transit duration, $t_D$ [h] & 4.327 $\pm$ 0.022 & 2.429 $\pm$ 0.060 & 3.08 $\pm$ 0.12 \\
Radius ratio, $R_p/R_s$ & 0.03195 $\pm$ 0.00032 & 0.01453 $\pm$ 0.00040 & 0.01440 $\pm$ 0.00044 \\
Radius, $R_p$ [R$_\oplus$] & 3.410 $\pm$ 0.046 & 1.551 $\pm$ 0.045 & 1.538 $\pm$ 0.049 \\
Incident flux, $S_p$ [kW\,m$^{-2}$] & 74.7 $\pm$ 5.1 & 451 $\pm$ 31 & 203 $\pm$ 14 \\
Equilibrium temp., $T_{eq}$ [K] & 716 $\pm$ 12 & 1123 $\pm$ 19 & 920 $\pm$ 16 \\
Impact parameter, $b$ & 0.273 $\pm$ 0.072 & 0.514 $\pm$ 0.047 & 0.443 $\pm$ 0.090 \\
\hline
RV semi-amplitude $K$ [m\,s$^{-1}$] & $2.63^{+0.20}_{-0.19}$ & 2.04 $\pm$ 0.11 & 1.91 $\pm$ 0.12 \\
Mass $M_p$ [M$_\oplus$] & $9.10^{+0.82}_{-0.79}$ & 4.50 $\pm$ 0.32 & 5.14 $\pm$ 0.41 \\
\hline
\end{tabular}
\label{tab:planetary_params}
\end{table*}
\renewcommand{\arraystretch}{1}

TESS observed \starname \ in two sectors, once in December 2018 (Sector 6) and once in December 2020 (Sector 33), at a two-minute cadence on camera 2. In addition, \starname \ was then observed by CHEOPS (program CH\_PR100031) as part of the Guaranteed Time Observations for a total of 231.5 hours, split into 17 visits between 2022-11-13 and 2023-02-08 during which 3 transits of planet~b, 9 transits of planet~c and 8 transits of planet~d were observed. An observation log of all available CHEOPS data is provided in Table \ref{tab:file_keys}. The undetrended CHEOPS light curves are visualised in Figure \ref{fig:undetrended_CHEOPS} in the appendix.

We used the publicly available code \texttt{chexoplanet}\footnote{\url{https://github.com/hposborn/chexoplanet}} which jointly models both the TESS and CHEOPS photometry using the \texttt{exoplanet} library \citep{exoplanet:joss}.
For the TESS photometry, we used the PDCSAP photometry with two-minute cadence. 
Long-timescale trends, either systematic or from stellar rotation, were further removed using a cubic spline with breakpoints spacing of 0.9 days and in-transit data masked.

Since CHEOPS is in a nadir-locked orbit around the Earth, its field of view rotates around the target star once per 99-minute orbit. In combination with each background star having strongly asymmetric PSF \citep{Benz+2021}, this can introduce flux modulation with roll angle. To reduce this modulation we used PIPE\footnote{\url{https://github.com/alphapsa/PIPE}} \citep{bra24} to extract PSF photometry \citep[also see][for more details on PIPE]{mor21, sza21, ser22}. To further take into account residual modulation, we used \texttt{chexoplanet} with two systematic removal steps.
The first is to model linear and quadratic decorrelation of the flux with respect to various hyperparameters.
In order to determine which decorrelation parameters are justified, \texttt{chexoplanet} first fits each individual CHEOPS light curve using all available hyper-parameter timeseries, which are normalised to have a mean $\mu=0.0$ and a standard deviation $\sigma=1.0$.
This includes measured x and y centroid position, the first three aliases of the $\cos$ and $\sin$ of the roll-angle \citep[for a detailed description of the roll-angle treatment see e.g.][]{Lendl+2020,Bonfanti+2021}, on-board temperature $T$, background flux, major residuals of the PSF fit as found via PCA, and time itself (to account for flux trends in time due to e.g. stellar activity).
In the case of variables with noisy measurement which could add white noise to the flux such as the change in temperature $\Delta T$, their variation with time is fitted with cubic basis splines and both the high- and low-frequency variation are used as separate decorrelation parameters.
A first step of Bayesian model comparison reveals which decorrelation parameters improve the model (we require a Bayes Factor >1), and a second step determines if these decorrelation parameters are similar across all observations (using leave-one-out comparison).
The result is a series of linear and quadratic parameters which modify the CHEOPS flux according to the variation of a hyperparameter either for individual observations or globally. These are co-modelled with the transits.

Although both the PSF modelling and the detrending can remove the majority of the impact of systematics on the CHEOPS flux, the "spiky" PSF combined with the rapid field rotation often also results in sharp variations in flux as a function of roll-angle.
However, such variation is typically extremely repeatable, and as hundreds of orbits of CHEOPS observations exist for HIP\,29442, we can co-model these shorter-frequency variations with the transit model.
To this end, \texttt{chexoplanet} uses a cubic spline with breakpoints spaced every 9 degrees (40 total splines for continuous data) common across all observations.
The flux at each spline position therefore becomes an independent parameter to be co-modelled with the exoplanetary transits.

The transits were modelled using \texttt{exoplanet}, which samples the parameter space using the Hamiltonian Monte Carlo implementation, PyMC3 \citep{exoplanet:pymc3}.
The model used circular orbits, and limb-darkening parameters for TESS and CHEOPS were constrained to the expected theoretical values derived by \citet{claret2017limb} and \citet{claret2021limb} respectively.
Planetary radius ratios were fitted using a broad log-Normal prior.
Impact parameters were fitted using the prior of \cite{exoplanet:espinoza18}.
To account for additional white noise, jitter terms were also included for each instrument (TESS and CHEOPS) with broad log-Normal priors.
Initial fits to TESS and CHEOPS photometry revealed inconsistencies in the depth, duration and timing, especially for planet~d.
As a compromise between a full photodynamical TTV model (which struggles due to the low per-transit S/N of the planets) and a linear ephemeris model, we modelled periods split across the TESS and CHEOPS data for all three planets.
The resulting photometric fits can be seen in Figure \ref{fig:transits}, while the derived planetary parameters are listed in Table \ref{tab:planetary_params}.

\subsubsection{Radius discrepancies with Damasso et al. 2023}
\citet{Damasso+2023} used TESS photometry to reveal radii of $3.48\pm0.07$, $1.58\pm0.10$ and $1.37\pm0.11$ R$_\oplus$ for planets b, c and d respectively.
This suggested the inner-most planets c and d were significantly ($\sim$2$\sigma$) different in radius, with the middle planet being the smallest. 
Our analysis based on a higher-quality dataset of TESS and CHEOPS photometry found radii of $3.410\pm0.046$, $1.551\pm0.045$ and $1.538\pm0.049$ R$_\oplus$.
Hence, we found <1$\sigma$ agreement for planets b and c.
For planet d, our value and error are $\sim$3.4$\sigma$ larger than the median value of \citet{Damasso+2023}, although only $\sim$1.5$\sigma$ using their larger uncertainties. The resulting difference in mass and radius is also visualised in Figure~\ref{fig:m_r_diagram}.
We note that \citet{Damasso+2023} were aware of the limitations of their radius measurements given the low per-transit S/N in the TESS light curves, and adopted a 3$\sigma$ confidence interval for example for their internal structure modelling.
The 8 transits observed by CHEOPS more than double the number of transits of planet~d observed by TESS, greatly boosting the S/N.
We are also helped by observing these transits within a single observing season, while the TESS observations were separated by two years and therefore were more at risk of ephemeris drift due to transit timing variations.
Although an analysis using each transit time as a free parameter did not reveal any significant TTVs with amplitudes above 1 hr, our split period modelling between TESS and CHEOPS did reveal a $1.5\sigma$ drift in period between the two datasets.
We also found such a split period model produced a better phase-folded transit fit to the TESS and CHEOPS datasets.
We attempted to split the model into three parts - one for each of the TESS sectors plus CHEOPS but found that doing so resulted in no clear detection of a periodic transit signal at the expected ephemeris in the first TESS sector, due the lower S/N from only three transits.
\starname{}\,d therefore shows that caution is required when determining the radii of small planets from multi-year TESS photometry when the per-transit S/N is low and there are significant gaps between observations.

\subsection{Radial velocities}
\begin{figure}
    \centering
    \includegraphics[width=0.86\linewidth]{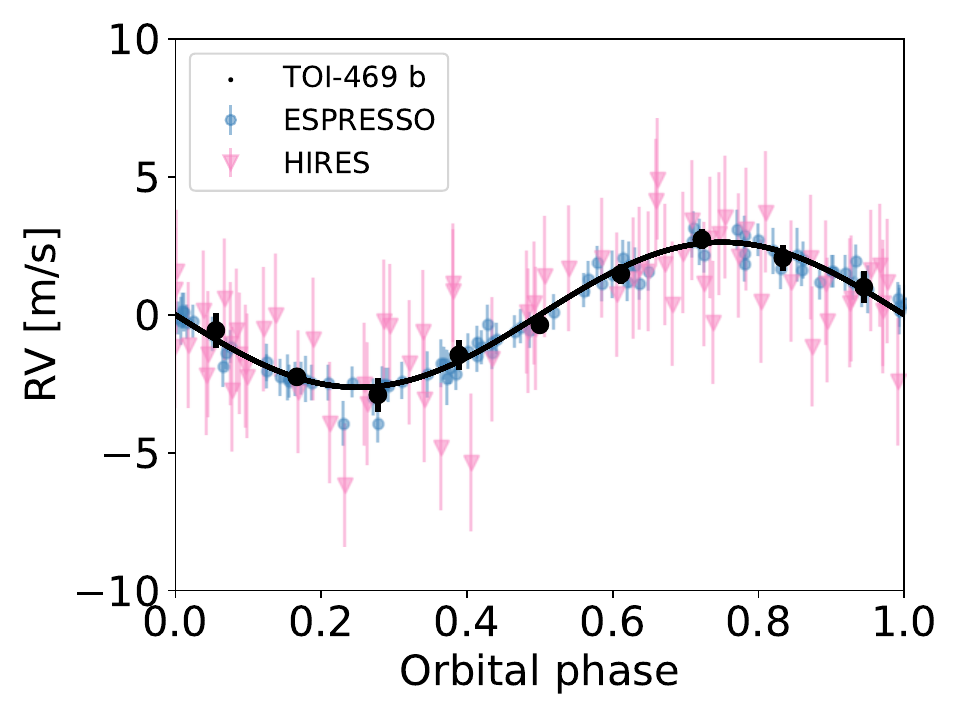}\\
    \includegraphics[width=0.86\linewidth]{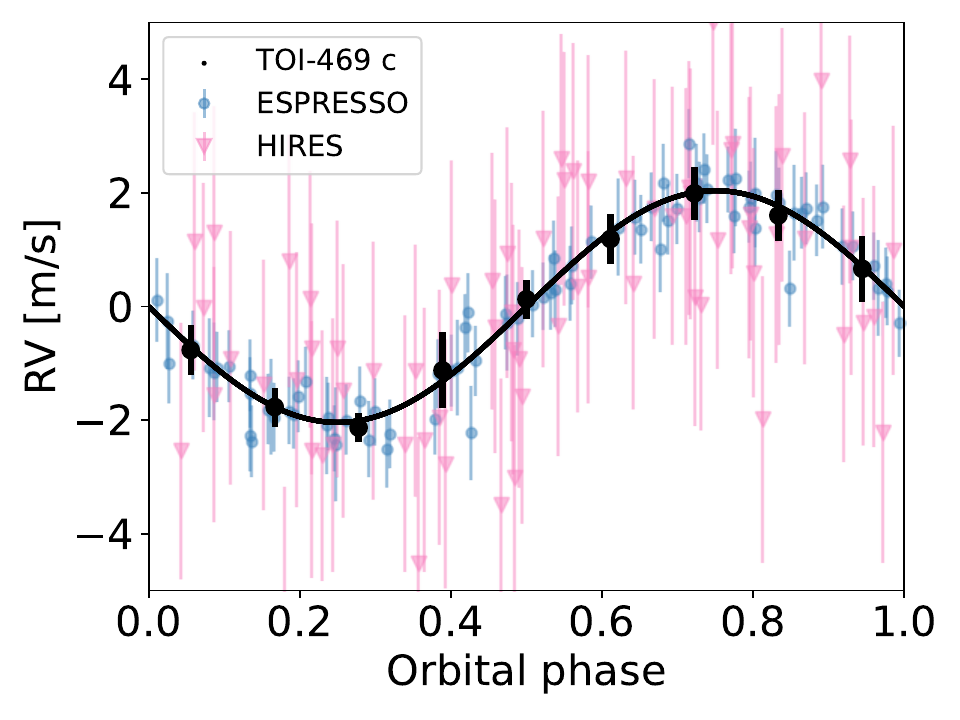}\\
    \includegraphics[width=0.86\linewidth]{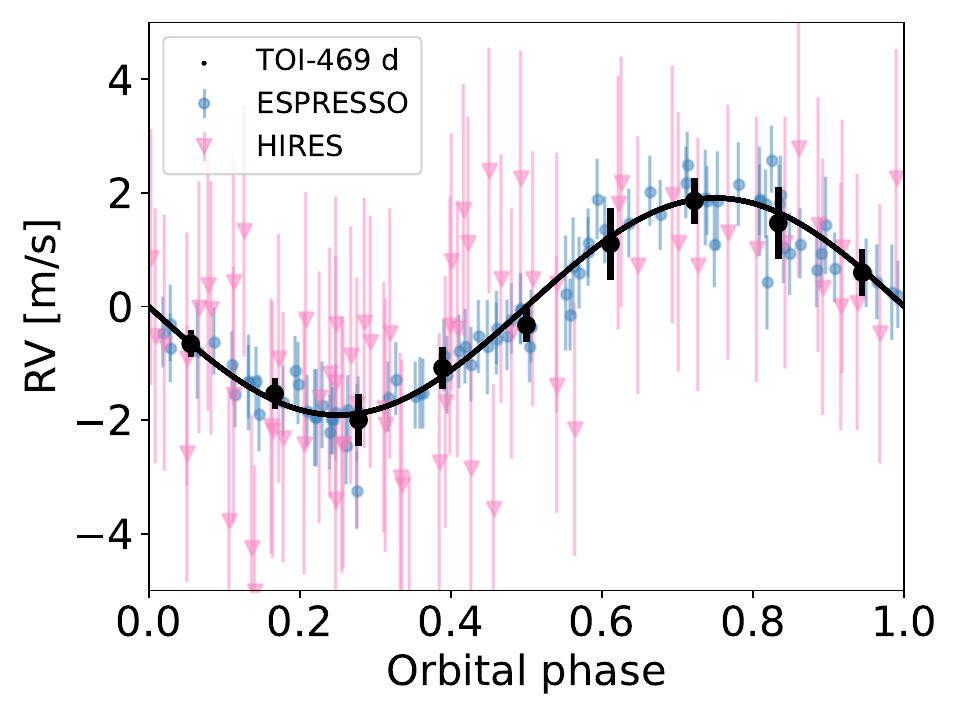}
    \caption{Radial velocities versus orbital phase for \starname\ b (top), c (middle) and d (bottom), including both the dataset from ESPRESSO \citep[][round markers in blue]{Damasso+2023} and from HIRES \citep[][triangle markers in pink]{AkanaMurphy+2023}. The best-fit model is shown in black.}
    \label{fig:rvs}
\end{figure}

\cite{Damasso+2023} analysed 83 spectra of \starname~collected with ESPRESSO, reporting RV semi-amplitudes of 2.8$\pm$0.2, 2.1$\pm$0.1 and 1.9$\pm$0.1~ms$^{-1}$ and masses of 9.6$\pm$0.8, 4.5$\pm$0.3 and 5.1$\pm$0.4~M$_\oplus$ for \starname~b, c and d respectively. There are also 71 additional Keck-HIRES RVs presented by \cite{AkanaMurphy+2023}, who report an RV semi-amplitude of 1.62$\pm$0.67~ms$^{-1}$ and a mass of 5.8$\pm$2.4~M$_\oplus$ for planet~b. We re-ran the model from \cite{Damasso+2023} using our updated ephemeris and including both the ESPRESSO and Keck-HIRES RVs. The best fit RV model for all three planets is plotted in Figure~\ref{fig:rvs} along with the RV Doppler signals from both datasets. The error bars are inflated, accounting for the uncorrelated jitter terms. The resulting RV semi-amplitudes and planetary masses are presented in Table~\ref{tab:planetary_params}. They are in good agreement with the values presented in \cite{Damasso+2023}.

\subsection{The planets of the \starname \ system within the exoplanet population}
\label{sec:demographics}
\begin{figure}
    \centering
    \includegraphics[width=\linewidth]{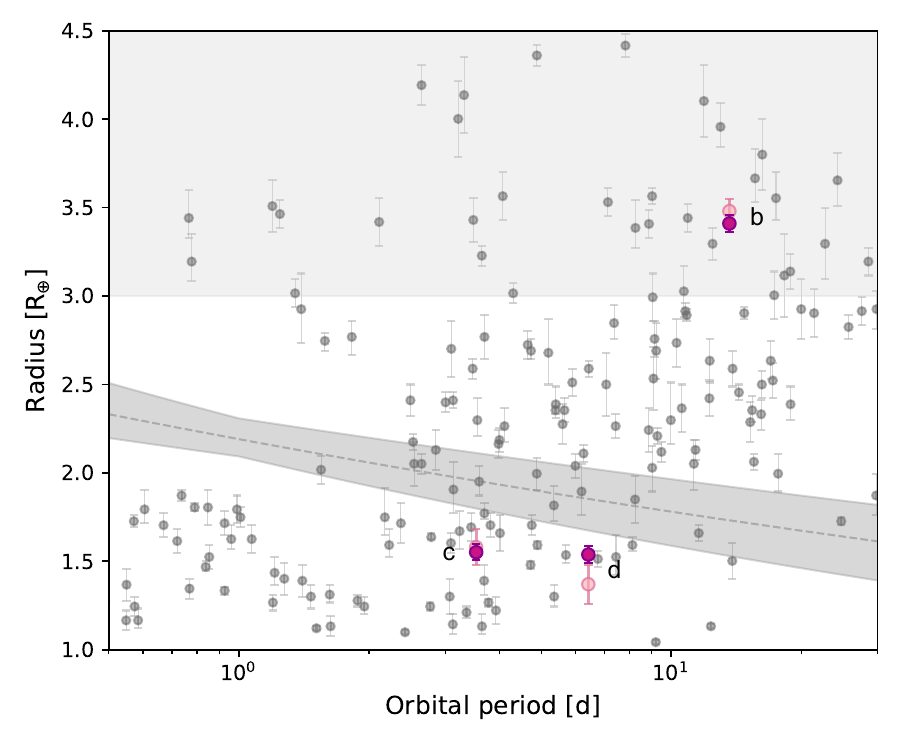}
    \caption{Location of the \starname \ planets (purple markers) in the orbital period-radius diagram in relation to the known exoplanet population, plotted using the PlanetS exoplanet catalogue. The transparent pink points show the planetary parameters derived by \cite{Damasso+2023}. The location of the radius valley is indicated in dark grey using the fit from \cite{Ho+2023}. The area shaded in light grey is the radius cliff beyond 3 R$_\oplus$.}
    \label{fig:a_r_diagram}
\end{figure}

Figure \ref{fig:a_r_diagram} shows the planets of the \starname \ system in the context of the population of small exoplanets, in terms of their planetary radii and orbital periods. This plot was generated using the updated version of the PlanetS catalogue of transiting exoplanets with reliable mass and radius values \citep[initially introduced by][available on DACE\footnote{\url{https://dace.unige.ch/exoplanets}}]{Otegi+2020a}. We can see that \starname~c and d lie below the radius valley \citep[plotted using the fit from][for the stellar mass of \starname]{Ho+2023}, while \starname~b not only lies above the radius valley but actually beyond the radius cliff found at around 3 R$_\oplus$ in the Kepler data \citep{Fulton+2018}. One of the most popular explanations for the radius cliff in the literature is the fugacity crisis hypothesis presented by \cite{Kite+2019}. They showed that starting from planetary radii of around 3 R$_\oplus$, the pressure at the base of a hydrogen-dominated atmosphere becomes high enough for the hydrogen to dissolve into the magma below, thus inhibiting further planet growth. Other works also suggested more weakly bound atmospheres at higher planetary masses \citep{Owen+Wu2017} and the timing of atmospheric accretion \citep{Lee+2016} as possible explanations. More recently, \cite{Burn+2024} also reproduced the radius cliff using a combined planet formation and evolution model for planets with fully mixed H/He and water envelopes.

\section{Internal structure analysis}
\label{sec:internal_structure}
\begin{figure*}
    \centering
    \includegraphics[width=\textwidth]{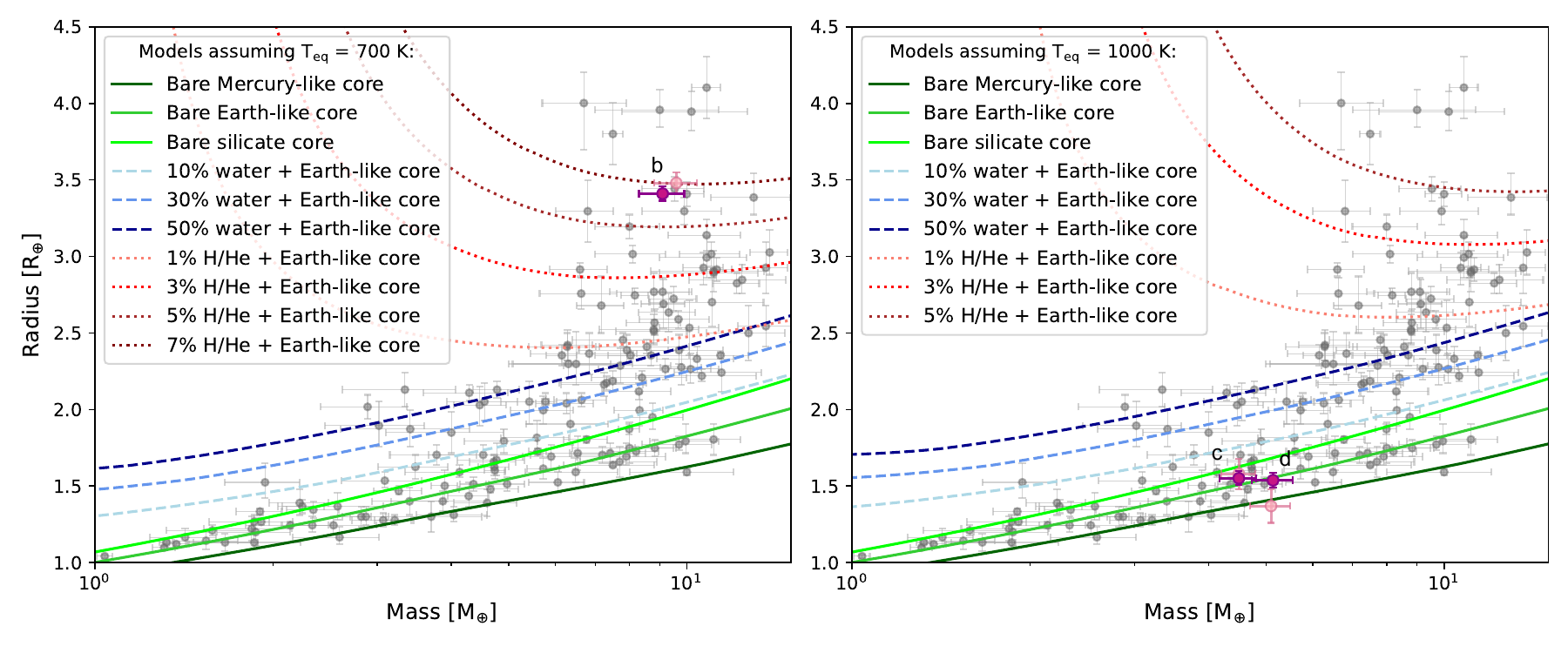}
    \caption{Mass-radius diagrams showing the locations of \starname~b (left panel) and \starname~c and d (right panel), in relation to the known exoplanet population. The transparent pink points show the planetary parameters derived by \cite{Damasso+2023}. Mass-radius relations for fixed compositions and equilibrium temperatures were generated using BICEPS \citep{Haldemann+2024}. Shown are three different compositions for bare cores (solid lines in different shades of green), Earth-like cores with different amounts of water (dashed lines in different shades of blue) and with different amounts of H/He (dotted lines in different shades of red). Models in the left panel were calculated assuming an equilibrium temperature of 700 K (similar to \starname~b), while models in the right panel were calculated assuming an equilibrium temperature of 1000 K (similar to \starname~c and d).}
    \label{fig:m_r_diagram}
\end{figure*}

As a first step in analysing the interior composition of the planets in the \starname\ system, we look at their location in a mass-radius diagram as shown in Figure \ref{fig:m_r_diagram}. We find that planets~c and d lie close to the mass-radius relation for a bare Earth-like core (right panel). Meanwhile, planet~b lies well above the 50\% water line, generated assuming an Earth-like core and an equilibrium temperature similar to \starname~b, and therefore seems to host an envelope with at least some H/He (left panel). All mass-radius relations were generated using the forward model of the BICEPS code \citep{Haldemann+2024}. However, as determining the internal structure of an exoplanet from its radius and mass is a highly degenerate problem, we will run a full Bayesian inference model for all three planets in the following section. The high precision of the radii and masses we have derived in Section~\ref{sec:data+analysis} provides us with ideal circumstances to do so.

\subsection{Previous work}
\label{sec:internal_structure -- old_model}
In past work \citep[e.g.][]{Leleu+2021, Lacedelli+2022, Wilson+2022, Luque+2023}, we have used an inverse method based on the work of \cite{Dorn+2015,Dorn+2017} to derive posterior distributions for the internal structure parameters of each planet, which was introduced and reviewed in detail in \cite{Leleu+2021}. As in \cite{Dorn+2015,Dorn+2017}, our framework is split into two main parts: a forward model that calculates the radius of a planetary structure with a given mass, composition and external conditions, and a Bayesian inference algorithm used to sample the parameter space of possible internal structure parameters. These sampled parameters are then passed to the forward model, which allows us to compare the calculated radii of the sampled planetary structures to the observationally derived radius, or more precisely to the derived transit depth. This general approach is also used in other works, for example by \cite{Acuna+2021} and \cite{Haldemann+2024}.

To make the use of such an inverse method feasible even with a quite computationally expensive forward model, these other works use Markov chain Monte Carlo (MCMC) or nested sampling as their Bayesian inference algorithms. What differentiates our internal structure modelling framework \texttt{plaNETic}\footnote{\url{https://github.com/joannegger/plaNETic}} from them is the use of a deep neural network (DNN) that is trained on data generated with the chosen forward model, a method first introduced in this context by \cite{Alibert+2019}. This DNN is then used to replace the forward model in the Bayesian inference scheme, taking the same input parameters as the forward model and using them to calculate the radii of the sampled planetary structures, but speeding up the calculation by more than four orders of magnitude. This extensive decrease in computation time allows us to use a brute-force full grid sampling approach for the Bayesian inference scheme, while still speeding up the computation of the posterior distributions of the internal structure parameters for the individual planetary systems. Other works such as \cite{Baumeister+2020} and \cite{Haldemann+2023} also use neural networks to speed up the calculation of planetary internal structures. The difference to our model is that they replace the entire Bayesian framework with a neural network and not just the forward model. This means that their methods can compute posteriors for the internal structure of a planet faster than ours, but they also do not allow for changes in the used inference method, such as the used priors.

The new version of \texttt{plaNETic} introduced in this work is publicly available and can be adapted for any choice of forward model, all that is necessary is training a DNN on the desired model. By default, \texttt{plaNETic} uses DNNs trained on the planetary structure model of BICEPS \citep{Haldemann+2024}, as described in detail in Section \ref{sec:internal_structure -- new_model}. In the following, we will first briefly summarise the specifics of the model used in past work.

In the preliminary version of \texttt{plaNETic} introduced in \cite{Leleu+2021}, a simple structure model is used as the forward model. Each planet is assumed to be built up of four distinct layers. The innermost layer consists of an inner iron core with a molar fraction of up to 19\% of sulphur, used as a placeholder for any lighter elements in the core \citep[modelled using the equations of state (EoS) from][]{Hakim+2018}. On top of this core, a silicate mantle made up of oxidised silicon, magnesium and iron \citep[EoS from][]{Sotin+2007} is added. Next, we add a condensed water layer \citep[EoS from][]{Haldemann+2020} with fixed outer boundary conditions of $P=1\ \mathrm{bar}$ and $T=300\ \mathrm{K}$. On top of this structure, a pure H/He envelope is modelled separately according to the fit from \cite{Lopez+Fortney2014}. A DNN with 6 hidden layers of 2048 units each was then trained on a large database with 5 million points generated with this forward model. The accuracy of this DNN was investigated in detail in \cite{Leleu+2021}, with the study showing an error on the predicted radius that is lower than 0.4\% in 99.9\% of the cases.

For each planet in the system, 100 million different combinations of internal structure parameters are sampled, with a uniform prior for the mass fractions of the inner iron core, silicate mantle and water layer (on the simplex on which they add up to unity) and a log-uniform prior for the mass of H/He added on top. Additionally, we introduced an upper limit of 0.5 for the water mass fraction in accordance with \cite{Thiabaud+2014} and \cite{Marboeuf+2014}. All planets in the system are then modelled simultaneously in order to leverage the correlation between the observed properties of the different planets in the system, which stems from the fact that both the planetary mass and radius are measured in relation to the host star. This interdependency of the mass and radius values of planets in the same system is also considered by \cite{Dorn+2018}, who run a resampling scheme after inferring the structure of the individual planets as opposed to modelling all planets simultaneously. Furthermore, we followed \cite{Thiabaud+2015} and assumed that the Si/Mg/Fe ratios of all planets in the system are identical and match the ones calculated from the stellar abundances exactly for this previous version of the model.

\subsection{Internal structure modeling framework}
\label{sec:internal_structure -- new_model}

\subsubsection{Forward model}
\label{sec:internal_structure -- forward model}
In this work, we now introduce a new and improved version of \texttt{plaNETic}. As a first change compared to the previous version, we use the planetary structure model of BICEPS \citep{Haldemann+2024} as our forward model, which allows for a significant increase in physical accuracy compared to the previously used structure model. BICEPS does include not only this planetary structure model but also an MCMC-based Bayesian inference code. However, this part of the code is not used here.

As before, each planetary structure is built up of an inner core, mantle and volatile layer. For the core and mantle layers, the structure model of BICEPS uses an expansive collection of EoS that allows for both solid and molten phases. In principle, BICEPS allows for any combination of a pure water layer with a possibly water-enriched H/He envelope. For this work, we will use a model setup where the volatile layer is a uniformly mixed envelope of water and H/He with no pure water layer present. This volatile layer is modelled self-consistently with the rest of the planet and is split into two distinct sub-layers, an irradiated outer atmosphere and an envelope with almost all stellar flux already absorbed in the layer above. The irradiated atmosphere is modelled using the non-grey analytical atmosphere model of \cite{Parmentier+Guillot2014} and \cite{Parmentier+2015} together with the opacity grid of \cite{Freedman+2014}. As the used atmosphere model depends on the intrinsic luminosity of the modelled planet, a quantity that especially for small planets usually cannot be measured directly, we use the age-luminosity relation from \cite{Mordasini2020}. This relation is a fit of coupled formation and evolution calculations as presented in \cite{Mordasini+2012}, but also includes the cooling model of the solid core from \cite{Linder+2019}. Whether the envelope follows a convective or radiative profile is determined self-consistently using the Schwarzschild criterion. Condensation is not taken into account, since the planets of the \starname\ system are on close-in orbits and therefore have high equilibrium temperatures. Finally, the transit radius is defined as the radius where the chord optical depth is $\tau_\textrm{ch} =$ 2/3 \citep{Guillot2010}. For a more detailed description of the planetary structure model of BICEPS including the full list of used EoS, we refer the reader to \cite{Haldemann+2024}.

In the subsequent sections, we will use the following notation to refer to the internal structure parameters introduced above: w$_\textrm{layer}$ denotes the mass fraction of a given layer with respect to the entire planet, while Z$_\textrm{envelope}$ is defined as the mass fraction of water in the volatile layer. Finally, x$_\textrm{element,layer}$ denotes the molar fraction of a given element in the specified layer.

\subsubsection{Inference algorithm and priors}
\label{sec:internal_structure -- inverse model}
The inference method implemented in \texttt{plaNETic} is a full-grid acceptance-rejection sampling algorithm. As a first step, 10000 synthetic stars are sampled from the previously observationally derived distributions of the relevant stellar properties (R$_\star$, M$_\star$, t$_\star$, T$_\mathrm{eff}$, [Si/H], [Mg/H] and [Fe/H]), listed in Table \ref{tab:stellar_params} for \starname. For each of these synthetic stars, 10000 synthetic planetary systems are then sampled. This is done by sampling RV semi-amplitudes and orbital periods from the observationally derived distributions, as summarised in Table \ref{tab:planetary_params} for the \starname\ planets. We also simultaneously sample all relevant internal structure parameters from the chosen priors for each planet in the system. Based on these sampled parameters for both the star and the planets in the system, the neural network trained on the forward model is then used to calculate the transit depths that these synthetic planets would have. Each synthetic planet is then compared to the derived distribution of the transit depth for the observed equivalent planet and either accepted or rejected based on the calculated likelihood of the two agreeing. By modelling all planets in the system simultaneously and using transit depths and RV semi-amplitudes instead of radii and masses directly, we ensure that all synthetic planets orbit exactly the same star (not just within the error intervals) and include the correlation of the planetary radii and masses of planets in the same system.

For the priors of the mass fractions of the different layers (inner core, mantle and envelope) and the composition of these layers (molar fractions of Fe and S for the inner core; oxidised Si, Mg and Fe for the mantle; water and H/He for the envelope), we consider different scenarios. In the first case (hereafter case A), we use a prior for the mass fractions of the inner core layer, mantle layer and the total water mass fraction in the planet that is uniform on a simplex on which they add up to 1, while still assuming an upper limit of 0.5 for the accreted water \citep{Thiabaud+2014, Marboeuf+2014}. Additionally, we then use a log-uniform prior for the accreted H/He. After sampling the mass fraction of accreted H/He, the mass fractions of the inner core, mantle and water are then rescaled so that the sum of these four mass fractions equals unity. These priors are motivated by the assumption that the planet has formed outside the iceline, which means that it has accreted water not only through the accreted gas but has also accreted icy planetesimals. Even though this scenario assumes that water is accreted in the form of ice, it is assumed that the total amount of water sampled is located in the envelope and uniformly mixed with the sampled amount of H/He for the calculation of the internal structure of the planet.

Conversely, for the second case (case B), we assume that the planet formed inside the iceline. This means that any water present comes from the gas the planet has accreted. In this case, we use a uniform prior to sample the mass fraction of the inner core and calculate the corresponding mass fraction of the mantle layer as 1 minus the core mass fraction (both with respect to the total mass of refractories in the planet). We also apply a log-uniform prior for the mass fraction of the accreted gas envelope. Finally, we sample the mass fraction of water with respect to the total mass of the gaseous envelope using a Gaussian prior with a mean of 0.5\% and a standard deviation of 0.25\%, in accordance with the molecular abundance of water in the Solar nebula \citep{Mousis+2009, Lodders2003}. This means that an envelope composition of 1$\times$ solar \citep{Fortney+2013} is enclosed within the 3 sigma interval. We then again re-scale the mass fraction of the inner core and mantle layers so the sum of all three layers add up to unity.

In a second step, we also consider different options for sampling the planetary Si/Mg/Fe ratios (for both case A and case B described above). On the one hand, \cite{Thiabaud+2015} used a combined chemical and planet formation model and found that the elemental Si/Mg/Fe ratios in planets are essentially identical to the ones of their host star. Similarly, \cite{Pelletier+2023} measure abundances of refractory elements in the giant planet WASP-76~b and find them mostly stellar-like. This assumption is also frequently used in other internal structure modelling frameworks, such as \cite{Dorn+2017}, \cite{Acuna+2021} and our own past work \citep[e.g.][]{Leleu+2021, Lacedelli+2022, Wilson+2022, Luque+2023}. On the other hand, \cite{Adibekyan+2021} also find a correlation between the composition of rocky exoplanets and their host stars. However, according to their study planets seem to be enriched in iron compared to their host stars. Other studies indicate similar findings \citep[e.g.][]{Liu+2023}. \cite{Guimond+2024} also provide a thorough review on the processes through which stellar and planetary compositions are related. In the following, we use three different options for our compositional priors: (1) assuming the planetary Si/Mg/Fe ratios to match the stellar ones exactly, (2) implementing the iron-enriched planetary Si/Fe and Mg/Fe ratios using the relation by \cite{Adibekyan+2021} while keeping the stellar Si/Mg ratio, and (3) sampling the elemental abundances of Si, Mg and Fe in the planet uniformly from a simplex, without taking into account stellar abundances. For the last option, we add an upper limit of 0.75 for the amount of Fe compared to Si and Mg in the planet, similar to \cite{Acuna+2021}. The reasoning behind this is that a planet made up of pure iron with a volatile layer on top is not physical from a planet formation perspective. We do note, however, that this would not apply if one were to model a super-Mercury like planet \citep[such as GJ~367~b,][]{Goffo+2023}.

\subsubsection{Training DNNs to replace the forward model}
\label{sec:internal_structure -- dnns}

As for the previously used model version, we trained DNNs for this improved forward model to reduce the necessary computation time. As in the previous version of \texttt{plaNETic}, the DNNs used here are conventional feed-forward neural networks with a layer of input neurons, multiple hidden layers and one output neuron. The output of such a neural network is calculated layer by layer, starting with the input layer which receives a set of physical values as input. The value of each neuron in the next layer is then determined as a linear combination of the values and weights of the neurons in the previous layer. To also allow for non-linearity, an activation function is then used to calculate the neuron's value. Training a neural network means the process of iteratively improving the weights of all neurons based on a set of training data, so that the value of the output neuron approaches the expected value for each set of input parameters. The discrepancy between the values that the output neuron gives and the target values is measured with the so-called loss function, which is minimised during training \citep{Bishop2006, Goodfellow+2016}.

In contrast to this, \cite{Baumeister+2020} and \cite{Haldemann+2023} use more complex forms of neural networks (more specifically mixture density networks and conditional invertible neural networks). This is necessary because they solve the inverse problem directly using their neural networks, while we only replace our forward model.

To account for the increased complexity of the structure model with this new version of \texttt{plaNETic} and still reach a similar accuracy as before, we now trained separate DNNs for different mass regimes: (M1) 0.5 to 6 M$_\oplus$ and (M2) 6 to 15 M$_\oplus$. We also trained DNNs for a third mass regime (M3) that will be used in future work for planets with masses between 15 and 30 M$_\oplus$. For each mass regime, we generated two different large databases of 15 million internal structure models, one corresponding to an envelope with high water-enrichment (consistent with case A above) and one corresponding to a H/He envelope enriched with 0-1\% of water in mass (consistent with case B above). To create the databases, the internal structure parameters and boundary conditions were randomly sampled using the same set of priors as will be used when later inferring the internal structure of a planet. The transit radii of the sampled structures are then calculated using the planetary structure model of BICEPS \citep{Haldemann+2024}.

Each database was then split randomly into 80\% training data, 10\% validation data and 10\% test data. Using the Python library \texttt{tensorflow} \citep{tensorflow}, we then used each database to train a DNN. To this end, the input parameters of the training data set were first normalised feature-wise, i.e. each input parameter was scaled by dividing by the mean and subtracting the standard deviation of the distribution of the parameter in question. The individual layer mass fractions and molar fractions describing the composition of the core and mantle layers respectively are not independent of each other but always need to add up to 1. For this reason, only the layer mass fractions of the inner core, water and H/He are used for training the neural networks, as well as the molar fractions of sulphur in the core and silicon and magnesium in the mantle (but not the mantle mass fraction and the molar fractions of iron in the core and in the mantle). Overall, the neural network therefore has nine input neurons, corresponding to the equilibrium temperature of the planet, the atmospheric water mass fraction, the intrinsic luminosity of the planet (derived from the stellar age), the mass of the planet, the mass fractions of the inner core and the envelope with respect to the total planet mass, the molar fraction of sulphur in the inner core and the molar fractions of silicon and magnesium in the mantle. For both the envelope mass fraction and the intrinsic luminosity, we look at the quantities in log-space. Conversely, each DNN has a single output neuron with a linear output activation function that estimates the transit radius of the sampled structure.

For each database, we tried a range of different network architectures, which we then compared using the validation data. We chose the best DNN based on the number of data points in the validation data with a prediction error, defined as $\frac{R_{DNN} - R_{BICEPS}}{R_{BICEPS}}$, of more than 3\% (the radius precision usually reached when characterising a planet with CHEOPS). The final network architectures for each mass regime and water accretion option are summarised in Table \ref{tab:network_architectures} in the appendix. To allow for non-linearity, the standard ReLU function (given by $\textrm{ReLU}(x) = \max(x,0)$; \citeauthor{Nair+Hinton2010} \citeyear{Nair+Hinton2010}) was used as an activation function for each unit. Each DNN was trained for a maximum of 10000 training epochs with early stopping activated if no improvement was seen for 100 epochs to avoid overfitting. For all DNNs, this happened after training for between 400 and 1100 epochs. We used the mean square error of the transit radius as our loss function and fit the DNNs using the ADAM optimiser \citep{King+Ba2015} with a learning rate starting at 0.001 that is reduced by a factor 2 each time the validation loss reaches a plateau for 30 training epochs. Since our databases are too large to fit into memory at once, we used a data generator for both the training and validation data that randomly selects batches of 1024 data points that are then used for training and validation.

\subsubsection{DNN performance}
\label{sec:internal_structure -- dnn_performance}

\begin{figure}
    \centering
    \includegraphics[width=0.95\linewidth]{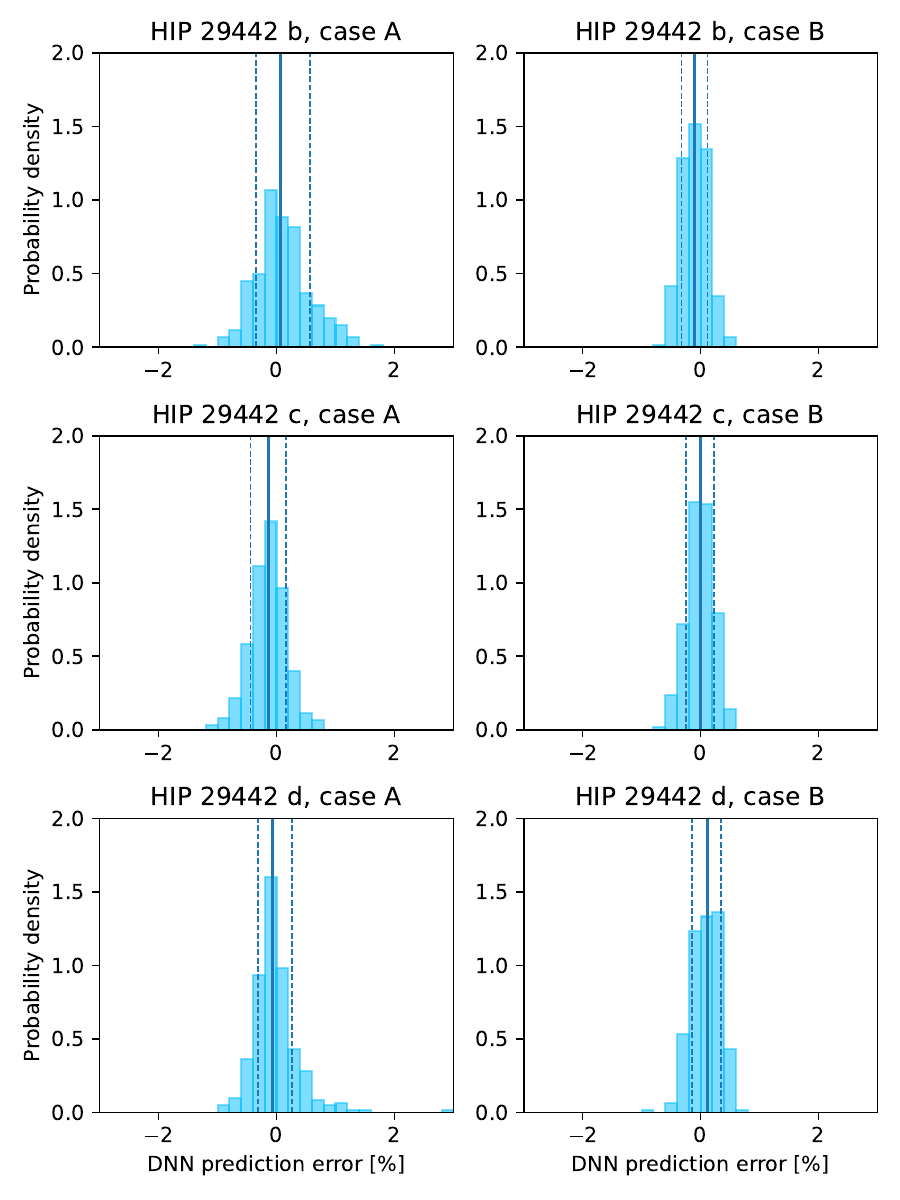}
    \caption{Histograms of the DNN prediction error of the transit radius for 300 randomly sampled points in the posteriors, for each planet (rows) and both of the water prior options (columns). The vertical lines show the median and 16$^\textrm{th}$ and 84$^\textrm{th}$ percentiles of each distribution.}
    \label{fig:dnn_pred_error}
\end{figure}

We now used this new version of \texttt{plaNETic} to calculate posterior distributions of the internal structure parameters for \starname~b, c and d, using both water priors (cases A and B) and all three compositional priors for the elemental Si/Mg/Fe ratios (options 1, 2 and 3) discussed above. However, before discussing the results of these models, we test the performance of our DNNs by randomly choosing 100 structures from each posterior of each planet and running the full forward model of BICEPS for them. The resulting distribution of the prediction error of the DNNs on the transit radius of each structure is shown in Figure \ref{fig:dnn_pred_error}, for each planet and water prior separately, but combined for the three Si/Mg/Fe prior choices. In all cases, the median values of the distributions lie between $-0.13$ and $0.11\%$ with the 16$^\textrm{th}$ percentiles above $-0.40\%$ and the 84$^\textrm{th}$ percentiles below $0.60\%$. These prediction errors are well below the observational errors on the radii.

We also tested how much the use of DNNs speeds up the calculation of the transit radii of the same randomly sampled structures as compared to the full forward model. The full results of this test can be found in Table \ref{tab:speedup_DNN} in the appendix. To briefly summarise our findings, we can say that with the DNN, calculating the radii of 100 sampled structures takes around 0.035 seconds\footnote{Run on a 2021 MacBook Pro with an Apple M1 Pro chip.}, while with BICEPS, calculating the same radii takes on average about 23 minutes for model A and 38 minutes for model B\footnote{Run on a node of the HORUS cluster of the Theoretical Astrophysics and Planetary Science group at the University of Bern using 2 x 14-Core Intel Xeon Gold 6132 @ 2.6 GHz.}. This is a speed-up of around a factor 40000 for model A and a factor 63000 for model B. What needs to be taken into account with this calculation, however, is of course also the computation time needed to generate the database necessary for training each DNN, which is around 5 days on 280 CPUs.

\subsection{Results for \starname~b, c and d}
\label{sec:internal_structure -- results}

\begin{table*}
\renewcommand{\arraystretch}{1.2}
\caption{Overview of the different versions of \texttt{plaNETic} used to model the internal structure of \starname~b, c and d.}
\centering
\begin{tabular}{c|cccc}
\hline\hline
Model & EoS & Volatile layer & Water prior & Si/Mg/Fe prior \\
\hline
A1 & \multirow{3}{*}{\shortstack{Full set of EoS \\ from BICEPS}} & \multirow{3}{*}{\shortstack{Uniformly mixed H/He and water envelope}} & \multirow{3}{*}{\shortstack{Accreted as \\ solids}} & stellar\\
A2 &  &  &  & iron-enriched\\
A3 &  &  &  & free\\
\hline
B1 & \multirow{3}{*}{\shortstack{Full set of EoS \\ from BICEPS}} & \multirow{3}{*}{\shortstack{Uniformly mixed H/He and water envelope}} & \multirow{3}{*}{\shortstack{Accreted as \\ gas only}} & stellar\\
B2 &  &  &  & iron-enriched\\
B3 &  &  &  & free\\
\hline
\multirow{3}{*}{PREV} & \multirow{3}{*}{\shortstack{Limited set of EoS \\ {[1]}, [2], [3]}} & \multirow{3}{*}{\shortstack{Condensed water layer, \\ separately modelled H/He envelope [5]}} & \multirow{3}{*}{\shortstack{Accreted as \\ solids}} & \multirow{3}{*}{stellar}\\
& & & & \\
& & & & \\
\hline
\end{tabular}
\label{tab:model_versions}
\tablebib{
[1]~\citet{Hakim+2018}; [2]~\citet{Sotin+2007}; [3]~\citet{Haldemann+2020}; [5]~\citet{Lopez+Fortney2014}.
}
\end{table*}
\renewcommand{\arraystretch}{1.0}

\begin{figure*}
    \centering
    \includegraphics[width=\textwidth]{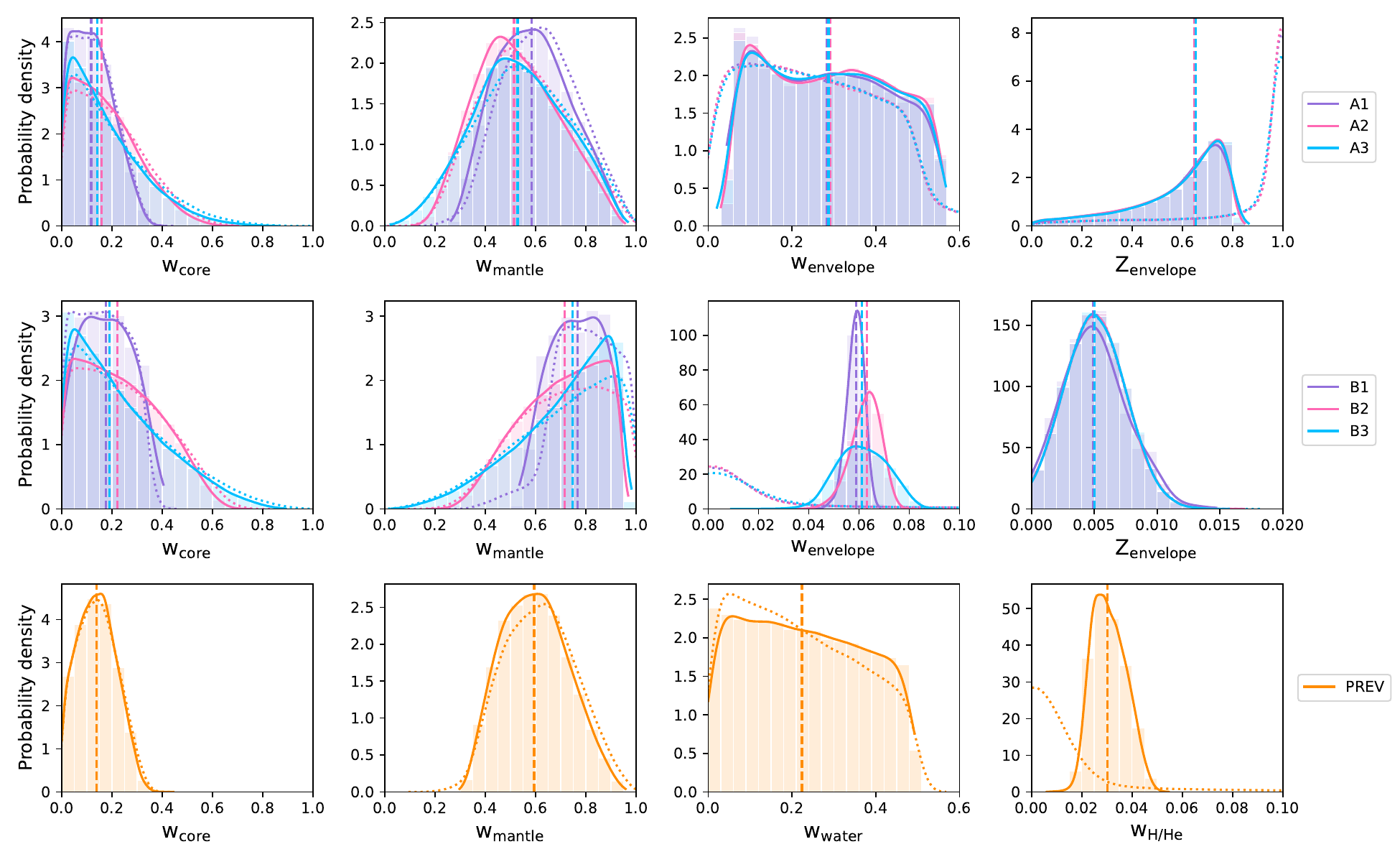}
    \caption{Posterior distributions of the most important internal structure parameters for \starname~b, namely the mass fractions of an inner core, mantle and envelope in the planet as well as the mass fraction of water in the envelope. The top row shows the posteriors assuming a prior consistent with the planet forming outside the iceline (case A), while the middle row uses a prior that is consistent with the planet forming inside the iceline (case B). For both water priors, we show the posteriors for three different compositional priors: first assuming that the planet has stellar Si/Mg/Fe ratios (purple), second assuming that the planetary bulk Si/Mg/Fe ratios are iron-enriched compared to the host star (pink) and third sampling the planetary bulk Si/Mg/Fe uniformly from the simplex on which the molar fractions add up to 1 (blue). The bottom row shows the posterior distributions of the mass fractions of an inner core, mantle, condensed water layer and separately modelled H/He envelope with respect to the total planet mass when applying the previously used version of the model assuming stellar Si/Mg/Fe ratios. The dashed vertical lines show the median value of each distribution. The dotted lines show the priors.}
    \label{fig:internal_structure_results_b}
\end{figure*}

\begin{figure*}
    \centering
    \includegraphics[width=\textwidth]{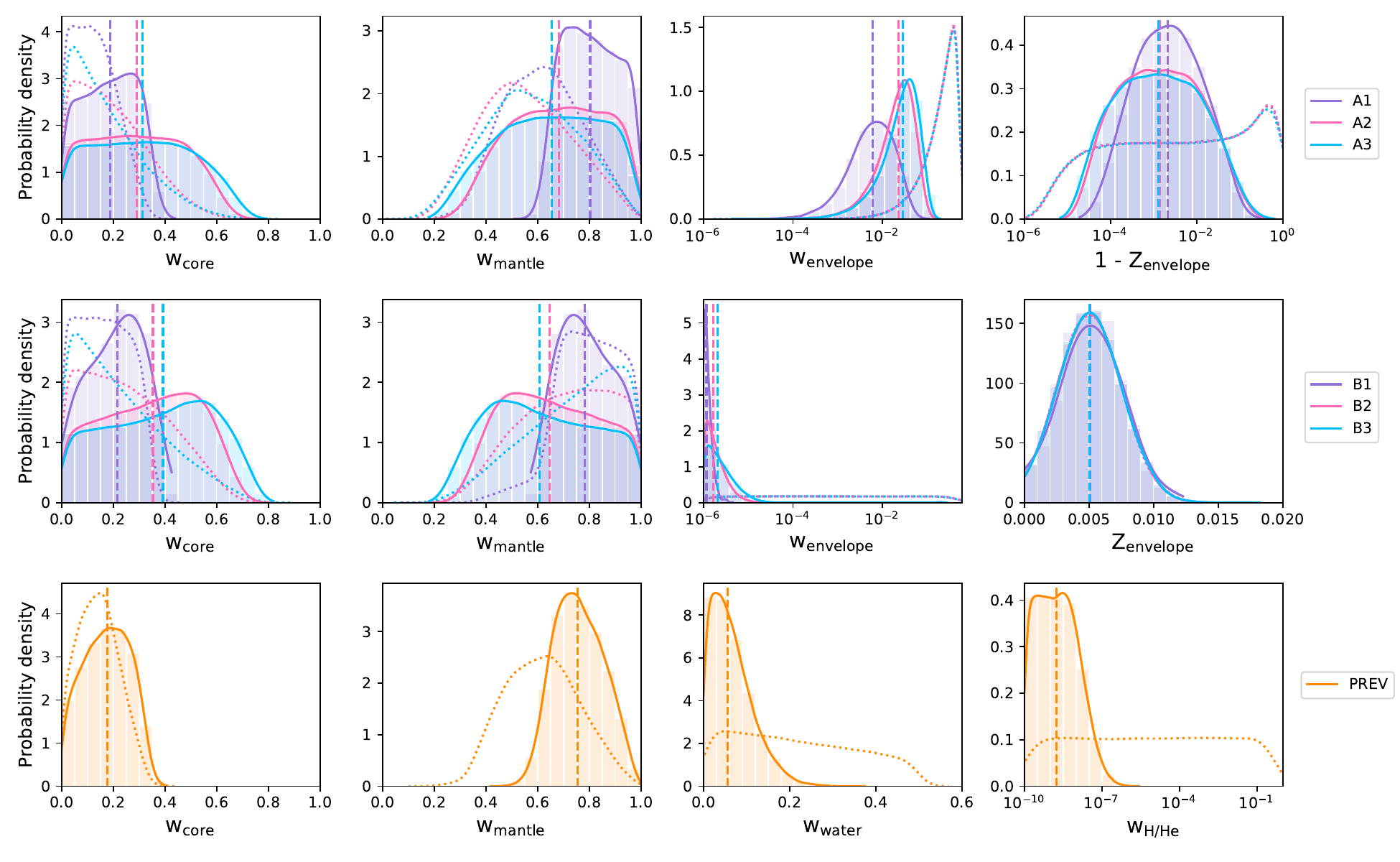}
    \caption{Same as Figure \ref{fig:internal_structure_results_b} but for \starname~c. Here, the rightmost panel in the top row shows the H/He mass fraction instead of the water mass fraction in the envelope.}
    \label{fig:internal_structure_results_c}
\end{figure*}

\begin{figure*}
    \centering
    \includegraphics[width=\textwidth]{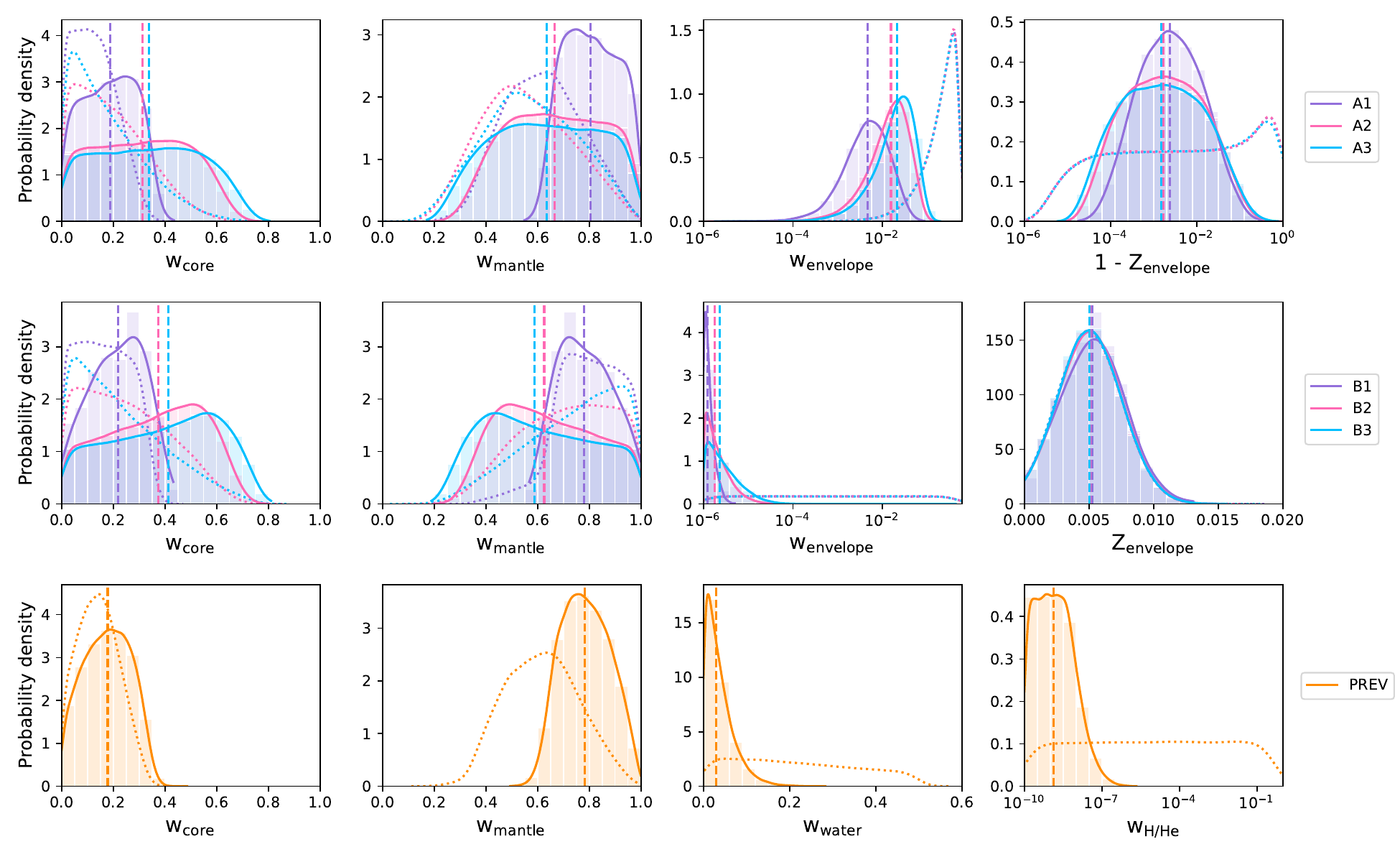}
    \caption{Same as Figure \ref{fig:internal_structure_results_b} but for \starname~d. Here, the rightmost panel in the top row shows the H/He mass fraction instead of the water mass fraction in the envelope.}
    \label{fig:internal_structure_results_d}
\end{figure*}

For each planet, we ran a total of seven different interior models, six using the new version of \texttt{plaNETic} with the different priors for the water content and elemental Si/Mg/Fe ratios described in Section~\ref{sec:internal_structure -- new_model} and one using the old version of the code, which assumes a condensed water layer. Table \ref{tab:model_versions} gives an overview of the different model versions used. The posterior distributions of the most important internal structure parameters for \starname~b, c and d are visualised in Figures \ref{fig:internal_structure_results_b}, \ref{fig:internal_structure_results_c} and \ref{fig:internal_structure_results_d} respectively.
The median values and one sigma errors of the posteriors of a wider range of parameters calculated for both the new and the old version of \texttt{plaNETic} are also listed in Tables \ref{tab:internal_structure_results_b}, \ref{tab:internal_structure_results_c}, \ref{tab:internal_structure_results_d} and \ref{tab:internal_structure_results_old} in the appendix.

For \starname~b, the results of the internal structure modelling depend significantly on whether we assume that the planet formed inside or outside the iceline. However, for one given water prior, the posteriors have similar median values for all three compositional priors for the planetary Si/Mg/Fe ratios. This is in agreement with the findings of \cite{Otegi+2020b}, who show that using stellar Si/Mg/Fe ratios as a proxy for the planetary composition does not always help to better constrain the internal layer mass fractions, especially for sub-Neptunes. Meanwhile, the spread of the distributions differs for the core and mantle mass fractions and, in the case of water prior B, to some extent also for the envelope mass fraction. However, for the core and mantle mass fractions, the posterior distributions are very close to the respective chosen priors (dotted lines). If we assume that \starname~b was formed outside the iceline (case A), we find a broad posterior distribution for the planet's envelope mass fraction spanning almost the full range from 0 to 60\%. The distribution of the envelope water mass fraction is a skewed Gaussian distribution that ranges from 0 up to almost 90\%, with a mode of around 75\%. If we assume that \starname~b formed inside the iceline (case B), the posterior for the envelope mass fraction is much more tightly constrained with a peak at around 6\%, which was to be expected as we significantly reduce the allowed compositional degeneracy in the envelope. 

We can now compare this to the results from the previous version of our internal structure model, which assumes the planet to have a condensed water layer and models a pure H/He envelope independently of the rest of the planet. We limit this comparison to model A1 to ensure that both models use the same priors, both in terms of water content and Si/Mg/Fe ratios. We also use water and H/He mass fractions with respect to the total planetary mass for both models, instead of using the total envelope mass fraction and envelope metallicity as shown in Figure~\ref{fig:internal_structure_results_b} for models A1. For model A1, the water mass fraction for planet~b is $18.4^{+16.1}_{-14.0}$\%, while the H/He mass fraction is $9.8^{+1.4}_{-3.0}$\%. For the previous model version, these values are $22.4^{+17.3}_{-15.6}$\% for the water mass fraction and $3.0^{+0.8}_{-0.6}$\% for the H/He mass fraction. The resulting water mass fractions for \starname~b are therefore similar but slightly lower with the new and more physically robust model. Both distributions allow for a broad range of values. At the same time, we infer higher H/He mass fractions with the new model. The inferred core and mantle mass fractions are very similar for both model versions.

The resulting posterior distributions for the interiors of \starname~c and d are very similar. Here, the posteriors of the core mass fractions are showing a trend of leaning towards the higher values covered by the priors for all three Si/Mg/Fe ratio options. For the mantle mass fractions, the same is true in the case of water prior A, while for water prior B, we observe the opposite effect for the iron-enriched and free Si/Mg/Fe priors. For both planets, we find that if they formed outside the iceline (case A), then their envelopes will mostly consist of water, with median values as high as 99.9\% for the water mass fraction in the envelope and total envelope mass fractions between 0 and 20\% depending on the chosen Si/Mg/Fe prior. In the formation scenario inside the iceline (case B), we find very small envelope mass fractions with median values between $10^{-6}$ and $10^{-5}$. If we again compare with the results from the previous model version, we can see that for both planet~c and d, we also get H/He mass fractions that are negligibly small, while the water mass fraction is considerably larger than for case~A1, which also uses a stellar Si/Mg/Fe prior and assumes that the planet accreted icy planetesimals outside the iceline. This result is expected, as we assume the water to be condensed and fix the temperature to 300~K at the water to H/He boundary in the previous model version. In conclusion, we find that planets~c and d cannot have a H/He dominated envelope, as mass fractions as small as the ones our posteriors showcase would have quickly been lost through evaporation. If the planets do have an envelope, it needs to be of higher metallicity.

\section{Hydrodynamic modelling}
\label{sec:hydro_modelling}

\subsection{Model setup}
To analyse what impact the different atmospheric compositions considered in Section\,\ref{sec:internal_structure} would have on the upper atmosphere's parameters and atmospheric mass loss, and hence also on the atmospheric evolution, we employ the 1D hydrodynamic model by \citet[][Cloudy e Hydro Ancora INsieme; CHAIN]{kubyshkina2024}. This code combines a 1D hydrodynamic upper atmosphere model based on that of \citet{kubyshkina2018grid} with the non-local thermodynamical equilibrium (NLTE) photoionisation and radiative transfer code Cloudy \citep{Ferland2017}. The former is responsible for modelling the hydrodynamic outflow, while the latter solves the detailed (photo)chemistry and level populations for elements up to zinc, providing realistic heating and cooling functions. Given the flexibility of Cloudy in terms of gas compositions (the code supports switching on and off any element except for hydrogen and adjusting their abundances, allowing or prohibiting the formation of non-homogeneous molecules, and including dust), the CHAIN code allows modelling of a wide range of hydrodynamically escaping atmospheres. The hydrodynamic part assumes the atmosphere to be well-mixed within the simulation domain. This is an adequate assumption while the exobase level lies above the sonic point and is likely met by the majority of close-in exoplanets. We verify this during the simulation. Further, the atmosphere is treated as a uniform outflow with the mean molecular weight set by the employed composition. 

The code accounts for the detailed spectral energy distribution (SED) of the host star and treats X-ray, extreme ultraviolet (EUV), and visible plus infrared (VIS+IR) parts of the spectra as separate sources set by their flux-wavelength dependence and the total flux within the given wavelength interval. To the best of our knowledge, no X-ray/EUV data is available for HIP 29442. Therefore, for our simulations, we employ the X-ray and EUV values predicted by the stellar evolution code Mors \citep{Johnstone2021,Spada2013} for the age and mass of the star given in Table~\ref{tab:stellar_params}; we further scale the VIS-IR flux to the $L_{\rm bol}$ of HIP 29442. For the shape of the spectrum, we employ that of a star of similar spectral type, i.e. HD\,97658, provided by the Measurements of the Ultraviolet Spectral Characteristics of Low-mass Exoplanetary Systems survey \citep[MUSCLES{;}][]{france2016,Youngblood2016,Youngblood2017,Loyd2016,Loyd2018}. We briefly discuss the implications of choosing this specific spectrum at the end of the section.

Details concerning the code implementation and testing for hydrogen-helium atmospheres (with and without heavier elements in atomic form) can be found in \citet{kubyshkina2024}. In this study, we focus on comparing atmospheres with hydrogen-helium and water-rich compositions. As the simulations with CHAIN (particularly for water-rich atmospheres) are computationally expensive, we cannot perform them for the whole range of parameters predicted in Sections \ref{sec:data+analysis} and \ref{sec:internal_structure}. Therefore, we employ the mean values from Table~\ref{tab:planetary_params} for the radii, masses, orbital separations, and equilibrium temperatures of the planets in the \starname \ system. We note that these parameters were constrained well in the present study, and the variations within $1\sigma$ for planetary radius and orbital parameters are not expected to have any considerable effect on the predictions of CHAIN; the variations in mass can lead to changes in the predicted atmospheric mass loss rates of about $\sim$10\%. For each of the planets, we further consider two limiting cases: pure hydrogen-helium atmospheres with stellar metallicities (further referred to as H-He atmospheres) and the posteriors of model A1 (water accreted in the form of ice and stellar  Si/Mg/Fe ratios; further referred to as H$_2$O atmospheres). More specifically, the latter denotes atmospheric water mass fractions of $\sim$65\% for the outermost planet~b and of $\sim$99\% for the inner planets c and d. We tested our models for different metallicities and found that for all the planets in the \starname \ system, the variations in Si/Mg/Fe ratios considered in Section\,\ref{sec:internal_structure} have a minor effect on the results; therefore, we only consider here the case of atmospheres with stellar metallicity.

\subsection{Atmospheric mass loss rates}
\begin{figure*}
    \centering
    \includegraphics[width=0.9\linewidth]{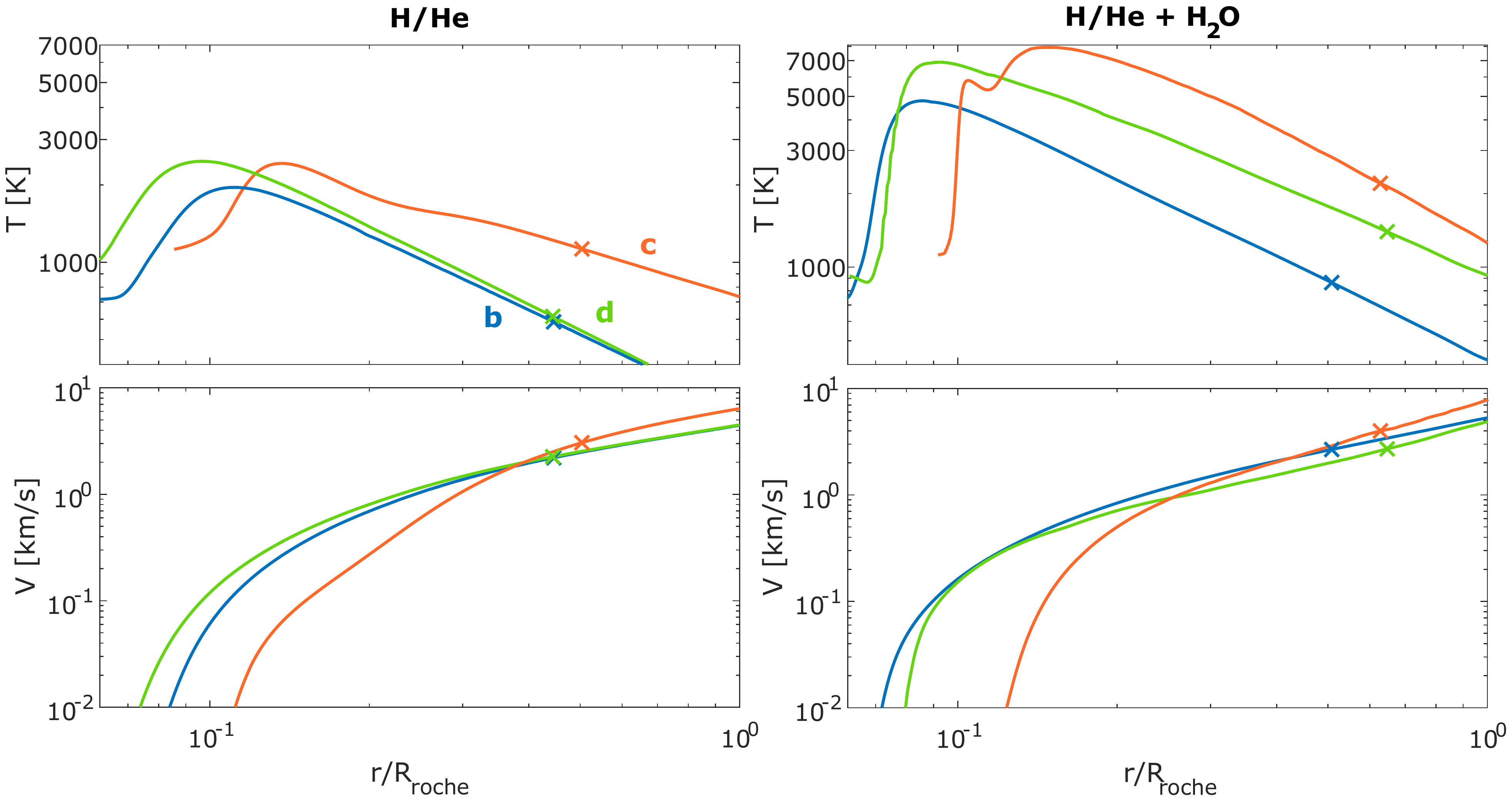}
    \caption{Atmospheric temperature (left Y-axis, solid lines) and bulk outflow velocity (right Y-axis, dashed lines) against radial distance normalised to the Roche radii of the planets, as predicted by CHAIN models. The top panel shows results for the H-He atmospheres and the bottom panel for H$_2$O atmospheres; different colours correspond to planets b (blue), c (orange), and d (green). Crosses denote the positions of sonic points.}
    \label{fig:hydro_TV_profiles}
\end{figure*}

From the hydrodynamic simulations, we found that in the case of H-He atmospheres, the general picture of the atmospheric outflow is rather similar for the two outer planets b and d: despite the density of planet~b being $\sim$4.25 times lower than that of planet~d, in terms of the atmospheric escape, this is compensated by the planet's lower temperature and lesser X-ray/EUV fluxes at its orbit. Planet~c, due to its shorter orbit, experiences stronger XUV heating and more intense outflow. 
We summarise the information about atmospheric escape from the planets in the \starname \ system in Table\,\ref{tab:hydro_escape} and show the height profiles of their atmospheric temperatures and bulk outflow velocities in Figure\,\ref{fig:hydro_TV_profiles}, for both H-He and H$_2$O atmospheric compositions. One can notice that the inclusion of large water fractions into the atmosphere does not have the same effect for different planets, and the effect does not solely depend on the total water fraction. 
\begin{table}
\renewcommand{\arraystretch}{1.5}
\caption{Atmospheric loss rates predicted for planets \starname~b, \starname~c and \starname~d by CHAIN.}
\centering
\begin{tabular}{lcccc}
\hline\hline
Planet & Model$^*$ & X-ray+EUV & escape rate & mass loss rate \\
       &           & [${\rm erg/s/cm^2}$] & [$10^{33}$/s] & [$10^9$g/s]  \\
\hline
b & H-He & 703.6 & 2.21        & 4.80         \\
b & H$_2$O  & 703.6 & 0.43        & 2.01         \\
c & H-He & 4250  & 13.95        & 30.27         \\
c & H$_2$O  & 4250  & 0.40        & 4.66         \\
d & H-He & 1917  & 0.66        & 1.43         \\
d & H$_2$O  & 1917  & 0.023        & 0.27         \\
\hline
\end{tabular}
\label{tab:hydro_escape}
\tablefoot{$^*$H-He -- hydrogen-helium atmospheres; H$_2$O -- atmospheres with large water fraction (model A1). The Si/Mg/Fe ratio is set to stellar values in both cases.}
\renewcommand{\arraystretch}{1.0}
\end{table}

The change in the atmospheric escape is the smallest for the outermost planet~b, which also has the smallest water fraction in the atmosphere according to the structure models. For better clarity, we include in Table\,\ref{tab:hydro_escape} both the escape rate (i.e., the number of particles escaping per unit time) and the mass loss rate (i.e., mass escaping per unit time). Thus, for planet~b, the escape rate reduces with the inclusion of water by a factor of 4.72. However, due to the increasing fraction of massive particles (mainly oxygen atoms) in the escaping material, the mass loss rate decreases by about a factor of two, which can be considered negligible from the point of view of atmospheric evolution \citep[e.g., see the discussion in][]{kubyshkina2024}. 

For the compact atmosphere of planet~d, despite the outflow parameters being very similar to those of planet~b in the case of H-He atmospheres, the inclusion of water leads to significant changes. Namely, the escape rate drops by about 30 times, and the mass loss rate decreases by a factor of 5.3, which can effectively cease the atmospheric loss.
For the innermost planet~c, the water mass fraction for the H$_2$O atmosphere predicted by internal structure models is almost equivalent to that of planet~d, and their masses and radii are also similar. However, due to the higher temperature and XUV flux at the orbit of planet~c, the photodissociation of water in its upper atmosphere becomes considerably more effective. This implies that despite a large water reservoir in the lower atmosphere, the water molecule fraction in the region where photoionisation heating occurs remains minor relative to hydrogen and helium (maximum H$_2$O/H$^0$ number fraction is comparable to that of planet~b and roughly 80 times lower than for planet~d). In contrast, fractions of ions are $\sim$3 times higher than for planet~d. Thus, water molecule cooling becomes less effective; meanwhile, the number density of atomic oxygen released from water molecules in photodissociation reactions increases and the radiative oxygen line cooling, acting at higher altitudes, becomes gradually more prominent with increasing atmospheric temperature (see the dark-yellow dotted line in Figure~\ref{fig:hydro_cooling}). Therefore, despite the lesser number of water molecules in the upper atmosphere, the reduction in the escape becomes more significant for planet c than in the case of planet~d (about a factor of 35 for the escape rate and a factor of about 6.5 for the mass loss rate).

\subsection{Cooling rates}
\begin{figure}
    \centering
    \includegraphics[width=\linewidth]{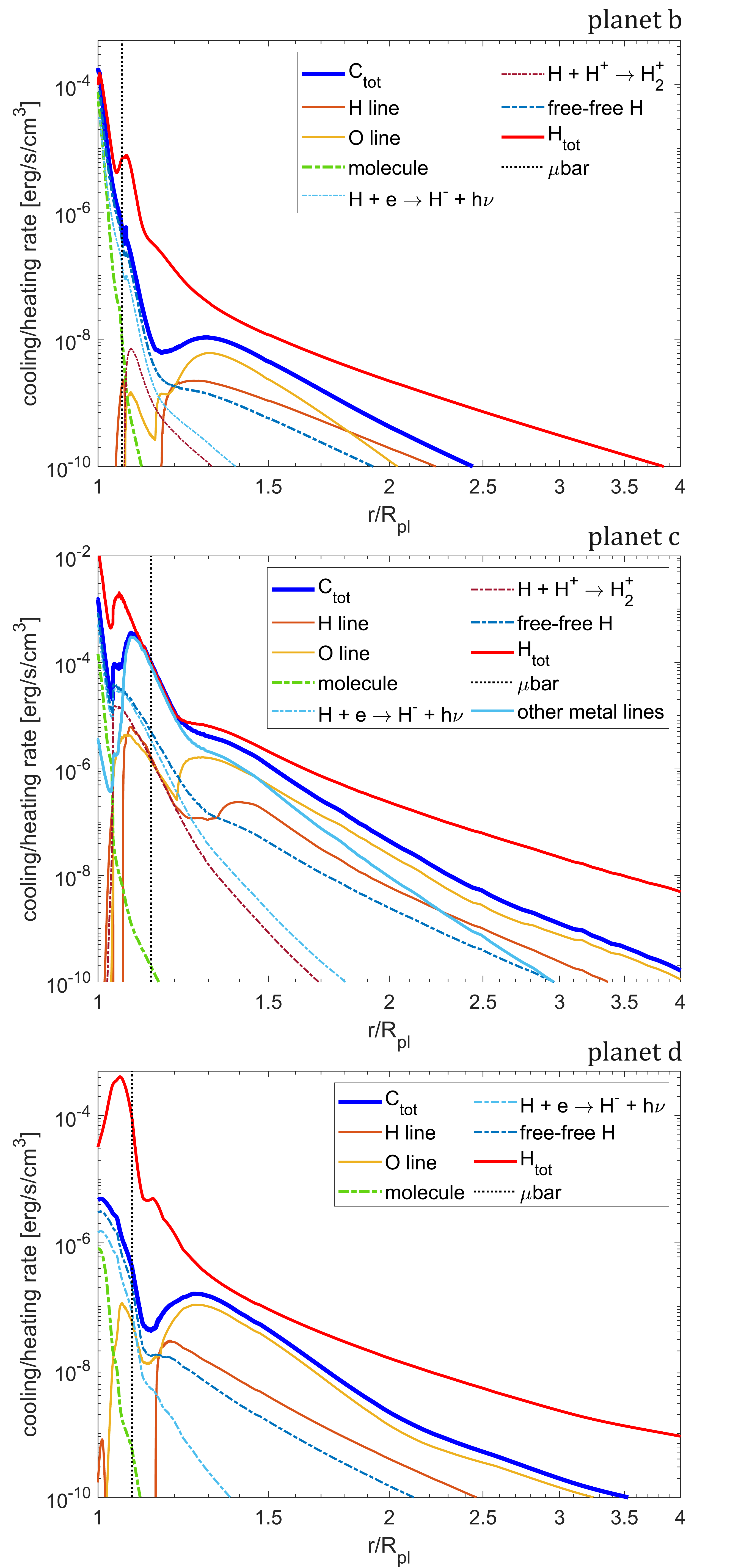}
    \caption{Radiative cooling/ heating rates against radial distance for the CHAIN runs of the H$_2$O atmospheres of planets b (top panel), c (middle panel), and d (bottom panel). The rates are shown between 1 and 4 $R_{\rm pl}$, where most of the photochemistry takes place; higher altitudes are dominated by adiabatic expansion. For individual cooling processes, we only show those that contribute at least 5\% to the total cooling rates (at some specific altitude). The line styles are explained in the legends, where the notation is the following: $C_{\rm tot}$ -- total cooling rate (the sum of all radiative processes); $H_{\rm tot}$ -- total heating rate; $\mu$bar -- the altitude corresponding to 1 $\mu$bar pressure level; X line -- line cooling of the X element; molecule -- the sum of molecule cooling processes (dominated by ${\rm H_2O}$ molecule cooling); free-free H -- bremsstrahlung cooling from hydrogen and helium.}
    \label{fig:hydro_cooling}
\end{figure}

We show the dominant (radiative) cooling processes in the H$_2$O atmospheres of planets~b, c, and d in Figure~\ref{fig:hydro_cooling}. Molecular (H$_2$O) cooling contributes up to 10\% of the total cooling in the lowermost atmospheric layers of planet~c,  and up to 43\% and 27\% for planets b and d, respectively. At higher altitudes, near the maximum of the photoionisation heating (the second peak of H$_{\rm tot}$), the cooling is dominated by the line cooling of H and O, bremsstrahlung cooling (deceleration of free electrons due to the electromagnetic interaction with H ions, also known as free-free H radiation), production of H$^-$, and recombination of H$_2^+$, similarly to the H-He case. For planet c, the hottest in the system, other metal lines also contribute considerably in this region. Despite the input from individual lines being not that high, their joint contribution dominates the cooling between $\sim1.06-1.22$\,$R_{\rm pl}$; the largest input, exceeding 10\% of the total cooling rate each, comes from magnesium and calcium lines.
The thorough testing of our models shows that the exact picture of the heating/cooling processes acting in the atmospheric region below the 1\,$\mu$bar level (shown by the black dotted line in Figure~\ref{fig:hydro_cooling}) depends strongly on a range of factors including the chosen lower boundary conditions and the shape of the SED employed in the simulations (in particular, the visible and infrared wavelengths), while the processes at the higher altitudes show a negligible dependence on these factors. However, we also find that the changes in heating/cooling rates in the densest atmospheric parts have a limited effect on the atmospheric loss rates; therefore, comparing the H-He and H$_2$O atmospheres remains valid as long as one keeps the boundary and model parameters consistent for different atmospheric compositions.

\subsection{Abundances of different atmospheric constituents}
\begin{figure}
    \centering
    \includegraphics[width=\linewidth]{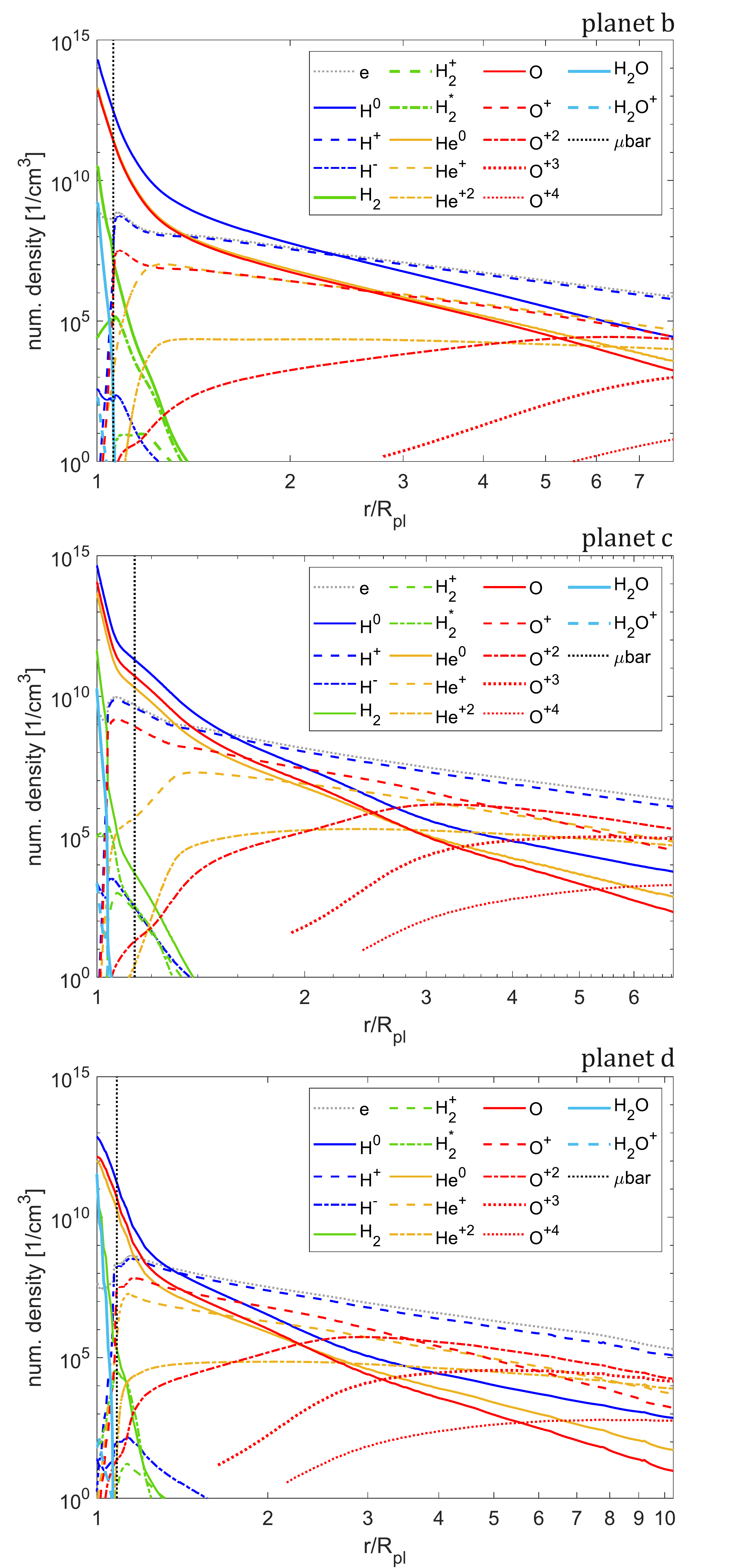}
    \caption{Numerical densities of relevant elements against the radial distance from the CHAIN runs of the H$_2$O atmospheres of planets b, c and d. The different line styles correspond to different species, as explained in the legend. The $\Sigma$H entry corresponds to the summed density of all hydrogen species (H, H$^+$, H$^-$, H$_2$, H$_2^+$, H$_2^*$, and H$_3^+$).}
    \label{fig:hydro_pops}
\end{figure}

Besides the differences in the total atmospheric loss rates and temperatures, the H-He and H$_2$O models differ significantly in terms of the species that are being lost. We show the height abundances of different atmospheric constituents in H$_2$O models in Figure~\ref{fig:hydro_pops}, and summarise the number and mass fractions of different atoms and ions in the escaping material (taken at the sonic point) for all models in Table~\ref{tab:hydro_efractions}. We note that the total fraction of specific elements (sum of free neutral atoms, ions, and constituents of different molecules) at all altitudes is fixed due to our model assumptions (see above) and a more accurate treatment requires a multi-fluid approach \citep[such as e.g.][]{Schulik_Booth2023}. However, the ion/molecule fractions are defined with a detailed chemistry framework.
\begin{table*}
\renewcommand{\arraystretch}{1.5}
\caption{Elemental fractions at the sonic point for planets \starname~b, \starname~c, and \starname~d predicted by CHAIN.}
\centering
\begin{tabular}{lccccccccc}
\hline\hline
Planet & Model & ${\rm H^0}$ & ${\rm H^+}$ & ${\rm He^0}$ & ${\rm He^+}$ &  ${\rm He^{+2}}$ & ${\rm O^0}$ & ${\rm O^+}$ & ${\rm O^{+2}}$ \\
       &           & [\%]        & [\%]        &     [\%]     & [\%]         &  [\%]            &  [\%]       &  [\%]       &  [\%]           \\
\hline
 b & H-He & 31.02/24.30 & 59.80/46.86 &  2.78/8.65  & 6.08/18.92 &  0.23/0.71        &  0.02/0.24 & 0.02/0.30 & $7.4\times10^{-4}$/0.01  \\
 b & H$_2$O  & 2.92/1.22 & 80.81/33.89 & 0.38/0.64 & 6.34/10.57 &  1.64/2.74          &  0.17/1.11 &  3.44/23.10   &  3.98/26.72    \\
 c & H-He & 54.67/42.84 & 36.15/28.32  & 5.03/15.67 & 3.99/12.42  &  0.06/0.19       &  0.03/0.38 & 0.01/0.18 & $5\times10^{-5}$/$6\times10^{-4}$  \\
 c & H$_2$O  & 0.37/0.11 & 73.03/22.20 & 0.05/0.06 & 4.33/5.23  & 2.96/3.58   &  0.01/0.07   &  2.17/10.55  &  11.97/58.20  \\
 d & H-He & 19.71/15.44 & 71.11/55.72  &  1.75/5.46  & 6.84/21.28  &  0.49/1.53        &  0.01/0.15 & 0.03/0.37 & 0.003/0.04  \\
 d & H$_2$O  & 0.44/0.15  & 72.95/24.70  &  0.03/0.04   & 3.26/4.39   &  4.05/5.44         &  0.01/0.05  &  1.36/7.38   &  10.68/57.84   \\
\hline
\end{tabular}
\label{tab:hydro_efractions}
\tablefoot{The fractions of the elements are given in the format (number fraction)/(mass fraction). Note, that the total abundancies of each element (neutral atoms and ions together) were set based on the interior models and do not vary with altitude; the ion/neutral and atom/molecule fractions are set by the chemical framework.}
\renewcommand{\arraystretch}{1.0}
\end{table*}

For all three planets, the abundance of water molecules in H$_2$O models maximises at low altitudes and drops steeply with increasing height, thus their fraction above $\sim$1.5\,$R_{\rm pl}$ becomes negligible. The same holds for H$_2$, OH, and CO molecules. It is notable that in the H-He atmospheres, H$_2$ is the most common hydrogen species below the $\mu$bar level, while in the H$_2$O atmospheres most of the H$_2$ is bound in water molecules. The same occurs for H$_2^+$. H$_2^+$, in turn, is necessary for the formation of H$_3^+$, which is therefore not present in H$_2$O atmospheres in significant amounts. Thus, at higher altitudes, the inclusion of water leads to an increase in abundance of neutral and ionised oxygen, and in the region where the atmospheric material can be considered escaped (as such we take the sonic point), the outflow consists mainly of hydrogen, helium, and oxygen ions. The fraction of neutral oxygen and helium atoms remains small in all cases. For oxygen, the number fraction at the sonic point constitutes 0.02--0.13\%, and the mass fraction 0.05--0.85\%, highest for planet~b; for neutral helium, the number and mass fractions in H$_2$O atmospheres vary between 0.03--0.32\% and 0.03--0.54\%, respectively, while for H-He atmospheres these ranges are 1.24--5.03\% and 3.85--15.67\%. The mass fraction of He$^+$ in the outflow varies between 2.97--24.03\% in all models, being higher for H-He models, while that of He$^{+2}$ is 0.19--3.64\%, being higher for H$_2$O atmospheres (in particular for hotter planets). The fraction of oxygen ions (O$^+$ and O$^{+2}$) at the sonic point in H$_2$O models dominates over H$^+$ in terms of mass, though not number, and O$^{+2}$ ions constitute more than half of all oxygen ions at the sonic point for the two inner planets. Finally, though not included in Table\,\ref{tab:hydro_efractions}, the fraction of O$^{+3}$ ions is non-negligible for the two inner planets (mass fractions of 2.9\% and 4.5\% for planets c and d, respectively), and fractions of O$^{+4}$ and O$^{+5}$ are non-zero.

Another result, potentially interesting for observations, is that the total fraction of neutral atoms at the sonic point and, in general, at high altitudes (where the escape can be probed in, e.g., Ly$\alpha$) is significantly higher in H-He atmospheres (by a factor of $\sim$7--64, increasing with equilibrium temperature). For 99\% water atmospheres, the number fraction of H$^0$ in the escaping material is less than 1\%, while most of the escaping hydrogen is ionised. This happens due to the increasing mean molecular weight of the atmosphere, making the gradients in density/pressure height profiles steeper, in particular at low altitudes, and, therefore, the ionisation front becomes narrower, though the ion fraction integrated over the whole simulation domain does not change as much. This can result in a decrease in the detectability of the escape in hydrogen/helium lines for the water-rich atmospheres compared to the pure hydrogen-helium atmospheres, even if the actual atmospheric mass loss rates are similar. This might lead to the non-detection of the atmospheric escape for some planets even if the theoretically predicted mass loss rates are high.
We note, however, that the simulations presented above were performed employing the model configuration not including the adiabatic wind effect into the advection term within Cloudy runs \citep[which is still the experimental part of Cloudy framework and therefore is not included in the `default' configuration of CHAIN, though it is physically relevant; see a detailed discussion in][]{kubyshkina2024}. The inclusion of this feature can lead to an increase of the fraction of neutral atoms in the outflow of up to a factor of a few, in particular for planet b with a puffier atmosphere. However, the outflow would remain ion-dominated.

\section{Possible formation and evolution pathways}
\label{sec:formation_evolution}

To discuss possible formation and evolution pathways of the \starname\ system, we compare the observed three-planet system to synthetic planetary systems generated using the Bern model \citep{Alibert+2005, Mordasini+2009, Emsenhuber+2021a, Emsenhuber+2021b}, a coupled formation and evolution model. In the following, we use a variation of the nominal population of the New Generation Planetary Population Synthesis \citep[NGPPS;][]{Emsenhuber+2021a, Emsenhuber+2021b} for Solar mass stars with a formation phase extended from 20 to 100~Myr \citep{Emsenhuber+2023} and an improved evolution model \citep{Burn+2024}. Each of the 1000 generated systems is given a system ID as an identifier and starts with 100 randomly distributed lunar-mass planetary embryos. Meanwhile, the initial conditions for the protoplanetary disks are sampled from observationally informed distributions using a Monte Carlo approach.

During the formation phase, the accretion of planetesimals and gas by these embryos is modelled along with the evolution of the gas disk, dynamical interactions between the embryos and their gas disk driven orbital migration. The long-term evolution of each planet is then modelled individually until an age of 5~Gyr, including atmospheric photoevaporation of both water and H/He, migration due to stellar tides and the cooling and contraction of the planetary interior. For the internal structure, it is assumed that the accreted water is uniformly mixed with any present H/He once the evolution phase of the model is reached.

Currently not yet included in this population are more recently studied effects such as hybrid pebble and planetesimal accretion \citep{Kessler+2023, Alibert+2018}, the dynamic formation of planetesimals and planetary embryos from dust and pebbles \citep{Voelkel+2020, Voelkel+2021}, planetesimal fragmentation \citep{Kaufmann+2023}, MHD-wind driven disk evolution \citep{Weder+2023}, the influence of structured disks \citep{Lau+2022, Jiang+2023} and a description of orbital migration based on torque densities \citep{Schib+2022}.

To identify possible formation and evolution pathways of \starname, we now search this population for synthetic analogues of the observed system, thereby following a similar approach as \cite{Ulmer-Moll+2023}.

\subsection{Finding analogues for the \starname \ system in the synthetic Bern model population}

\begin{figure*}
    \centering
    \includegraphics[width=0.89 \textwidth]{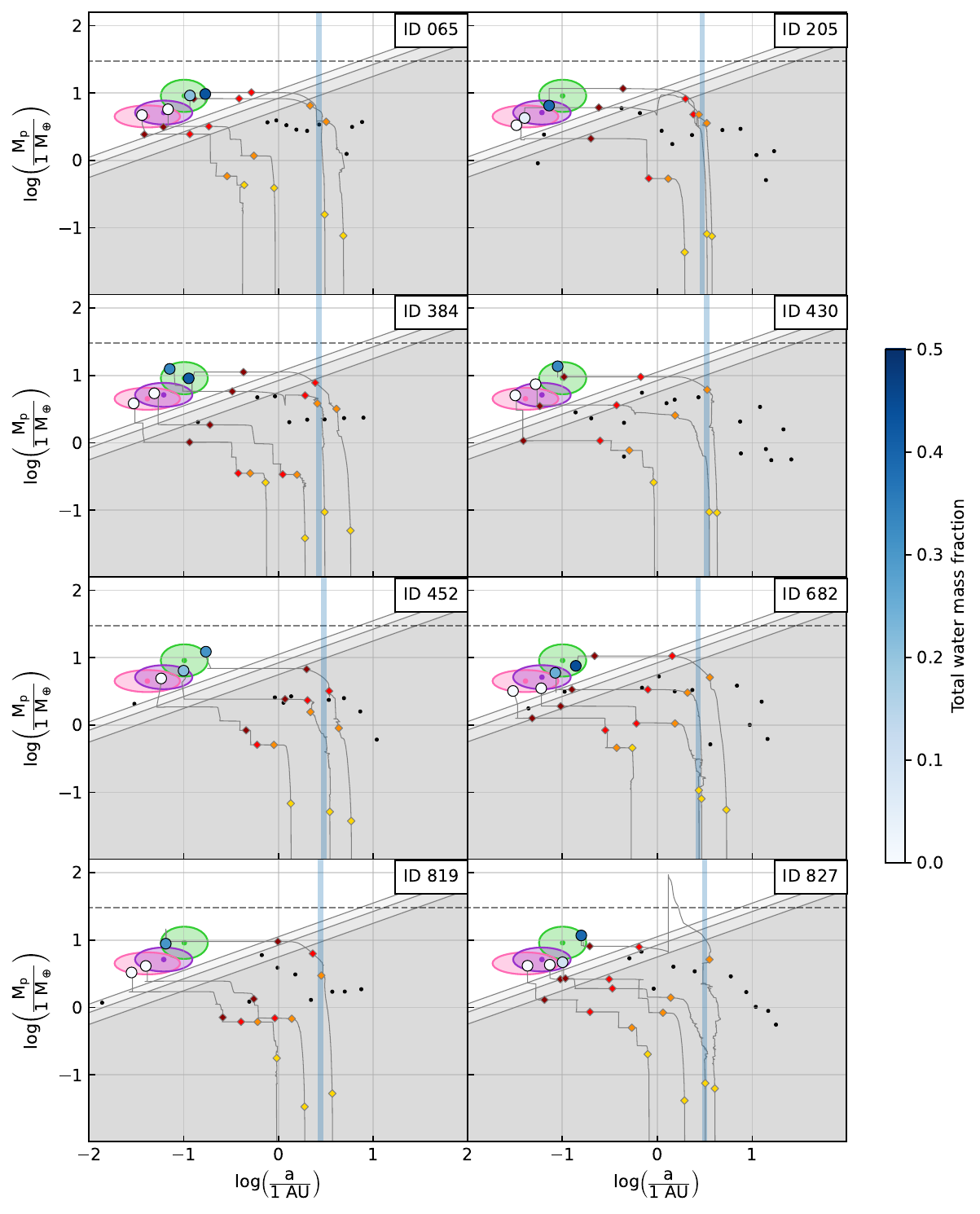}
    \caption{Planetary mass versus semi-major axis plots of eight synthetic planetary system analogues for the observed \starname \ system (one per panel, with the Bern model system ID indicated in the upper right corner). These planetary system analogues are taken from the nominal NGPPS population \citep{Emsenhuber+2021a, Emsenhuber+2021b} for Solar mass stars with an extended formation phase \citep{Emsenhuber+2023} and an improved evolution model \citep{Burn+2024}. The green, pink and purple ellipses show the observational values for planets b, c and d respectively, with a 25\% tolerance in log that was used when identifying synthetic analogues for each planet. The areas shaded in grey show simple RV biases of 0.50, 0.75 and 1.00 m/s respectively. Synthetic planets below the 1 m/s RV bias are likely not observable and are shown as black dots, down to a mass of 0.5~M$_\oplus$. For the synthetic planets above an RV bias of 1 m/s, which form the part of the planetary system that could be observed, the planet's total water mass fraction is indicated by the marker colour, while their formation tracks are shown as black lines. For each formation track, the temporal evolution is indicated by diamond-shaped markers, which show the position of the planet at 0.1~Myr (yellow), 1~Myr (orange), 2~Myr (red) and 3~Myr (darkred). The vertical light blue line indicates the location of the water ice line of each synthetic system, which is assumed to be constant. The dashed black line indicates a mass of 30~M$_\oplus$, corresponding to the upper limit for the critical core mass given by \cite{Bodenheimer+Pollack1986}.}
    \label{fig:bern_model_systems}
\end{figure*}

\begin{figure*}
    \centering
    \includegraphics[width=\textwidth]{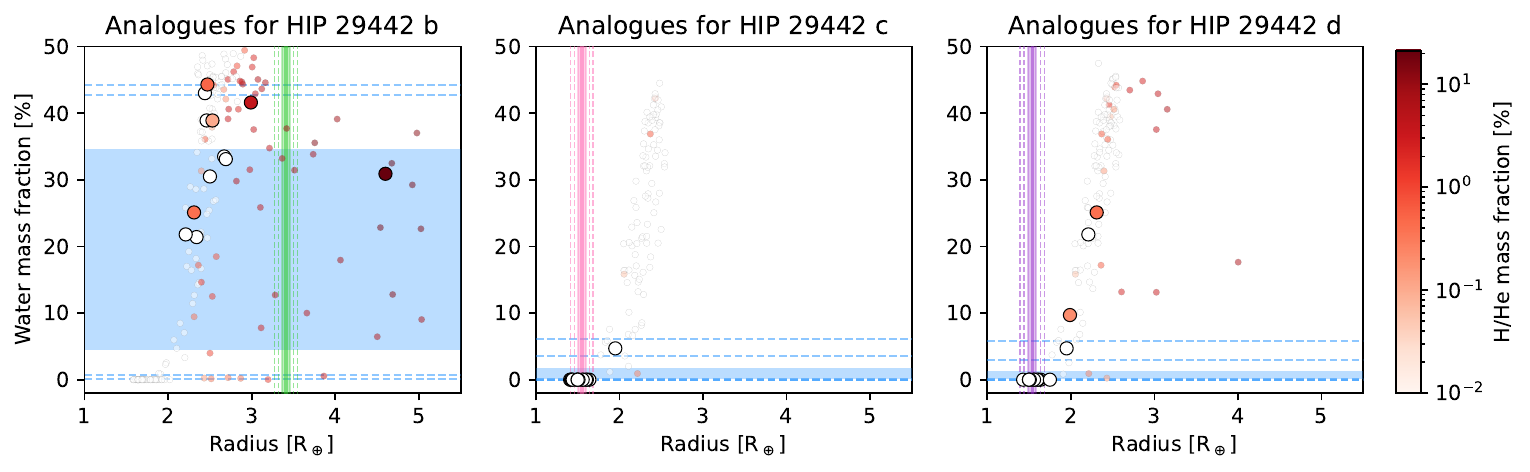}
    \caption{Water mass fraction versus planet radius with markers colour-coded according to the H/He mass fraction for the synthetic analogues of \starname~b, c and d. The large points are the synthetic analogues of planet~b, c and d considering the full planetary system, while the transparent, smaller points in the background are the synthetic analogues for the single planet system of each planet. The areas shaded in green, pink and purple respectively indicate the 1$\sigma$ confidence interval of the observational radius for each planet, while the areas shaded in blue are the water mass fractions inferred from the internal structure modelling (model A1 with stellar Si/Mg/Fe prior and assuming a formation outside the iceline). Dashed lines indicate the 2$\sigma$ and 3$\sigma$ values of the respective distributions. Synthetic planets that fall in the shaded areas are good synthetic matches for the observed planets.}
    \label{fig:bern_model_characteristics_planets}
\end{figure*}

To define how close a synthetic planet is to one of the observed planets, we choose the logarithmic distance in the semi-major axis versus mass plane as our metric, analogous to \cite{Kaufmann+2023}. Using this metric, we then define that a synthetic planet qualifies as an analogue for one of the observed planets if it lies in an elliptical area in the semi-major axis versus mass space with vertices and co-vertices at a 25\% variation in log of the semi-major axis and mass of the observed planet. As the size of the synthetic population is limited to 1000 systems, the dimensions of these elliptical areas were chosen rather liberally and they do overlap for the three \starname \ planets, meaning that in some cases a single synthetic planet can be an analogue for two of the observed planets at once. We now select synthetic systems with at least one analogue for each of the observed planets. From this subset of the synthetic population we eliminate systems that, after applying a simple RV bias of 1 m/s, contain less than two or more than four observable planets, systems where one or more of these observable planets are giants (M$_\textrm{p} > 30\  \textrm{M}_\oplus$), and also systems where the order of the synthetic planets does not match the observed system, due to the overlap of the elliptical areas used to identify synthetic analogues for the observed planets. This leaves us with the eight synthetic systems pictured in Figure \ref{fig:bern_model_systems}.

In addition, we also identify analogues for each of the three observed planets separately, which we will in the following refer to as synthetic single planet analogues. This allows us to make comparisons later on in this analysis between the planets of a certain kind present in the NGPPS population in general and the ones in systems like \starname. We find in total 200 synthetic single planet analogues for planet~b, 227 for planet~c and 238 for planet~d.

Figure \ref{fig:bern_model_characteristics_planets} summarises the radii, water and H/He mass fractions of the analogues for each of the three observed planets. Synthetic planets that fall in the areas shaded in colour are good synthetic matches for the observed planets with respect to their radii and water mass fractions inferred in Section \ref{sec:internal_structure}.

The radii of the synthetic planets (both for the system analogues and the single planet analogues) generally fit reasonably well with the ones derived in Section~\ref{sec:data+analysis} for \starname ~c and d, with some outliers with clearly larger radii especially for planet~d. These outliers tend to have water mass fractions larger than a few percent and are from the higher end of the defined mass range where a synthetic planet counts as an analogue for the observed planet. In contrast to planets~c and d, planet~b's observed radius lies more than 3 standard deviations above the average value for its synthetic system analogues. One possible explanation for this is the location of \starname ~b above the radius cliff, which also naturally emerges in the synthetic population \citep{Burn+2024} and means that the occurance rate of planets with radii above 3 R$_\oplus$ is much smaller than the one of planets between 2-3 R$_\oplus$. To find exact analogues of the \starname \ system, we would therefore likely need a larger synthetic population. This is to some extent confirmed when we also look at the single planet analogues for \starname ~b (small dots in Figure \ref{fig:bern_model_characteristics_planets}), as most of these synthetic planets have radii below 3 R$_\oplus$. However, there are also other possible explanations, for instance that the NGPPS population was generated for a fixed stellar mass of 1 M$_\odot$, which is slightly higher than the value we derived for \starname \ in Section~\ref{sec:star}. It is also possible that the H/He loss rate in the synthetic population is currently too high and the synthetic planets therefore are too small and have a H/He mass fraction that is too low. This of course would then also affect the single planet analogues. We did compare the present-day mass loss rates obtained in the previous section (see Table~\ref{tab:hydro_escape}) to the ones for the synthetic planets at an age of 5~Gyr, and found values of the same orders of magnitude. However, this does not exclude differences in the mass loss rates earlier in the stellar evolution, when the evaporation is much stronger. 

A large majority of the synthetic analogues of \starname ~c and d have lost their entire H/He envelope through photoevaporation. They also mostly do not contain any or only a small amount of water, up to a few percent in mass, with the exception of a few analogues for \starname ~d as discussed above. This is compatible with our results from Section~\ref{sec:internal_structure}, where we find envelope mass fractions of a few percent for a formation scenario outside the iceline with a very high envelope water mass fraction and very small envelope mass fractions of the order of $\sim$10$^{-6}$ for a water-poor formation. In the latter case, these envelopes would likely have been lost during the planet's evolution. The full system analogues of \starname ~b have also lost most if not all accreted H/He, with only two synthetic planets that have a mass fraction of more than 1\% of H/He. However, they all showcase water mass fractions between 20 and 45\%. Compared to the results from our internal structure analysis for a water-rich formation, the synthetic planets seem to have higher water mass fractions than what our internal structure analysis predicts, while at the same time more H/He seems to have been lost. This is in agreement with the scenario presented above of a H/He loss rate that is too high in the synthetic population, which could also explain lower planetary radii for the \starname~b analogues. Another possibility is that the evaporation of water is underestimated in the Bern model. It is also possible that some of the water would in reality be stored in the interior of the planets \citep[e.g.][]{Dorn+Lichtenberg2021}, which makes it more difficult to compare the water mass fractions for the synthetic Bern model planets with the ones derived in Section \ref{sec:internal_structure}. As is discussed further in Section \ref{sec:discussion}, the reason for this is that the water mass fractions that are inferred by \texttt{plaNETic} do not refer to the total water content of each planet, but only to the water that is not dissolved in the mantle and core.

\subsection{Patterns in the formation pathways of the Bern model analogues}
Looking at the formation tracks of these eight synthetic systems as plotted in Figure \ref{fig:bern_model_systems}, all of them would be classified as hybrids of classes I (compositionally ordered Earth and ice world systems) and II (migrated sub-Neptune systems) according to the system architecture classifications introduced in \cite{Emsenhuber+2023}. The corresponding analogues for \starname~b form at a distance between 3 and 6~AU from their host star outside the iceline. Through the accretion of planetesimals, they grow to a mass of around 5-10~M$_\oplus$. At this point, they start migrating inwards while still accreting mass. Once they reach the equality mass or the saturation mass \citep[the mass scales where migration becomes dominant over accretion; see][]{Emsenhuber+2023}, they keep migrating inwards at constant mass and never reach the critical core mass needed for rapid gas accretion. One exception is the analogue for planet~b in system~827, which does reach the critical core mass, subsequently undergoes rapid gas accretion and reaches a mass of almost 100~M$_\oplus$, but then loses its envelope through a giant impact. This is in agreement with the findings of \cite{Venturini+2020b}, who also predict based on a pebble-based formation model that a planet with a mass and radius similar to \starname~b should have formed beyond the iceline and be water-rich.

The formation tracks of the analogues for \starname~c and d are a bit more diverse. Most of them initially form inside the iceline, at a distance between 0.4 and 2 AU, while some analogues for planet~d also initially form just outside the iceline at around 3 or 4 AU. As is also outlined in \cite{Emsenhuber+2023} for planets in class I systems, these planets first grow in-situ through the accretion of planetesimals until they reach the planetesimal isolation mass. While only the more massive planets outside the iceline reach a mass high enough for significant inward migration, the inner planets are caught in resonant chains and pushed inwards. As all planetesimals inside the iceline have been accreted at this point, solid growth is no longer possible and the planets migrate inwards at constant mass. Once the gas disk disperses, the planets start to undergo collisions and can grow further through giant impacts.

\subsection{Common properties of the synthetic protoplanetary disks}

\begin{figure*}
    \centering
    \includegraphics[width=0.99\textwidth]{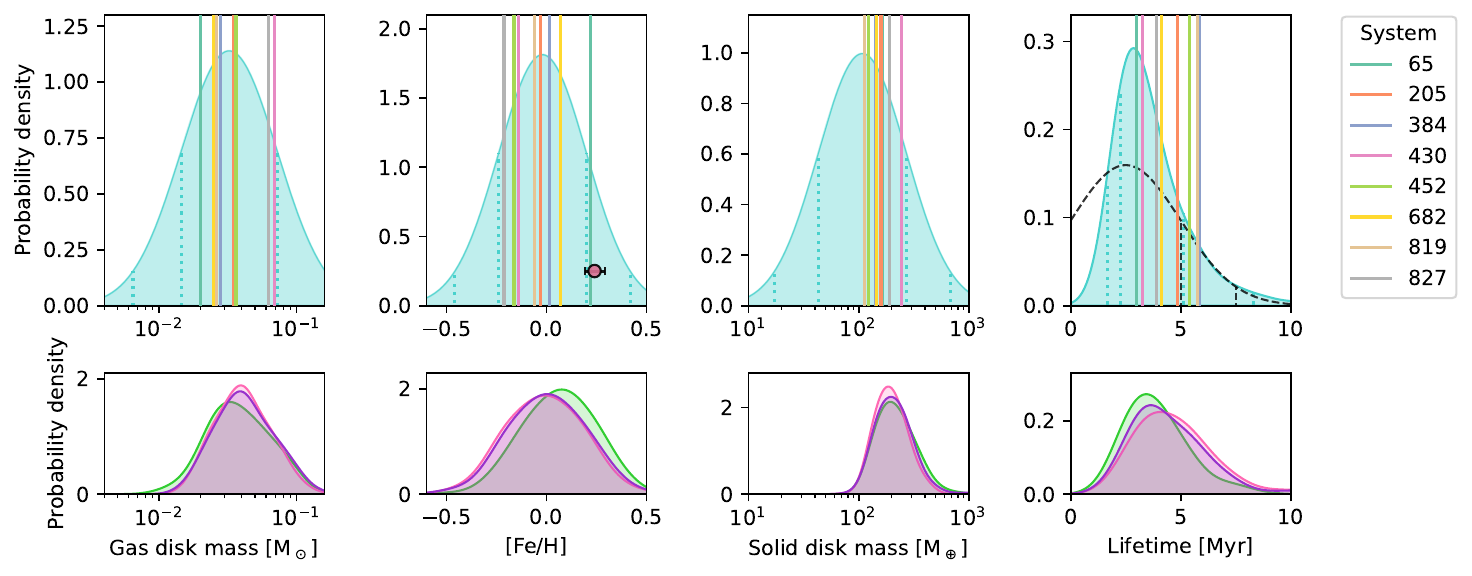}
    \caption{Properties of the protoplanetary disks of the synthetic system analogues for the \starname \ system (coloured lines) compared to the distributions they were sampled from \citep[turquoise curves, as specified by][]{Emsenhuber+2021b}. The dotted lines show the one and two sigma error intervals of the plotted distributions. The bottom row shows the distributions of the same disk properties for the single planet analogues for planets~b (green), c (pink) and d (purple). \textbf{Left:} Initial values of the gas disk masses with a probability density function fitted to \cite{Tychoniec+2018}. \textbf{Middle left:} Disk metallicities indicative of the disks' dust-to-gas ratios, sampled from \cite{Santos+2005}. The red point indicates the observationally derived metallicity of \starname. \textbf{Middle right:} Initial values of the solid disk masses, calculated from the sampled gas disk masses and dust-to-gas ratios and compared to the initial solid disk masses of all systems in the synthetic NGPPS population. \textbf{Right:} Disk lifetimes of the synthetic system analogues compared to the distribution of lifetimes calculated from the fraction of observed stars with disks at each age \citep[][black curve]{Mamajek2009} and the lifetimes of disks in the synthetic NGPPS population (turquoise).}
    \label{fig:bern_model_characteristics_disks}
\end{figure*}

As a next step, we study the initial conditions of the protoplanetary disks of the eight synthetic system analogues, as visualised in the upper panel of Figure \ref{fig:bern_model_characteristics_disks}, and compare them to the distributions they were sampled from (see \citeauthor{Emsenhuber+2021b} \citeyear{Emsenhuber+2021b} for details on the chosen initial conditions of the population synthesis) and to the distributions of the corresponding disk properties of the single planet analogues (lower panel in Figure \ref{fig:bern_model_characteristics_disks}). 

Most of the synthetic disks have initial gas disk masses between 0.020 and 0.036 M$_\odot$, while two of the systems have slightly more massive initial gas disks at 0.063 and 0.069 M$_\odot$. When comparing this to the probability density function fitted to the histogram of class I disks reported by \cite{Tychoniec+2018} from which the initial disk masses for the population were sampled, all values lie within one standard deviation from the mean and we do not see a clear pattern pointing towards high or low gas disk masses being required to form the system analogues of \starname \ (Figure \ref{fig:bern_model_characteristics_disks}, left). When looking at the distributions of the gas disk masses of the single planet analogues for \starname~b, c and d, we can however see that disks with gas masses at the lower end of the initial distribution do not seem to be able to produce planets of these masses that end up on such close-in orbits.

We now investigate the dust-to-gas ratios of the synthetic disks. For the model population, the relevant parameter that is sampled is the disk metallicity, which is assumed to match the one of the star exactly. The dust-to-gas ratio is then calculated as $\frac{f_{\textrm{D}/\textrm{G}}}{f_{\textrm{D}/\textrm{G},\odot}} = 10^{[\textrm{Fe}/\textrm{H}]}$, with $f_{\textrm{D}/\textrm{G},\odot} = 0.0149$ \citep{Lodders2003}. The stellar metallicities, in turn, are sampled from the distribution of \cite{Santos+2005} for the Coralie RV search sample. When we now look at the metallicities of the eight disks of our analogue systems (Figure \ref{fig:bern_model_characteristics_disks}, middle left), all but one are within one standard deviation of the mean of this distribution, and the one outside (system 65) is just above the one sigma upper bound. \cite{Emsenhuber+2023} find a positive correlation between the occurrence of sub-Neptune planets and the stellar metallicity, in agreement with observational results for the Kepler planets from \cite{Petigura+2018}. \cite{Mulders+2016} also find that hot exoplanets with orbital periods of less than 10 days are preferentially found around metal-rich stars. \starname \ is in fact a metal-rich star with a metallicity of [Fe/H]$=0.24 \pm 0.05$ (see Table \ref{tab:stellar_params}), which is in line with these studies. However, when looking at the stellar metallicity values of the system analogues for \starname, we do not find the same trend, which shows that a high stellar metallicity is not necessarily required to form a system like \starname \ in the synthetic population. However, the spectrally derived metallicity of \starname \ is compatible with the metallicity of system 65, but the metallicity values of all the other synthetic system analogues lie below the 3$\sigma$ lower bound of the observed value for \starname. In principle, we should restrict the search for planetary system analogues to initial conditions that also fit the observed stellar metallicity. However, this would require a larger synthetic population. We also note that the distribution of the metallicity values of the disks of the synthetic single planet analogues of planet~b is shifted slightly towards higher metallicity values compared the single planet analogues for planets~c and d.

While the initial mass of the solid disk is not a Monte Carlo variable in the population synthesis but is instead calculated as the product of the sampled gas disk mass and dust-to-gas ratio, it is one of the most relevant parameters for determining the type of planetary system formed by a disk \citep{Emsenhuber+2023}. As shown in the middle right panel of Figure \ref{fig:bern_model_characteristics_disks}, the solid disk masses of the eight system analogues of \starname \ all lie above the mean when compared to the entire NGPPS population, but within one sigma of the distribution. This is in agreement with the findings of \cite{Emsenhuber+2023}, who find that lower mass protoplanetary disks will not lead to the formation of sub-Neptunes, while disks with masses higher than this will likely lead to the formation of giant planets. We also see this when looking at the disks of the single planet analogues, which all have solid disk masses of at least 50~M$_\oplus$.

For the lifetimes of the synthetic protoplanetary disks, the relevant parameter that is sampled as part of the population synthesis in the Bern model is the external photoevaporation rate, in combination with the initial disk mass. The photoevaporation rate is sampled from an empirical distribution that leads to lifetimes of the synthetic disks that match the lifetimes of observed disks, with a mean value between 3 and 4 Myr. The right panel of Figure \ref{fig:bern_model_characteristics_disks} shows both the disk lifetimes derived from an exponential fit of observational data for the fraction of stars with disks at different ages \citep{Mamajek2009} and the distribution of the disk lifetimes of the systems in the NGPPS population. When comparing these distributions to the lifetimes of the eight synthetic disks in our sample, there seems to be a tentative trend towards disks with longer lifetimes that are needed to form a system like \starname. The lifetimes of all eight synthetic disks lie above the mean value of \cite{Mamajek2009}, with three of them also lying above the 1$\sigma$ upper bound. Also this correlation aligns with the conclusions drawn by \cite{Emsenhuber+2023}, who point out that disks with longer lifetimes are more likely to form systems containing migrated sub-Neptunes as the planets outside the iceline then have more time to reach a mass of about 10 M$_\oplus$ and migrate inwards before the gas disk disperses.

\subsection{Resonances}

\begin{figure}
    \includegraphics[width=0.45\textwidth]{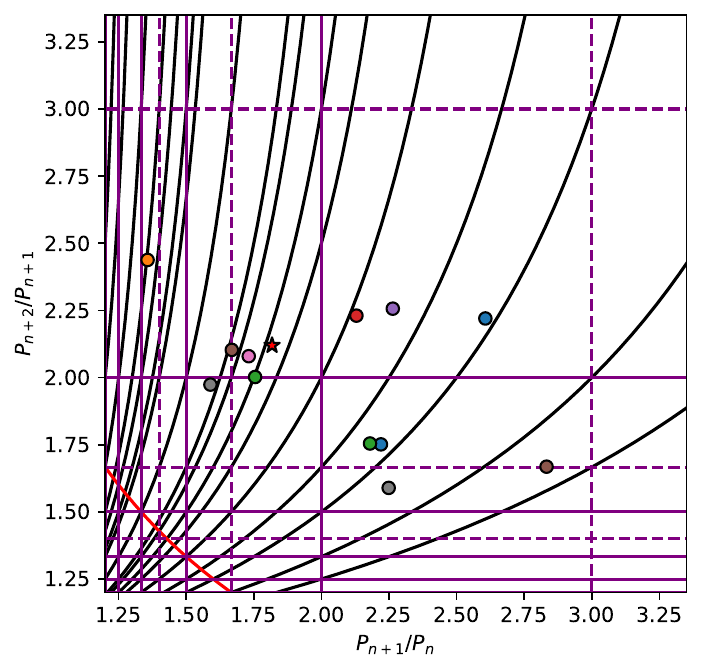}\\
    \includegraphics[width=0.45\textwidth]{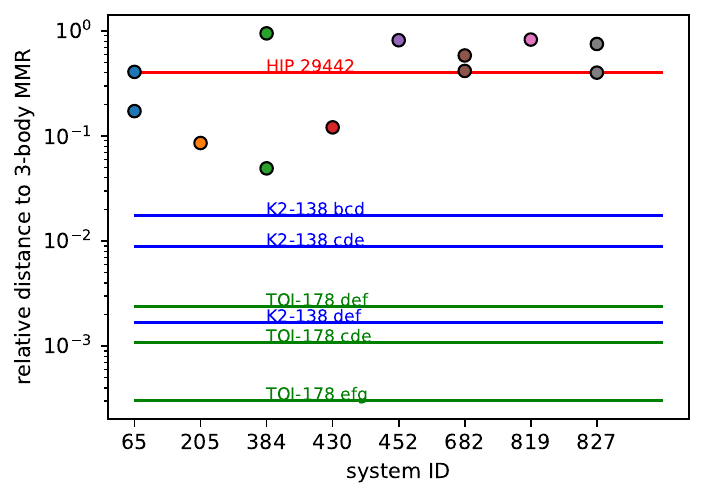}
    \caption{Resonance behaviour of the observed \starname\ system and the synthetic system analogues. The top panel shows the \starname \ system (red star) with respect to the web of three-body MMRs (in black) and two-body MMRs (in purple and red). The coloured circles show the position of the triplets of the analogue systems from the synthetic NGPPS population, the system ID corresponding to each colour is given on the x axis of the bottom panel. The bottom panel shows the same triplets with their relative distance to the nearest three-body MMR. The coloured lines show the distance to 3-body MMRs for the observed systems \starname, K2-138 and TOI-178.} 
    \label{fig:3bMMR}
\end{figure}

The upper panel of Figure \ref{fig:3bMMR} shows the position of the triplet of \starname \ and the selected NGPPS systems with respect to the web of mean motion resonances (MMRs). We denote $P_n$ as the orbital period of the n$^\mathrm{th}$ planet in the chain, counting from the innermost planet to the outermost. The 2-planet MMRs of the form $P_{n+1}/P_{n}=(k+q)/k$ (with $k$ and $q$ integers) are shown in purple, with $q=1$ for continuous lines and $q=2$ for dashed lines, the $P_{n+2}/P_{n}=2$ resonance is shown in red. Black lines show 3-body MMRs of the form $k/P_n - (k+q)/P_{n+1} + q/P_{n+2}=0$ with $k$ and $q$ integers in the 1 to 6 range. For architecture comparisons, the usage of 3-body MMRs is generally preferred, as the 2-planet MMR resonant state depends on the eccentricity of the planets, which is not always easily constrained in exoplanetary systems \citep{Cerioni+2022}. The first result of this graph concerns the orbital spacing between the planets of each system: The NGPPS systems have a spacing comparable to \starname, with most of the period ratios above the 3:2 2-planet MMR, which is the most common MMR in resonant chains \citep[e.g.][]{Lopez+2019,Leleu+2021,Luque+2023}.
The lower panel shows the relative distance of each triplet with respect to the closest 3-planet MMR, divided by the relative distance to the second-closest 3-planet MMR. This rescaling is used in order to account for the areas of the resonant web that are more densely populated. One can see that \starname, along with the NGPPS systems, are quite far from any 3-body resonance when compared to systems that are known to be in resonant chains, such as TOI-178 and K2-138 \citep{Lopez+2019,Leleu+2021,Delrez+2023}. 
If the \starname\ system initially formed as a resonant chain, it grew unstable during or after the protoplanetary disk phase \citep[e.g. ][]{Izidoro+2017}.

\subsection{Other architecture frameworks}
As discussed above, \cite{Emsenhuber+2023} classifies the architecture of synthetic planetary systems based on the planetary formation tracks. There are also other works that study the architecture of planetary systems, most of which are based on the directly observable, present-day properties of the planets in the system to be classified \citep[e.g.][]{Lissauer+2011,Fabrycky+2014,Millholland+2017,Weiss+2018,Mishra+2023a,Mishra+2023b,Davoult+2024}.

We will now apply one of these architecture frameworks, the one from \cite{Davoult+2024}, to the \starname\ system. Their framework is based on the four architecture classes introduced in \cite{Mishra+2023a,Mishra+2023b}, so each planetary system is classified as having a similar, mixed, ordered or anti-ordered architecture. \cite{Davoult+2024} base their architecture classification on a Principal Component Analysis of the planetary systems with respect to the logarithmic distance of the planets from the star and the logarithmic masses of the planets. In this framework, \starname\ is a system of similar architecture, which is characterised by \cite{Davoult+2024} as consisting of small planets, i.e. Earth-sized planets, super-Earths and sub-Neptunes. As the study points out, the formation of such systems has been investigated in detail by multiple authors, such as \cite{Izidoro+2017}, \cite{Goldberg+2022} and \cite{Batygin+2023}.

\section{Discussion}
\label{sec:discussion}

\subsection{Interior modelling sensitivity}
In Section~\ref{sec:internal_structure}, we have modelled the internal structure of all three planets in the \starname \ system, using the planetary radii and masses derived in Section~\ref{sec:data+analysis}. For each planet, we consider that it could have formed inside the iceline, assuming that water was only accreted as part of the accreted gas, as well as a formation scenario outside the iceline, where we assume that the planet also accreted water in solid form. We find that the mean density of all three planets can be explained with either of the two assumptions. In the case of a formation scenario inside the iceline, we find that \starname~b can be explained by an envelope mass fraction of a few percent, while the median values for the inferred envelope mass fractions for \starname~c and d lie at the order of $10^{-6}$ and would almost certainly have been lost through evaporation. For all three planets, the water mass fractions in the envelope are, by construction, very low in this scenario. Instead, if we assume that the planets formed outside the iceline and also accreted water in solid form, we find a wide range of possible envelope mass fractions up to almost 60\% for \starname~b, with water mass fractions in the envelope of up to almost 90\%. For \starname~c and d, we find envelope mass fractions of a few percent in this scenario, with envelopes that are almost entirely made up of water. In Section~\ref{sec:hydro_modelling}, we then used a hydrodynamic modelling approach to study the influence of our inferred atmospheric compositions, both for a water-rich and water-poor formation scenario, on the conditions in the upper atmospheres of such planets and the expected atmospheric mass loss rates. 

There are of course a number of limitations to this modelling approach. One aspect is that the calculated radius of a given planetary structure is still very model-dependent, especially for planets with larger envelope mass fractions. \cite{Haldemann+2024} studied in detail how the use of updated EoS influences the calculated transit radius of a planetary structure, including a comparison of the forward model of BICEPS with the commonly used isocomposition curves from \cite{Zeng+2019}, and found differences of a few percent for predominantly rocky exoplanets and of 10\% or larger for volatile-rich planets. \cite{Aguichine+2021} present mass-radius relations for water-rich planets and find steam envelopes that are significantly larger compared to the transit radii calculated with BICEPS, with differences of up to $\sim$20\% for planets with small masses and high equilibrium temperatures. These differences can be caused by discrepancies in the chosen irradiation model, opacities and definition of the transit radius. This is also discussed by \citeauthor{Venturini+2024} (\citeyear{Venturini+2024}, their Section C.2), who point out that the key difference between BICEPS and the models of \cite{Aguichine+2021} is that the latter assume the core-envelope boundary to be extremely hot based on evolution calculations accounting for the runaway greenhouse effect of water, an effect which more recent works point out might have been overestimated in the past \citep{Selsis+2023}. \cite{Nixon+2021} also study planetary structures where the water and H/He components of the planet's atmosphere are not mixed and find significantly different radii for the two cases. This is an additional source of degeneracy that we do not consider in our inverse modelling. However, in this case also the runaway greenhouse effect needs to be considered, as hydrogen and supercritical water behave close to an ideal mixture \citep{Soubiran+2015} and separate water and H/He layers would therefore not be stable under certain conditions \citep{Turbet+2019, Turbet+2020, Pierrehumbert2023, Innes+2023}.

A second important aspect is that we do not currently consider the geophysical evolution of each planet in our model. \cite{Dorn+Lichtenberg2021} show that large quantities of water can be stored in magma oceans of super-Earths and sub-Neptunes. Meanwhile, \cite{Vazan+2022} study the interiors of ice-rich planets and find that ice and rock are expected to stay mixed even after billions of years of thermal evolution. In both cases, the water mass fractions inferred in Section~\ref{sec:internal_structure} would not refer to the total water content of each planet, but to the amount of water not dissolved in the interior, as also pointed out by \cite{Haldemann+2024}. It is also possible that interactions between volatiles and a magma ocean produce additional, endogenic water \citep[e.g.][]{Kite+2020}, which makes the interpretation of a planet's water content even more complicated. 

\subsection{The \starname\ system in the context of the radius valley}

\begin{figure}
    \includegraphics[width=0.49\textwidth]{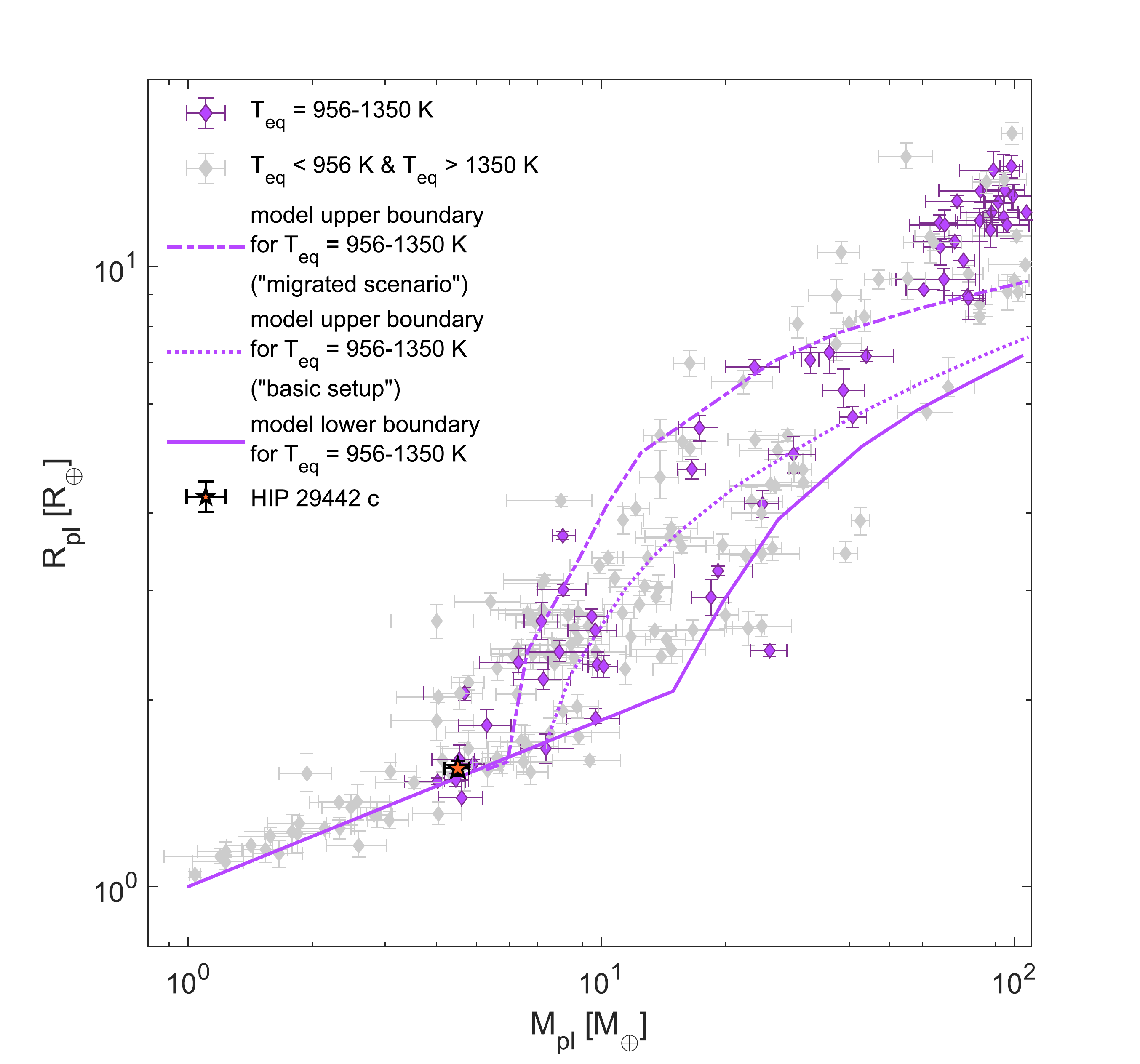}\\
    \includegraphics[width=0.49\textwidth]{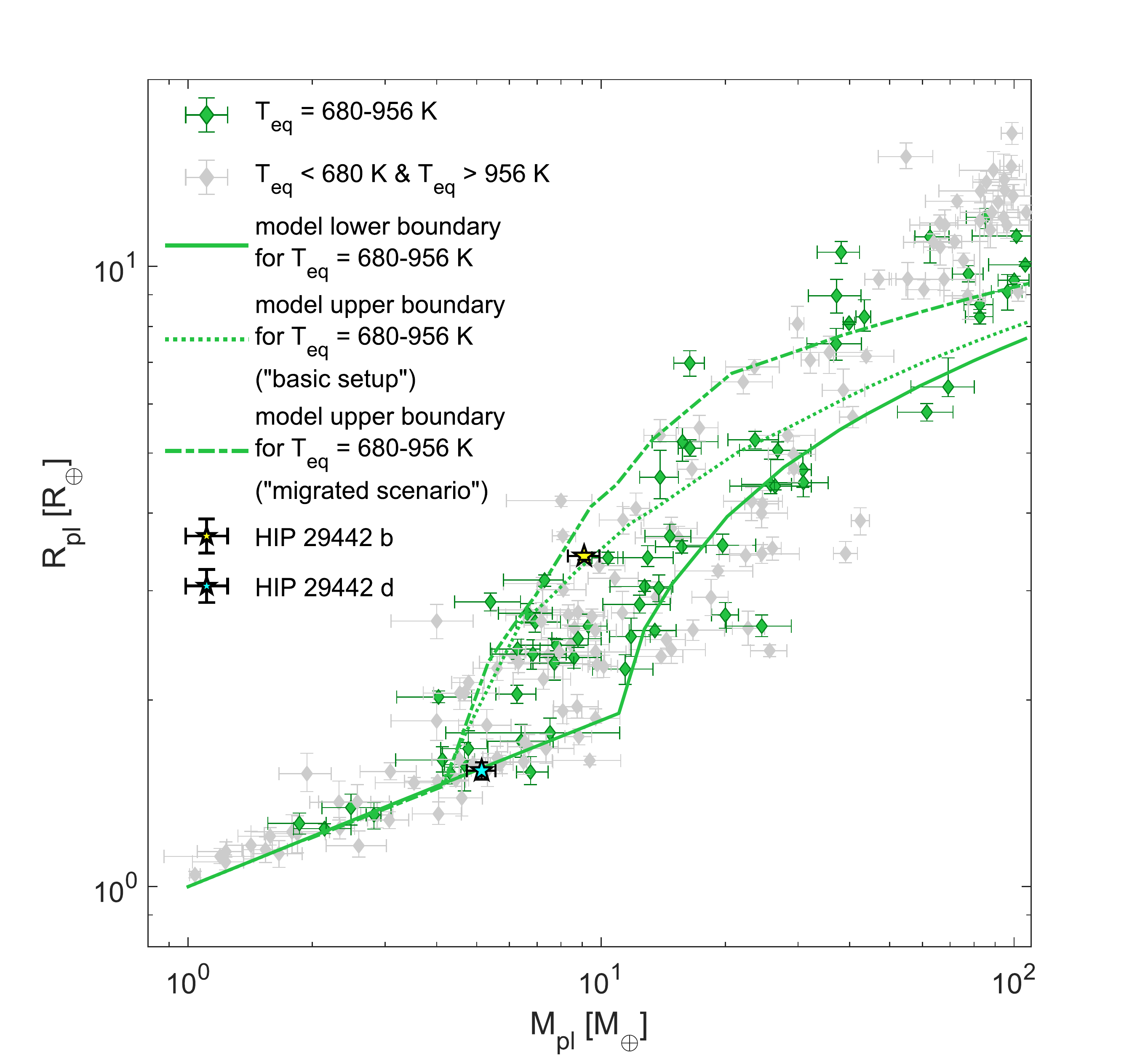}
    \caption{Location of \starname~c (upper panel) and \starname~b and d (lower panel) in the mass radius diagram, in relation to the expected radius spread at an age of 5~Gyr according to \cite{Kubyshkina+2022}, who ran a database of evolutionary tracks for planets with pure H/He envelopes. The solid lines show the lower model boundary, the short dashed lines (labelled as "basic setup" in the legend) the upper model boundary assuming initial H/He mass fractions calculated according to \cite{Mordasini2020} and the long dashed lines (labelled as "migrated scenario") assuming massive initial H/He mass fractions.}
    \label{fig:radius_valley_hhe_evaporation}
\end{figure}

As a next step, we explore how the \starname \ system fits in with two of the most popular explanations for the radius valley in the literature, on the one hand the loss of pure H/He envelopes either through photoevaporation \citep[e.g.][]{Owen+Wu2017,Jin+Mordasini2018} or core-powered mass loss \citep{Ginzburg+2018, Gupta+Schlichting2019} and on the other hand a scenario in line with coupled planet formation and evolution models, which predict water-rich sub-Neptunes \citep{Venturini+2020b,Burn+2024,Venturini+2024}. For the first scenario, we note that the present-day accuracy and completeness of observations remains insufficient to distinguish between photoevaporation and core-powered mass loss \citep{Rogers+2021,Gupta+2022}. For both mechanisms, we therefore consider the framework from \cite{Kubyshkina+2022}, which is a coupled interior-hydrodynamic atmospheric escape model. The underlying atmospheric mass loss model of this framework reproduces both the photoevaporation and the core-powered mass loss regime and predicts radius valley parameters for both regimes \citep{Affolter+2023A&A...676A.119A}.

Using this framework, \cite{Kubyshkina+2022} calculated model boundaries in the mass-radius space by combining a database of evolutionary tracks run up to 10~Gyr for planets with pure H/He envelopes with different masses, at different orbital separations and for 1~M$_\odot$ stars with different rotation histories. These boundaries then predict the expected radius spread of the exoplanet population. Figure~\ref{fig:radius_valley_hhe_evaporation} shows a mass-radius diagram with the observed planetary parameters for \starname~c in the upper and \starname~b and d in the lower panel, along with the model boundaries for the relevant temperature ranges corresponding to the equilibrium temperatures of the observed planets. Two different initial atmospheric mass fractions are considered for calculating the upper model boundaries (short and long dashed lines), based on the analytical fit from \cite{Mordasini2020} (`basic setup') and assuming massive initial envelopes (`migrated scenario'), respectively. In both cases, the initial atmospheric mass fraction depends on the planetary mass and, in the case of the basic setup, also on the orbital separation (hence, temperature). Therefore, the whole range of initial atmospheric mass fractions of the models in Figure~\ref{fig:radius_valley_hhe_evaporation} ranges between 0.5--70\% of the total planetary mass. For planets with parameters similar to the planets in the \starname \ system, they are $\sim$10\% or $\sim$26\% for planet~b, $\sim$3\% or $\sim$15\% for planet~c, and $\sim$5\% or $\sim$17\% for planet~d, with the larger values in each case corresponding to the migrated scenario.

We can see that the H/He models predict \starname~c to have lost any H/He envelope at an age of 5~Gyr independently of the size of its initial envelope and the rotation history of its host star. This is in agreement with the observed radius values and our internal structure modelling results for a prior consistent with a formation inside the iceline. 
\starname~d, in turn, lays at the edge of the radius valley, among the planets that can keep or lose their atmospheres upon the initial parameters and the irradiation history (near the fork made by the boundary lines in Figure~\ref{fig:radius_valley_hhe_evaporation}). Here, the upper model boundaries are given by the modelled planets at the lower edge of the temperature range and evolving around an inactive star, while the lower boundary corresponds to the higher edge of the temperature range and an active host star.
For the equilibrium temperature of planet~d, the planet would, in fact, need a mass of $\sim$6-7 M$_\oplus$ to keep a H/He envelope even around a slowly rotating star, which is $\sim$5--30\% higher than the mass we derived observationally. 
Meanwhile, \starname~b lies close to the model upper boundary for initial H/He mass fractions calculated using \cite{Mordasini2020} and therefore well within the expected radius spread. 
According to the evolution models by \cite{Kubyshkina+2022}, planets in this parameter range can lose $\sim$45--80\% of their initial H/He envelopes depending on their stellar rotation history, implying \starname~b had to start with an atmospheric mass fraction of $\sim$10--20\% if we consider the results of our internal structure models for a formation scenario inside the iceline. The observed radii of the \starname \ system are therefore in agreement with the interpretation of the radius valley based on the hydrodynamic escape of H/He envelopes.

However, if we consider the analysis of possible formation and evolution tracks we conducted in Section~\ref{sec:formation_evolution}, we do not find any synthetic planetary system analogues where the analogues for all three planets form inside the iceline. Of course, this depends on some fundamental assumptions underlying the model used to generate the synthetic population. Including different physical processes such as MHD-wind driven disks (where there could be less orbital migration than for the currently included turbulent disks) or pebble drift (which could bring a large amount of refractory solids into the interior part of the disk) could potentially allow to form the planets more or less in situ.

This brings us to the second scenario. In our analysis of possible formation and evolution tracks of the observed system, we identified a set of eight synthetic systems in the considered variation of the nominal NGPPS population that match the properties of the observed system well with respect to the masses and semi-major axes of the planets. From our tests, we found that the radii, water mass fractions, and H/He mass fractions of these synthetic planets match the ones derived in Sections \ref{sec:data+analysis} and \ref{sec:internal_structure} for the observed planets well for planets c and d, while the synthetic analogues for planet~b have radii that are too low and water mass fractions that are slightly too high compared to the values for the observed planet. 
As we have pointed out, this could potentially be explained by an overestimated H/He and an underestimated water loss rate in the model. The higher water mass fractions could also be explained by some of the water being dissolved in the interior of the planet, as the water mass fractions we inferred as part of our internal structure analysis only consider the water stored in the envelope of the planet. 
Furthermore, we have also tested the resonance behaviour of the observed and the synthetic systems and found them to be similar in that they are all quite far from any 3-body resonance.
When we consider these eight synthetic systems, all analogues for planet~b formed beyond the iceline, while the analogues for planets~c and d mostly formed inside the iceline with some that formed just outside it.

For the water-rich atmospheres inferred by our internal structure models using a prior consistent with a formation outside the iceline, a model for estimating the atmospheric loss self-consistently throughout the evolution does not currently exist. However, we can use the results obtained in Section~\ref{sec:hydro_modelling} to make some preliminary estimates. For present-day parameters with the inclusion of water into H/He atmospheres, we get a reduction in the mass loss by about an order of magnitude for planets~c and d and by about a factor two for planet~b. This implies that \starname~d, which already lies close to the borderline of evaporation for pure H/He atmospheres, would likely hold on to (some of) its atmosphere. 
In turn, \starname~c could lose $\sim$0.4--0.9~M$_{\oplus}$ of volatiles even in the case of water-rich atmospheres. This assumes the mass loss rates in the first Gyr of the evolution to be 10--30 times higher than at present, which takes into account higher XUV but ignores the possible planetary radius inflation at early ages. 
Finally, for \starname~b, we can consider that H/He is more easily lost than water, which means that the atmospheric water mass fraction will only increase during evolution. The present-day value for the reduction in the escape rates between H/He and water-rich atmospheres is therefore an upper limit. This means that we can expect the primordial atmosphere of \starname~b to be at least $\sim$30\% heavier than the present-day one. From these results we can therefore conclude that the observed properties of the \starname \ system are also in agreement with this second explanation for the bimodality of the radius distribution. 

Overall, we have shown that the observationally derived parameters of the planets in the \starname\ system can be explained on the one hand by the classical evaporation scenario of pure H/He envelopes, either through photoevaporation or core-powered mass loss, but on the other hand also by the water-rich sub-Neptune scenario predicted by coupled formation and evolution models. While it is well established that most planetary systems straddling the radius valley can be explained using the photoevaporation model \citep[e.g.][]{Owen+CamposEstrada2020}, this has not been investigated systematically for the water-rich sub-Neptune hypothesis so far, which makes our findings especially interesting. Although it is not possible to clearly distinguish between the two scenarios based on the currently available observations, our analysis provides a strong case to search for water in the atmospheres of the planets in the \starname\ system with JWST.

\section{Conclusions and summary}
\label{sec:conclusion}

In this study, we presented the results of an extensive photometric follow-up campaign of the three planets transiting the evolved K0 star \starname \ on 3.5 (planet~c), 6.4 (planet~d) and 13.6 (planet~b) day orbits with CHEOPS. We obtained a total of 17 visits with a combined observing time of 9.6 days and containing 20 transits, which, together with the data from TESS sectors 6 and 33, allowed us to derive precise radii of $3.410\pm0.046$, $1.551\pm0.045$ and $1.538\pm0.049$ R$_\oplus$ for \starname~b, c and d respectively. These radii are compatible within <1$\sigma$ with the values previously derived by \citet{Damasso+2023} for planets~b and c, while for planet~d, our value and error are $\sim$3.4$\sigma$ larger than their median value ($\sim$1.5$\sigma$ using their larger uncertainties). This changed the density of this middle planet from one compatible with a bare Mercury-like core to one that is approximately compatible with a bare Earth-like core instead, and lead us to conclude that it is advisable to be cautious when determining the radii of small planets from multi-year TESS photometry when there are significant gaps between observations and the per-transit S/N is low. We also re-ran the RV model from \cite{Damasso+2023} including not only their ESPRESSO RVs but also additional public Keck-HIRES data from \cite{AkanaMurphy+2023}. Our derived planetary masses of $9.10^{+0.82}_{-0.79}$, $4.50\pm0.32$ and $5.14\pm0.41$ M$_\oplus$ for \starname~b, c and d respectively are in good agreement with the values presented in \cite{Damasso+2023}. Overall, this makes \starname\ one of the most precisely characterised multiplanetary systems spanning the radius valley to date.

These very precise radii and masses then provided us with an ideal starting point for analysing the internal structure of the three planets and the formation history of the system, which is especially interesting as \starname\ is a multiplanet system spanning the radius valley. For this purpose, we introduced a new version of our neural network-based Bayesian internal structure modelling framework (\texttt{plaNETic}), which is publically available. We found that the mean densities of all three planets can be explained using compositional priors informed by both a water-poor and a water-rich formation scenario. From our models, we expect \starname~b to host a H/He dominated envelope of a few percent in mass with a very low envelope water mass fraction if it formed inside the iceline, while in case of the planet forming outside the iceline, a wide range of compositions is possible, with envelope mass fractions ranging up to almost 60\% and water mass fractions in the envelope of up to almost 90\%. For \starname~c and d, we found very low envelope mass fractions at the order of $10^{-6}$ that would almost certainly have evaporated in the water-poor formation scenario, and envelopes of up to a few percent in mass made up almost entirely of water for a water-rich formation outside the iceline. We therefore conclude that if planets~c and d do have an envelope, it cannot be pure H/He but needs to be of higher metallicity.

Further, we identified possible formation and evolution pathways of the system by identifying system analogues in the synthetic NGPPS population and found that according to the Bern model for planet formation and evolution, \starname~b likely formed at a distance between 3 and 6 AU outside the iceline and then migrated inward, while planets c and d likely formed inside or just barely outside the iceline. We also found that according to our analysis, a protoplanetary disk with a disk mass between 100 and 300 M$_\oplus$ and longer than average lifetime is necessary to form a system like \starname. 

Finally, we showed that the derived observational parameters of this system are compatible with both of the most popular explanations for the radius valley in the literature, sub-Neptunes with pure H/He envelopes undergoing atmospheric escape either through photoevaporation or core-powered mass loss or a scenario with water-rich sub-Neptunes as predicted by planet formation and evolution models.

\begin{acknowledgements}
    We thank the anonymous referee for valuable comments that helped improve the manuscript.
    We further thank Nicolas Kaufmann, Andrin Kessler, Jeanne Davoult, Jesse Weder and Remo Burn for productive discussions about the Bern model.
    CHEOPS is an ESA mission in partnership with Switzerland with important contributions to the payload and the ground segment from Austria, Belgium, France, Germany, Hungary, Italy, Portugal, Spain, Sweden, and the United Kingdom. The CHEOPS Consortium would like to gratefully acknowledge the support received by all the agencies, offices, universities, and industries involved. Their flexibility and willingness to explore new approaches were essential to the success of this mission. CHEOPS data analysed in this article will be made available in the CHEOPS mission archive (\url{https://cheops.unige.ch/archive_browser/}). 
    This work has been carried out within the framework of the NCCR PlanetS supported by the Swiss National Science Foundation under grants 51NF40\_182901 and 51NF40\_205606. 
    YAl acknowledges support from the Swiss National Science Foundation (SNSF) under grant 200020\_192038. 
    MNG is the ESA CHEOPS Project Scientist and Mission Representative, and as such also responsible for the Guest Observers (GO) Programme. MNG does not relay proprietary information between the GO and Guaranteed Time Observation (GTO) Programmes, and does not decide on the definition and target selection of the GTO Programme. 
    ML acknowledges support of the Swiss National Science Foundation under grant number PCEFP2\_194576. 
    ABr was supported by the SNSA. 
    TWi acknowledges support from the UKSA and the University of Warwick. 
    S.G.S. acknowledge support from FCT through FCT contract nr. CEECIND/00826/2018 and POPH/FSE (EC). 
    The Portuguese team thanks the Portuguese Space Agency for the provision of financial support in the framework of the PRODEX Programme of the European Space Agency (ESA) under contract number 4000142255. 
    J.H. acknowledges the support from the Swiss National Science Foundation under grant 200021\_204847 'PlanetsInTime'. 
    The Belgian participation to CHEOPS has been supported by the Belgian Federal Science Policy Office (BELSPO) in the framework of the PRODEX Program, and by the University of Liège through an ARC grant for Concerted Research Actions financed by the Wallonia-Brussels Federation. 
    L.D. thanks the Belgian Federal Science Policy Office (BELSPO) for the provision of financial support in the framework of the PRODEX Programme of the European Space Agency (ESA) under contract number 4000142531. 
    LBo, GBr, VNa, IPa, GPi, RRa, GSc, VSi, and TZi acknowledge support from CHEOPS ASI-INAF agreement n. 2019-29-HH.0. 
    TZi acknowledges NVIDIA Academic Hardware Grant Program for the use of the Titan V GPU card and the Italian MUR Departments of Excellence grant 2023-2027 “Quantum Frontiers”. 
    We acknowledge financial support from the Agencia Estatal de Investigación of the Ministerio de Ciencia e Innovación MCIN/AEI/10.13039/501100011033 and the ERDF “A way of making Europe” through projects PID2019-107061GB-C61, PID2019-107061GB-C66, PID2021-125627OB-C31, and PID2021-125627OB-C32, from the Centre of Excellence “Severo Ochoa” award to the Instituto de Astrofísica de Canarias (CEX2019-000920-S), from the Centre of Excellence “María de Maeztu” award to the Institut de Ciències de l’Espai (CEX2020-001058-M), and from the Generalitat de Catalunya/CERCA programme. 
    We acknowledge financial support from the Agencia Estatal de Investigación of the Ministerio de Ciencia e Innovación MCIN/AEI/10.13039/501100011033 and the ERDF “A way of making Europe” through projects PID2019-107061GB-C61, PID2019-107061GB-C66, PID2021-125627OB-C31, and PID2021-125627OB-C32, from the Centre of Excellence “Severo Ochoa'' award to the Instituto de Astrofísica de Canarias (CEX2019-000920-S), from the Centre of Excellence “María de Maeztu” award to the Institut de Ciències de l’Espai (CEX2020-001058-M), and from the Generalitat de Catalunya/CERCA programme. 
    S.C.C.B. acknowledges support from FCT through FCT contracts nr. IF/01312/2014/CP1215/CT0004. 
    C.B. acknowledges support from the Swiss Space Office through the ESA PRODEX program. 
    A.C.-G. is funded by the Spanish Ministry of Science through MCIN/AEI/10.13039/501100011033 grant PID2019-107061GB-C61. 
    ACC acknowledges support from STFC consolidated grant numbers ST/R000824/1 and ST/V000861/1, and UKSA grant number ST/R003203/1. 
    A.C., A.D., B.E., K.G., and J.K. acknowledge their role as ESA-appointed CHEOPS Science Team Members.
    P.E.C. is funded by the Austrian Science Fund (FWF) Erwin Schroedinger Fellowship, program J4595-N. 
    This project was supported by the CNES. 
    This work was supported by FCT - Funda\c{c}\~{a}o para a Ci\^{e}ncia e a Tecnologia through national funds and by FEDER through COMPETE2020 through the research grants UIDB/04434/2020, UIDP/04434/2020, 2022.06962.PTDC. 
    O.D.S.D. is supported in the form of work contract (DL 57/2016/CP1364/CT0004) funded by national funds through FCT. 
    B.-O. D. acknowledges support from the Swiss State Secretariat for Education, Research and Innovation (SERI) under contract number MB22.00046. 
    This project has received funding from the Swiss National Science Foundation for project 200021\_200726. It has also been carried out within the framework of the National Centre of Competence in Research PlanetS supported by the Swiss National Science Foundation under grant 51NF40\_205606. The authors acknowledge the financial support of the SNSF. 
    MF and CMP gratefully acknowledge the support of the Swedish National Space Agency (DNR 65/19, 174/18). 
    DG gratefully acknowledges financial support from the CRT foundation under Grant No. 2018.2323 “Gaseousor rocky? Unveiling the nature of small worlds”. 
    M.G. is an F.R.S.-FNRS Senior Research Associate. 
    CHe acknowledges support from the European Union H2020-MSCA-ITN-2019 under Grant Agreement no. 860470 (CHAMELEON). 
    J.K. acknowledges the Swedish Research Council (VR: Etableringsbidrag 2017-04945).
    K.W.F.L. was supported by Deutsche Forschungsgemeinschaft grants RA714/14-1 within the DFG Schwerpunkt SPP 1992, Exploring the Diversity of Extrasolar Planets. 
    This work was granted access to the HPC resources of MesoPSL financed by the Region Ile de France and the project Equip@Meso (reference ANR-10-EQPX-29-01) of the programme Investissements d'Avenir supervised by the Agence Nationale pour la Recherche. 
    PM acknowledges support from STFC research grant number ST/R000638/1. 
    This work was also partially supported by a grant from the Simons Foundation (PI Queloz, grant number 327127). 
    NCSa acknowledges funding by the European Union (ERC, FIERCE, 101052347). Views and opinions expressed are however those of the author(s) only and do not necessarily reflect those of the European Union or the European Research Council. Neither the European Union nor the granting authority can be held responsible for them. 
    A. S. acknowledges support from the Swiss Space Office through the ESA PRODEX program. 
    GyMSz acknowledges the support of the Hungarian National Research, Development and Innovation Office (NKFIH) grant K-125015, a a PRODEX Experiment Agreement No. 4000137122, the Lend\"ulet LP2018-7/2021 grant of the Hungarian Academy of Science and the support of the city of Szombathely. 
    V.V.G. is an F.R.S-FNRS Research Associate. 
    JV acknowledges support from the Swiss National Science Foundation (SNSF) under grant PZ00P2\_208945. 
    NAW acknowledges UKSA grant ST/R004838/1.
    AL acknowledge support of the Swiss National Science Foundation under grant number  TMSGI2\_211697.
    \\ \textit{Software.} The following software packages have been used for this publication: Python-numpy \citep{numpy}, Python-pandas \citep{pandas}, Python-matplotlib \citep{matplotlib}, Python-seaborn \citep{seaborn}, Python-tensorflow \citep{tensorflow}, Python-scipy \citep{scipy}.
\end{acknowledgements}

%
%

\bibliographystyle{aa} 
\bibliography{biblio} 

\onecolumn

\appendix

\section{Observations and data analysis}
\begin{figure}[H]
    \centering
    \includegraphics[width=\textwidth]{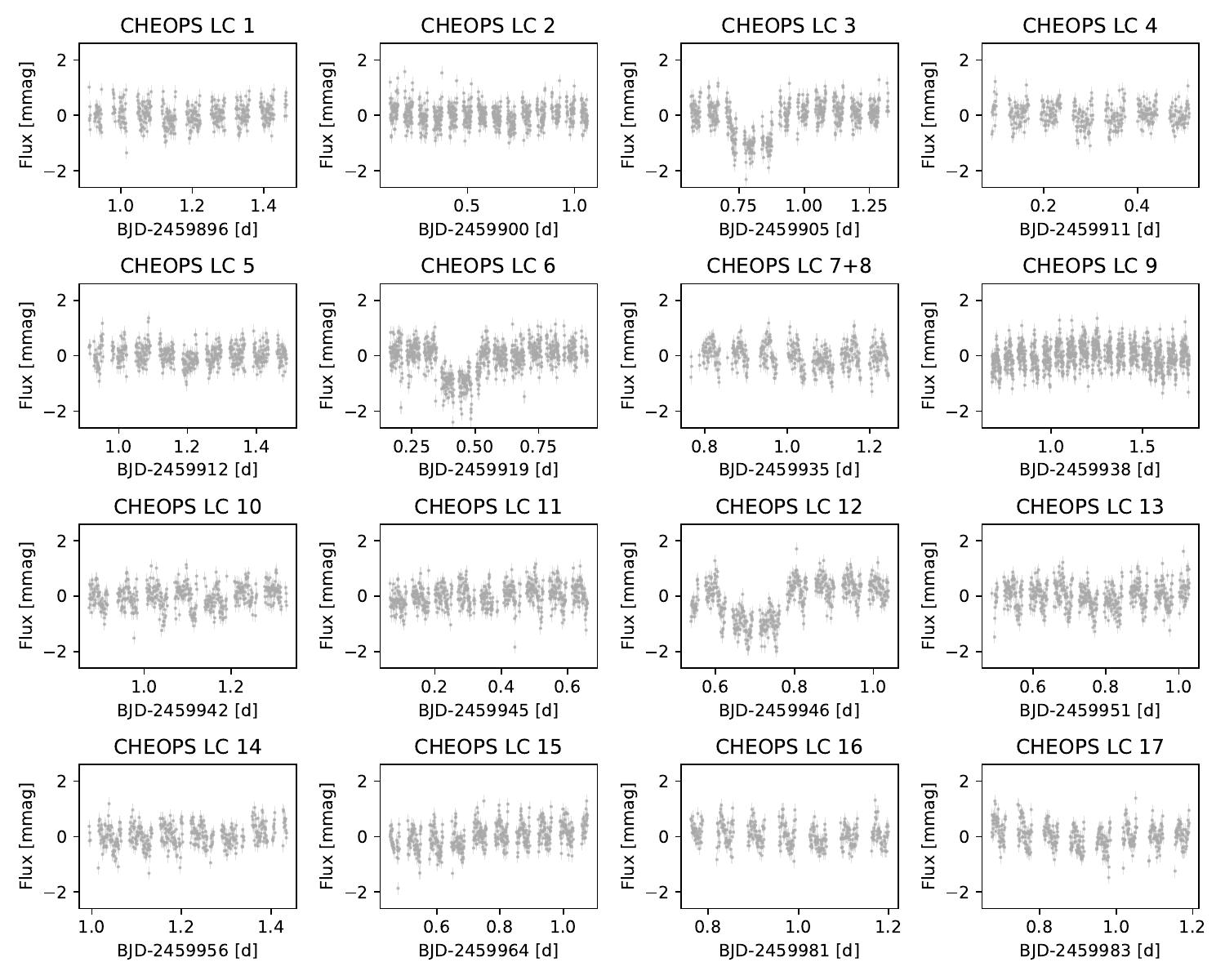}
    \caption{Undetrended CHEOPS light curves of \starname, identified by the IDs introduced in Table \ref{tab:file_keys}.}
    \label{fig:undetrended_CHEOPS}
\end{figure}

\section{Internal structure analysis}
\subsection{Chosen DNN architectures}
\begin{table}[H]
\renewcommand{\arraystretch}{1.5}
\caption{Network architectures of the best performing DNNs used as a forward model in the \texttt{plaNETic} code.}
\centering
\begin{tabular}{ll|rrrrrr}
\hline\hline
\multicolumn{2}{c|}{Model} & HL1 & HL2 & HL3 & HL4 & HL5 & HL6 \\
\hline
\multicolumn{2}{c|}{PREV} & 2048 & 2048 & 2048 & 2048 & 2048 & 2048 \\ 
\hline
A & M1 & 2048 & 2048 & 2048 & 2048 & 2048 & 2048 \\ 
A & M2 & 2048 & 2048 & 2048 & 2048 & 2048 & 2048 \\ 
\hline
B & M1 & 512 & 1024 & 2048 & 1024 & 512 & 256 \\ 
B & M2 & 2048 & 2048 & 2048 & 2048 & 2048 & 2048 \\ 
\hline
\end{tabular}
\label{tab:network_architectures}
\tablefoot{Network architectures are specified for both the old version of \texttt{plaNETic} (denoted as PREV) and the version that is newly introduced in this work. For the new version, separate DNNs were trained for different planetary mass ranges (M1 for planets from 0.5 to 6 M$_\oplus$, M2 for planets from 6 to 15 M$_\oplus$) and water prior options (A assuming a formation outside the iceline, B inside). The table shows the number of units in each of the hidden layers (HL) of the DNN in question.}
\end{table}
\renewcommand{\arraystretch}{1.0}

\subsection{DNN performance analysis}
\begin{table}[H]
\renewcommand{\arraystretch}{1.5}
\caption{Time needed to calculate the transit radii of 100 randomly sampled planetary structures from the posteriors of the internal structure models for \starname~b, c and d, both with the trained DNNs and the full planetary structure model from BICEPS.}
\centering
\begin{tabular}{c|ccc|ccc|ccc}
\hline\hline
\multirow{2}{*}{Model} & \multicolumn{3}{c|}{HIP 29442 b} & \multicolumn{3}{c|}{HIP 29442 c} & \multicolumn{3}{c}{HIP 29442 d} \\
& t$_\textrm{BICEPS}$ [s] & t$_\textrm{DNN}$ [s] & Speed-up & t$_\textrm{BICEPS}$ [s] & t$_\textrm{DNN}$ [s] & Speed-up & t$_\textrm{BICEPS}$ [s] & t$_\textrm{DNN}$ [s] & Speed-up \\
\hline
A1 & 1363 & 0.034 & 40088$\times$ & 1462 & 0.038 & 38474$\times$ & 1454 & 0.034 & 42706$\times$ \\
A2 & 1209 & 0.031 & 39000$\times$ & 1415 & 0.036 & 39306$\times$ & 1402 & 0.036 & 38944$\times$ \\
A3 & 1350 & 0.032 & 42188$\times$ & 1372 & 0.035 & 39200$\times$ & 1341 & 0.035 & 38314$\times$ \\
\hline
B1 & 2414 & 0.036 & 67056$\times$ & 2405 & 0.035 & 68714$\times$ & 2403 & 0.038 & 63237$\times$ \\
B2 & 2070 & 0.035 & 59143$\times$ & 2085 & 0.036 & 57917$\times$ & 2079 & 0.038 & 54711$\times$ \\
B3 & 2329 & 0.036 & 64694$\times$ & 2295 & 0.035 & 65571$\times$ & 2295 & 0.034 & 67500$\times$ \\
\hline
\hline
\end{tabular}
\label{tab:speedup_DNN}
\end{table}
\renewcommand{\arraystretch}{1.0}

\subsection{Internal structure modelling posteriors}
\begin{table}[H]
\renewcommand{\arraystretch}{1.5}
\caption{Median and one sigma errors for the posterior distributions of the internal structure modelling for \starname \ b using the new version of \texttt{plaNETic} described in Section \ref{sec:internal_structure -- new_model}.}
\centering
\begin{tabular}{r|ccc|ccc}
\hline \hline
Water prior &              \multicolumn{3}{c|}{Formation outside iceline (water-rich)} & \multicolumn{3}{c}{Formation inside iceline (water-poor)} \\
Si/Mg/Fe prior &           Stellar (A1) &       Iron-enriched (A2) &      Free (A3) &
                           Stellar (B1) &       Iron-enriched (B2) &      Free (B3) \\
\hline
w$_\textrm{core}$ [\%] &        $11.8_{-8.1}^{+9.2}$ &    $16.0_{-11.2}^{+15.2}$ &    $13.1_{-9.5}^{+16.0}$ &
                           $17.6_{-11.5}^{+11.8}$ &    $22.2_{-15.4}^{+19.2}$ &    $17.6_{-12.7}^{+20.8}$ \\
w$_\textrm{mantle}$ [\%] &      $58.4_{-14.8}^{+16.2}$ &    $51.4_{-15.6}^{+19.1}$ &    $54.4_{-16.8}^{+19.4}$ &
                           $76.6_{-11.9}^{+11.5}$ &    $71.6_{-19.5}^{+15.5}$ &    $76.5_{-21.3}^{+13.1}$ \\
w$_\textrm{envelope}$ [\%] &    $28.3_{-17.0}^{+17.5}$ &    $29.2_{-17.1}^{+17.2}$ &    $28.5_{-17.0}^{+17.4}$ &
                           $5.9_{-0.3}^{+0.3}$ &    $6.3_{-0.7}^{+0.5}$ &    $6.0_{-0.9}^{+0.9}$ \\
\hline
Z$_\textrm{envelope}$ [\%] &        $65.0_{-25.6}^{+10.5}$ &    $64.7_{-26.1}^{+10.7}$ &    $65.6_{-25.7}^{+10.5}$ &
                           $0.5_{-0.2}^{+0.3}$ &    $0.5_{-0.2}^{+0.2}$ &    $0.5_{-0.2}^{+0.2}$ \\
\hline
x$_\textrm{Fe,core}$ [\%] &     $90.4_{-6.4}^{+6.5}$ &    $90.4_{-6.4}^{+6.5}$ &    $90.5_{-6.5}^{+6.5}$ &
                           $90.2_{-6.1}^{+7.0}$ &    $90.4_{-6.4}^{+6.5}$ &    $90.4_{-6.4}^{+6.5}$ \\
x$_\textrm{S,core}$ [\%] &      $9.6_{-6.5}^{+6.4}$ &    $9.6_{-6.5}^{+6.4}$ &    $9.5_{-6.5}^{+6.5}$ &
                           $9.8_{-7.0}^{+6.1}$ &    $9.6_{-6.5}^{+6.4}$ &    $9.6_{-6.5}^{+6.4}$ \\
\hline
x$_\textrm{Si,mantle}$ [\%] &   $36.7_{-5.7}^{+6.5}$ &    $32.7_{-8.8}^{+9.2}$ &    $33.0_{-23.0}^{+29.6}$ &
                           $35.5_{-6.5}^{+7.2}$ &    $32.7_{-8.8}^{+9.2}$ &     $33.6_{-23.3}^{+29.5}$ \\
x$_\textrm{Mg,mantle}$ [\%] &   $44.9_{-6.4}^{+7.5}$ &    $38.5_{-10.3}^{+10.6}$ &    $36.3_{-24.4}^{+30.6}$ &
                           $45.1_{-7.4}^{+8.1}$ &    $38.6_{-10.3}^{+10.6}$ &    $36.1_{-24.5}^{+30.1}$ \\
x$_\textrm{Fe,mantle}$ [\%] &   $18.4_{-11.7}^{+9.2}$ &    $28.0_{-18.4}^{+18.5}$ &    $22.0_{-16.1}^{+24.2}$ &
                           $19.5_{-12.7}^{+9.4}$ &    $28.0_{-18.4}^{+18.6}$ &    $21.7_{-15.7}^{+24.0}$ \\
\hline
\end{tabular}
\label{tab:internal_structure_results_b}
\end{table}
\renewcommand{\arraystretch}{1.0}

\begin{table}[H]
\renewcommand{\arraystretch}{1.5}
\caption{Median and one sigma errors for the posterior distributions of the internal structure modelling for \starname \ c using the new version of \texttt{plaNETic} described in Section \ref{sec:internal_structure -- new_model}.}
\centering
\begin{tabular}{r|ccc|ccc}
\hline \hline
Water prior &              \multicolumn{3}{c|}{Formation outside iceline (water-rich)} & \multicolumn{3}{c}{Formation inside iceline (water-poor)} \\
Si/Mg/Fe prior &           Stellar (A1) &       Iron-enriched (A2) &      Free (A3) &
                           Stellar (B1) &       Iron-enriched (B2) &      Free (B3) \\
\hline
w$_\textrm{core}$ [\%] &        $18.7_{-12.3}^{+11.1}$ &    $29.2_{-19.5}^{+19.8}$ &    $31.4_{-21.2}^{+21.5}$ &
                           $20.4_{-12.9}^{+10.6}$ &    $35.4_{-22.5}^{+19.1}$ &    $39.2_{-25.6}^{+20.9}$ \\
w$_\textrm{mantle}$ [\%] &      $80.3_{-11.2}^{+12.5}$ &    $68.2_{-20.3}^{+19.5}$ &    $65.4_{-22.2}^{+21.3}$ &
                           $75.4_{-10.9}^{+14.3}$ &    $64.6_{-19.1}^{+22.5}$ &    $60.7_{-20.9}^{+25.6}$ \\
w$_\textrm{envelope}$ [\%] &    $0.6_{-0.4}^{+1.2}$ &    $2.3_{-1.5}^{+2.5}$ &    $2.9_{-2.0}^{+3.1}$ &
                           $\left(1.3_{-0.2}^{+0.3}\right)$ $10^{-4}$ &    $\left(1.7_{-0.5}^{+1.5}\right)$ $10^{-4}$ &    $\left(2.1_{-0.9}^{+2.8}\right)$ $10^{-4}$ \\
\hline
Z$_\textrm{envelope}$ [\%] &        $99.8_{-1.3}^{+0.2}$ &    $99.9_{-1.4}^{+0.1}$ &    $99.9_{-1.4}^{+0.1}$ &
                           $0.5_{-0.2}^{+0.3}$ &    $0.5_{-0.2}^{+0.2}$ &    $0.5_{-0.2}^{+0.2}$ \\
\hline
x$_\textrm{Fe,core}$ [\%] &     $90.3_{-6.3}^{+6.7}$ &    $90.6_{-6.6}^{+6.4}$ &    $90.5_{-6.5}^{+6.4}$ &
                           $90.0_{-6.6}^{+6.9}$ &    $90.3_{-6.4}^{+6.5}$ &    $90.4_{-6.4}^{+6.5}$ \\
x$_\textrm{S,core}$ [\%] &      $9.7_{-6.7}^{+6.3}$ &    $9.4_{-6.4}^{+6.6}$ &    $9.5_{-6.4}^{+6.5}$ &
                           $10.0_{-6.9}^{+6.6}$ &    $9.7_{-6.5}^{+6.4}$ &    $9.6_{-6.5}^{+6.4}$ \\
\hline
x$_\textrm{Si,mantle}$ [\%] &   $37.1_{-5.8}^{+6.8}$ &    $29.7_{-8.0}^{+10.5}$ &    $22.5_{-15.7}^{+25.4}$ &
                           $36.4_{-6.3}^{+6.3}$ &    $28.8_{-8.0}^{+10.4}$ &    $20.0_{-14.2}^{+22.9}$ \\
x$_\textrm{Mg,mantle}$ [\%] &   $45.6_{-6.4}^{+7.1}$ &    $35.1_{-9.6}^{+11.9}$ &    $33.8_{-19.7}^{+23.2}$ &
                           $45.9_{-7.2}^{+8.5}$ &    $34.0_{-9.5}^{+12.1}$ &    $33.9_{-18.7}^{+24.6}$ \\
x$_\textrm{Fe,mantle}$ [\%] &   $16.8_{-10.9}^{+9.9}$ &    $35.0_{-21.9}^{+16.9}$ &    $38.8_{-24.2}^{+20.2}$ &
                           $16.8_{-9.8}^{+10.9}$ &    $36.8_{-21.9}^{+17.0}$ &    $40.7_{-25.4}^{+20.1}$ \\
\hline
\end{tabular}
\label{tab:internal_structure_results_c}
\end{table}
\renewcommand{\arraystretch}{1.0}

\begin{table}[H]
\renewcommand{\arraystretch}{1.5}
\caption{Median and one sigma errors for the posterior distributions of the internal structure modelling for \starname \ d using the new version of \texttt{plaNETic} described in Section \ref{sec:internal_structure -- new_model}.}
\centering
\begin{tabular}{r|ccc|ccc}
\hline \hline
Water prior &              \multicolumn{3}{c|}{Formation outside iceline (water-rich)} & \multicolumn{3}{c}{Formation inside iceline (water-poor)} \\
Si/Mg/Fe prior &           Stellar (A1) &       Iron-enriched (A2) &      Free (A3) &
                           Stellar (B1) &       Iron-enriched (B2) &      Free (B3) \\
\hline
w$_\textrm{core}$ [\%] &        $18.7_{-12.2}^{+11.0}$ &    $31.4_{-20.9}^{+19.9}$ &    $33.9_{-22.8}^{+22.0}$ &
                           $21.9_{-12.5}^{+10.6}$ &    $37.4_{-23.4}^{+18.5}$ &    $41.3_{-26.6}^{+20.7}$ \\
w$_\textrm{mantle}$ [\%] &      $80.6_{-11.2}^{+12.3}$ &    $66.7_{-20.1}^{+20.9}$ &    $63.5_{-22.5}^{+22.8}$ &
                           $78.1_{-10.6}^{+12.5}$ &    $62.6_{-18.5}^{+23.4}$ &    $58.7_{-20.7}^{+26.6}$ \\
w$_\textrm{envelope}$ [\%] &    $0.5_{-0.3}^{+0.9}$ &    $1.6_{-1.1}^{+2.1}$ &    $2.1_{-1.5}^{+2.7}$ &
                           $\left(1.2_{-0.2}^{+0.5}\right)$ $10^{-4}$ &    $\left(1.8_{-0.6}^{+2.1}\right)$ $10^{-4}$ &    $\left(2.3_{-1.1}^{+4.2}\right)$ $10^{-4}$ \\
\hline
Z$_\textrm{envelope}$ [\%] &        $99.8_{-1.3}^{+0.2}$ &    $99.8_{-1.5}^{+0.2}$ &    $99.8_{-1.5}^{+0.1}$ &
                           $0.5_{-0.3}^{+0.2}$ &    $0.5_{-0.3}^{+0.2}$ &    $0.5_{-0.2}^{+0.2}$  \\
\hline
x$_\textrm{Fe,core}$ [\%] &     $90.4_{-6.4}^{+6.5}$ &    $90.6_{-6.5}^{+6.4}$ &    $90.5_{-6.5}^{+6.5}$ &
                           $89.1_{-5.5}^{+6.7}$ &    $90.4_{-6.3}^{+6.4}$ &    $90.4_{-6.4}^{+6.4}$ \\
x$_\textrm{S,core}$ [\%] &      $9.6_{-6.5}^{+6.4}$ &    $9.4_{-6.4}^{+6.5}$ &    $9.5_{-6.5}^{+6.5}$ &
                           $10.9_{-6.7}^{+5.5}$ &    $9.6_{-6.4}^{+6.3}$ &    $9.6_{-6.4}^{+6.4}$ \\
\hline
x$_\textrm{Si,mantle}$ [\%] &   $37.0_{-5.7}^{+7.0}$ &    $29.1_{-8.0}^{+10.7}$ &    $20.2_{-14.3}^{+23.6}$ &
                           $36.6_{-6.7}^{+7.3}$ &    $28.9_{-8.3}^{+10.5}$ &    $18.5_{-13.4}^{+22.2}$ \\
x$_\textrm{Mg,mantle}$ [\%] &   $45.5_{-6.2}^{+7.0}$ &    $34.4_{-9.6}^{+12.2}$ &    $33.9_{-18.8}^{+22.5}$ &
                           $46.5_{-7.4}^{+7.8}$ &    $34.1_{-9.7}^{+12.2}$ &    $34.5_{-18.4}^{+24.6}$ \\
x$_\textrm{Fe,mantle}$ [\%] &   $16.9_{-10.7}^{+9.7}$ &    $36.3_{-22.7}^{+16.9}$ &    $41.7_{-25.8}^{+18.9}$ &
                           $16.2_{-9.9}^{+10.4}$ &    $36.6_{-22.1}^{+17.6}$ &    $41.7_{-26.0}^{+19.8}$ \\
\hline
\end{tabular}
\label{tab:internal_structure_results_d}
\end{table}
\renewcommand{\arraystretch}{1.0}

\begin{table}[H]
\renewcommand{\arraystretch}{1.5}
\caption{Median and one sigma errors for the posterior distributions of the internal structure modelling for \starname \ b, c and d using the previous version of our internal structure model described in Section \ref{sec:internal_structure -- old_model}.}
\centering
\begin{tabular}{r|ccc}
\hline \hline
& \starname\ b & \starname\ c & \starname\ d \\
\hline
w$_\textrm{core}$ [\%] &        $13.8_{-8.11}^{+8.22}$ &    $17.7_{-10.3}^{+9.5}$ &    $17.9_{-10.3}^{+9.6}$ \\
w$_\textrm{mantle}$ [\%] &      $59.4_{-13.4}^{+14.3}$ &    $75.5_{-9.4}^{+10.9}$ &    $78.3_{-9.7}^{+10.6}$ \\
w$_\textrm{water}$ [\%] &       $22.4_{-15.6}^{+17.3}$ &    $5.7_{-3.8}^{+5.6}$ &    $2.9_{-2.1}^{+3.8}$ \\
w$_\textrm{H/He}$ [\%] &        $3.0_{-0.6}^{+0.8}$ &    $\left(1.7_{-1.5}^{+10.7}\right)$ $10^{-7}$ &    $\left(1.3_{-1.1}^{+7.0}\right)$ $10^{-7}$ \\
\hline
x$_\textrm{Fe,core}$ [\%] &     $90.3_{-6.3}^{+6.6}$ &    $90.3_{-6.4}^{+6.5}$ &    $90.4_{-6.4}^{+6.5}$ \\
x$_\textrm{S,core}$ [\%] &      $9.7_{-6.6}^{+6.3}$ &    $9.7_{-6.5}^{+6.4}$ &    $9.6_{-6.5}^{+6.4}$ \\
\hline
x$_\textrm{Si,mantle}$ [\%] &   $39.4_{-4.0}^{+5.3}$ &    $39.2_{-3.9}^{+5.2}$ &    $39.0_{-3.8}^{+5.1}$ \\
x$_\textrm{Mg,mantle}$ [\%] &   $44.9_{-6.9}^{+7.3}$ &    $44.7_{-6.7}^{+7.2}$ &    $44.5_{-6.6}^{+7.3}$ \\
x$_\textrm{Fe,mantle}$ [\%] &   $15.0_{-9.6}^{+9.3}$ &    $15.5_{-9.6}^{+9.0}$ &    $15.9_{-9.5}^{+8.6}$ \\
\hline
\end{tabular}
\label{tab:internal_structure_results_old}
\tablefoot{Note that all mass fractions are in relation to the total planet mass, including also the H/He envelope, to be more comparable to the results generated with the new version of \texttt{plaNETic}. However, this means that these results are not comparable with the ones reported in our own previous work where we employed the same model.}
\end{table}
\renewcommand{\arraystretch}{1.0}

\end{document}